\documentclass[10pt,preprint]{aastex}
\newenvironment{deluxetable*}{\begin{deluxetable}}{\end{deluxetable}}

\slugcomment{\textit{Astrophysical Journal} \textbf{713}:1-21: 2010}

\shorttitle{energy and helicity budgets of the flux injection
  hypothesis} \shortauthors{P. W. Schuck}

\usepackage{color}
\usepackage{eqnref}

\tablewidth{0pt}
\newcommand{\NR}{500}
\renewcommand{\phs}{\phm{-}}
\newcommand{\thetaCCW}{\psi}
\newcommand{\vsun}{617}
\newcommand{\zero}{\alpha}
\newcommand{\grad}{\mbox{\boldmath{$\nabla$}}}
\newcommand{\cross}{\mbox{\boldmath{$\times$}}}
\newcommand{\J}{\mbox{\boldmath{$J$}}}
\newcommand{\B}{\mbox{\boldmath{$B$}}}
\newcommand{\A}{\mbox{\boldmath{$A$}}}
\newcommand{\constraint}{{\mathcal{C}_1}}
\newcommand{\CONSTRAINT}{{\mathcal{C}_2}}
\newcommand{\Ap}{\mbox{\boldmath{$A$}}_\mathrm{R}}
\newcommand{\Bp}{\mbox{\boldmath{$B$}}_\mathrm{R}}
\newcommand{\BesselJ}{\mathrm{J}}
\newcommand{\E}{\mbox{\boldmath{$E$}}}
\newcommand{\vm}{\mbox{\boldmath{$v$}}}

\newcommand{\zhat}{\smash{\widehat{\zeta}}}

\newcommand{\Nd}{N}
\newcommand{\Np}{P}

\newcommand{\reality}{v}

\newcommand{\horder}{m}
\newcommand{\model}{\widehat{h}}
\newcommand{\veta}{\vectorfont{\eta}}
\newcommand{\neta}{\eta}

\newcommand{\vectorfont}[1]{\mbox{\boldmath{$#1$}}}

\newcommand{\rsun}{R_\sun}
\newcommand{\Rsun}{{R}_\sun}
\newcommand{\Dsun}{{D}_\sun}
\newcommand{\nhat}{\widehat{n}}
\newcommand{\TIME}{t}
\newcommand{\localz}{\zeta}
\newcommand{\localzhat}{\widehat{\zeta}}
\newcommand{\toroidal}{\phi}
\newcommand{\poloidal}{\theta}
\newcommand{\ac}{a_\mathrm{c}}
\newcommand{\rc}{r_\mathrm{c}}
\newcommand{\af}{a_\mathrm{p}}
\newcommand{\aA}{a_{*}}
\newcommand{\Height}{H}
\newcommand{\Helicity}{K}
\newcommand{\arclength}{\Theta}
\newcommand{\twist}{\vartheta}
\newcommand{\Lat}{\Theta}
\newcommand{\Lon}{\Phi}
\newcommand{\Major}{\mathcal{R}}
\newcommand{\Sf}{S_\mathrm{f}}
\renewcommand{\sf}{s_\mathrm{f}}
\newcommand{\eden}{\mathcal{U}}
\newcommand{\dindex}{i}
\newcommand{\parm}{\lambda}
\newcommand{\degree}{d}
\newcommand{\Hp}{240\,\mathrm{km}}

\newcommand{\Pph}{1.82\times10^{-4}\,\mathrm{g}\,\mathrm{cm}^{-1}\,\mathrm{s}^{-2}}
\newcommand{\Cs}{7.2\,\mathrm{km}\,\mbox{s}^{-1}}
\newcommand{\Rhoph}{5.85\times10^{-8}\,\mathrm{g}/\mathrm{cm}^3}

\newcommand{\BTp}{500}
\newcommand{\BTc}{4}
\newcommand{\AP}{6.7\times10^{8}}
\newcommand{\AC}{7.5\times10^{9}}

\begin{document}


\title{THE PHOTOSPHERIC ENERGY AND HELICITY BUDGETS OF THE FLUX-INJECTION HYPOTHESIS}


\author{P. W. Schuck\altaffilmark{\dag}}
\affil{NASA Goddard Space Flight Center}
\affil{Room 250, Building 21
Space Weather Laboratory, Code 674
Heliophysics Science Division
8801 Greenbelt Rd.
Greenbelt, MD 20771, USA}
\altaffiltext{\dag}{peter.schuck@nasa.gov} 



\begin{abstract}

The flux-injection hypothesis for driving coronal mass ejections (CMEs)
requires the transport of substantial magnetic energy and helicity flux
through the photosphere concomitant with the eruption. Under the
magnetohydrodynamics approximation, these fluxes are produced by twisting
magnetic field and/or flux emergence in the photosphere.  A CME trajectory,
observed 2000 September 12 and fitted with a flux-rope model constrains energy
and helicity budgets for testing the flux-injection hypothesis. Optimal
velocity profiles for several driving scenarios are estimated by minimizing
the photospheric plasma velocities for a cylindrically symmetric flux-rope
magnetic field subject to the flux budgets required by the flux-rope
model. Ideal flux injection, involving only flux emergence, requires
hypersonic upflows in excess of the solar escape velocity
$617\,\mbox{km}\mbox{s}^{-1}$ over an area of $6\times10^8\,\mathrm{km}^2$ to
satisfy the energy and helicity budgets of the flux-rope model. These
estimates are compared with magnetic field and Doppler measurements from
\textit{Solar Heliospheric Observatory}/Michelson Doppler Imager on 2000
September 12 at the footpoints of the CME. The observed Doppler
signatures are insufficient to account for the required energy and helicity
budgets of the flux-injection hypothesis.

\end{abstract}




\section{INTRODUCTION}
Over the last 10 years, the flux-rope model developed by
\cite{Chen1989,Chen1996} has been used to describe the dynamics of coronal
mass ejections (CMEs) observed by the Large Angle Spectrometric Coronagraphs
(LASCO) aboard the \textit{Solar and Heliospheric
  Observatory}\footnote{\textit{SOHO} is a project of international
  cooperation between ESA and NASA.} (\textit{SOHO})
\cite[]{Chen1997a,Chen2000a,Wood1999,Krall2001,Chen2006,Krall2006a}. However,
the photospheric flux injection paradigm used to initiate and drive the
eruption has been criticized because the surge of electromagnetic energy flowing
through the photosphere is ``difficult to reconcile with the extremely
tranquil conditions that exist during flares and CMEs''
\cite[]{Forbes2000,Forbes2001,Rust2001}. \cite{Chen2000a}, \cite{Chen2001}, \cite{Krall2001},
\cite{Chen2003} and \cite{Chen2010} have attempted to address these criticisms
and a previous study has examined the implications of uniformly twisting to
coronal footpoints of the flux rope \cite[]{Krall2000}.  However, no
quantitative comparisons between flux-injection hypothesis and detailed
photospheric observations have been considered by the formal literature. \par
The goal of this paper is to provide a framework for testing the
flux-injection hypothesis through the photospheric signatures implied by the
energy and helicity budgets of CMEs described by the flux-rope model. The
paper is organized as follows: Section~\ref{sec:model} describes the flux-rope
model of \cite{Chen1989} and a simple extension of the flux-rope model
magnetic field into the photosphere. \par
Section~\ref{sec:energy} develops the photospheric fluxes necessary to satisfy
the energy and helicity budgets of CMEs fitted with the flux-rope model.  The
photospheric magnetic field combined with the photospheric fluxes is used to
estimate minimum velocities necessary to satisfy the energy and helicity
budgets required by CME trajectories fitted by the flux-rope model under the
flux-injection hypothesis. Two examples of CME trajectories fitted with the
flux-rope model are used to constrain the photospheric velocities. 1) The
first event is the 2000 September 12 CME that erupted from decaying NOAA
active region 9163. The height-time data for the CME trajectory
\cite[]{Chen2006} were derived from measurements of the filament in absorption
observed by the Global H$\alpha$ Network at Kanzeh\"ohe Solar Observatory
(KSO) in Austria \cite[]{Steinegger2000}, and LASCO C2 and C3 observations of
the filament in emission \cite[]{Brueckner1995}. Complementary observations of
the filament were made in \ion{Fe}{12} 195 \AA~by the \textit{SOHO}/EUV
Imaging Telescope (EIT) instrument \cite[]{Delaboudiniere1995}. The eruption
occurred shortly after 11:30~UT and the filament first appeared in LASCO C2 at
12:30~UT. This CME was associated with an M1.0 class flare with distinct flare
ribbons that persisted for 2 hr.  Various aspects of this event, such as
morphology, timing, and reconnection rate have been discussed by
\cite{Vrsnak2003}, \cite{Schuck2004b} and \cite{Qui2004}. 2) The second event
is the 2003 October 28 CME which originated from large complex NOAA active
region 10468 at about 11:00 UT. The event was extremely fast, and thus the
height-time data for the leading edge of the CME consists of only one LASCO C2
image at about 11:30~UT and four subsequent LASCO C3 images. This event was
associated with an extremely powerful X17 flare and consequently has received
extensive attention in the literature
\cite[]{Seppala2004,Skoug2004,Woods2004,Zurbuchen2004,Bieber2005,Chi2005,Degenstein2005,Hu2005,Gopalswamy2005,Looper2005,Pallamraju2005,Tsurutani2005,Yurchyshyn2005,Krall2006a,Manchester2008}. \par
Section~\ref{sec:observations} compares the photospheric velocities implied by
the flux-rope CME trajectory event from 2000 September 12, against detailed
photospheric Doppler measurements from the Michelson Doppler Imager (MDI)
aboard \textit{SOHO} \cite[]{Scherrer1995}. Finally,
Section~\ref{sec:conclusions} compares these theoretical results and
Doppler observations with previous work. \par
\section{THE FLUX-ROPE MODEL\label{sec:model}}
\begin{figure*}[!t] 
\ifx\xfig\undefined
\centerline{\includegraphics[viewport=150 436 459 735,width=4.0in,clip=]{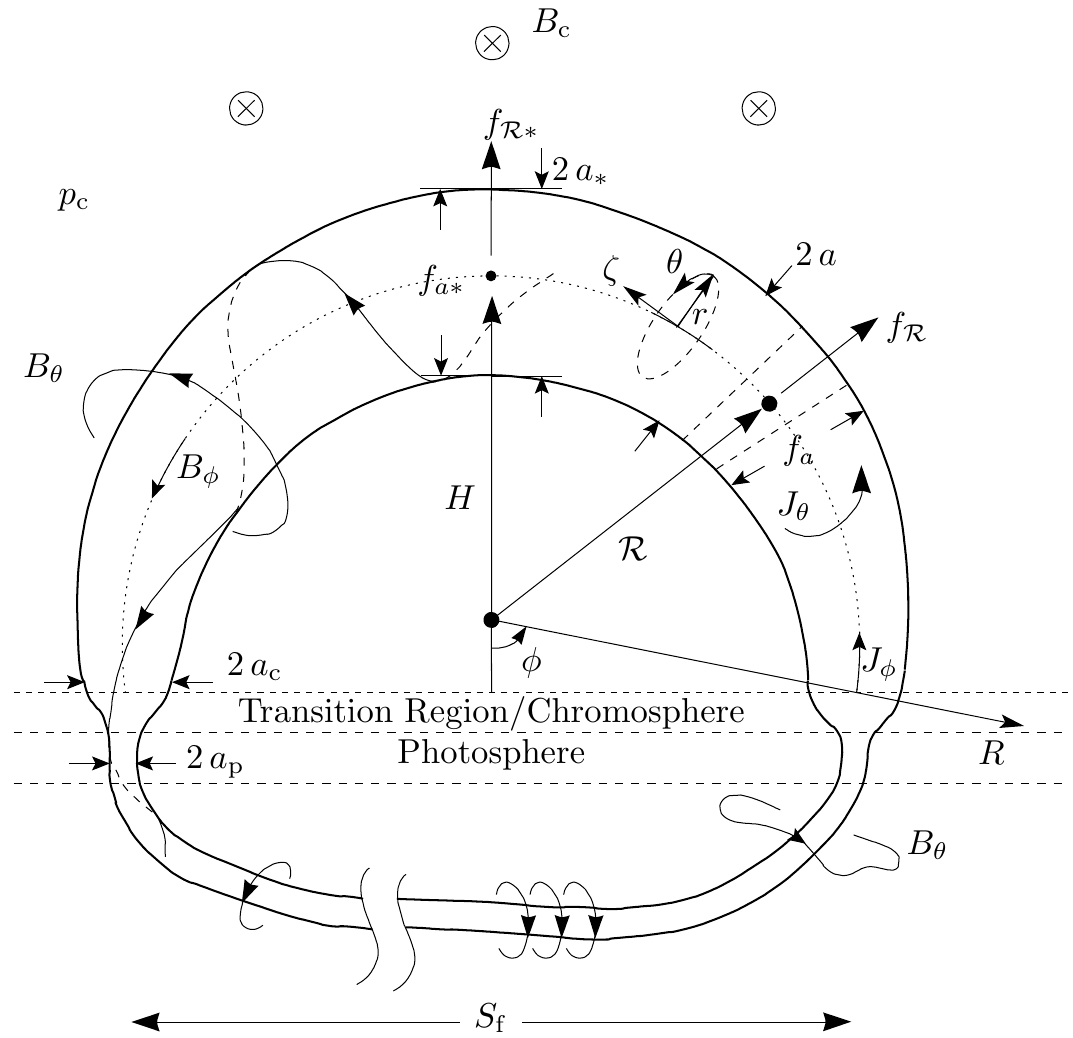}}
\else
\centerline{\input{flux-injection2.pstex_t}}
\fi
\caption{Schematic diagram of the flux-rope model current loop
  adapted from \cite{Chen1989,Chen1996}. The subscripts ``$\toroidal$'' and ``$\poloidal$'' refer to the
toroidal and poloidal directions in the cylindrical
$\left(R,\toroidal,Z\right)$ and the local polar $\left(r,\theta,\localz\right)$
coordinate systems for a circular torus where
$\widehat{Z}$ is out of the plane of the page. \label{fig:flux_injection}}
\end{figure*}
Figure~\ref{fig:flux_injection} shows a schematic diagram of the flux-rope CME
model current loop, adapted from
\cite{Chen1989,Chen1996}. \cite{Shafranov1966} derived the magnetohydrodynamic
(MHD) forces per-unit-length acting the major $\Major$ and minor $a$ radii of
a current carrying toroidal section
\begin{mathletters}
\label{eqn:shafranov}
\begin{equation} 
{f}_{R}\equiv{{I_\toroidal^2}\over{c^2\,\Major}}
\left[\ln\left({{8\,\Major}\over{a}}\right) + {\Delta\beta\over 2} -
{{\left\langle{B}_\toroidal\right\rangle^2}\over{2\,B_{\poloidal{a}}^2}} +{2\,\Major\over a}
{{B_\mathrm{c}}\over{B_{\poloidal{a}}}} - 1 + {{\xi}\over{2}}\right],\label{eqn:force}
\end{equation} 
\begin{equation} 
f_\mathrm{a}\equiv\frac{I_\toroidal^2}{c^2\,a}\,\left(\frac{\left\langle{B}_\toroidal\right\rangle^2}{B_{\theta{a}}^2}-1+\Delta\beta\right),\label{eqn:minor}
\end{equation} 
\end{mathletters}
where $I_\toroidal$ is the toroidal current,  $c$ is the speed of light,
$\left\langle{B}_\toroidal\right\rangle$ is the average toroidal field inside
the current channel, $B_{\mathrm{c}}$ is a prescribed function describing the
overlying coronal field perpendicular to the flux rope, $B_{\poloidal{a}}
\equiv B_\poloidal\left(a\right) =2\,I_\toroidal/a\,c$ is the poloidal
magnetic field at the edge of the current channel, and
\begin{equation}
\xi\equiv\frac{2\,\int_0^{a}{dr}\,{r}B_\poloidal^2\left(r\right)}{a^2\,B_{{\poloidal}a}^2}=\frac{c^2}{2\,I_\toroidal^2}\,\int_0^{a}{dr}\,{r}B_\poloidal^2\left(r\right)\sim \mathcal{O}\left(1\right),
\end{equation}
is the internal inductance.  The quantity $\Delta\beta \equiv
8\pi(\left\langle{p}\right\rangle -
p_{\mathrm{c}})/B_{\poloidal{a}}^2$ is the differential of plasma $\beta$
based on $B_{\poloidal{a}}$ where $\left\langle{p}\right\rangle$
is the average internal flux-rope pressure and $p_{\mathrm{c}}$ is the coronal
pressure. The subscripts ``$\toroidal$''
and ``$\poloidal$'' refer to the toroidal and poloidal directions in the
cylindrical $\left(R,\toroidal,Z\right)$ and the local polar
$\left(r,\theta,\localz\right)$ coordinate systems for a circular torus
with $R=\Major+r\,\cos\theta$, $Z=r\,\sin\theta$ and
$\localzhat=\widehat{\toroidal}$ where $\widehat{Z}$ is out of the plane of
the page.
\cite{Chen1989} applied Equation~(\ref{eqn:shafranov}) at the peak of the flux
rope to derive the dynamical equations
\begin{mathletters}
\label{eqn:apex}
\begin{equation} 
M\,\frac{d^2 \Height}{d t^2}={{I_\toroidal^2}\over{c^2\,\Major}}
\left[\ln\left({{8\,\Major}\over{\aA}}\right) + {\Delta\beta_*\over 2} -
{{\left\langle{B}_{\toroidal_*}\right\rangle^2}\over{2\,B_{\poloidal{\aA}}^2}} +{2\,\Major\over \aA}
{{B_\mathrm{c}}\over{B_{\poloidal{\aA}}}} - 1 + {{\xi_*}\over{2}}\right] + f_\mathrm{d} +
f_\mathrm{g},\label{eqn:apex:height}
\end{equation} 
and
\begin{equation} 
M\,\frac{d^2 \aA}{d t^2} =\frac{I_\toroidal^2}{c^2\,\aA}\,\left(\frac{\left\langle{B}_{\toroidal_*}\right\rangle^2}{B_{\theta{\aA}}^2}-1+\Delta\beta_*\right),\label{eqn:apex:minor}
\end{equation} 
\end{mathletters}
where $\Height>0$ is the height of the center of the current channel above the
photosphere henceforth referred to as the ``apex.'' The *'s in
Equation~(\ref{eqn:apex}) denote quantities evaluated over the cross section
at the apex where $M\equiv\left\langle{n_*}\right\rangle\,m\,\pi\,\aA^2$ is
the mass per-unit-length of the flux rope and $\left\langle{n_*}\right\rangle$
is the average density of the flux rope.  The additional terms
\begin{mathletters}
\begin{equation}
f_\mathrm{d}=c_\mathrm{d}\,n_\mathrm{c}\,m\,\aA\,\left(V_\mathrm{SW}-V\right)\,\left|V_\mathrm{SW}-V\right|
\end{equation}
and
\begin{equation}
f_\mathrm{g}=\pi\,\aA^2\,m\,g\,\left(n_\mathrm{c}-\left\langle{n_*}\right\rangle\right),
\end{equation}
\end{mathletters}
introduced in \cite{Chen1989}, represent the drag and gravitational forces
per-unit-length of the flux rope respectively where
$c_\mathrm{d}\sim\mathcal{O}\left(1\right)$ is the drag coefficient, 
$m=1.67\times10^{-24}\,\mbox{g}$ is the mass of hydrogen,
$n_\mathrm{c}$ is the ambient coronal density, $V\equiv{dH}/dt$,
$V_\mathrm{SW}$ is the velocity of the ambient solar wind, and the
gravitational acceleration is
\begin{equation}
g=\frac{g_\sun}{\left(1+H/R_\sun\right)^2},
\end{equation}
with $g_\sun=2.74\times10^{4}\,\mbox{cm}\mbox{s}^{-1}$ and
$R_\sun=6.74\times10^{10}\,\mbox{cm}$.
The flux-rope footpoints are separated by a distance $\Sf$ in the photosphere
and assumed to remain fixed throughout the evolution of the flux-rope CME by
the ``dense subphotospheric plasma'' \cite[]{Chen1989}. The current-channel
radius at the base of the corona $\ac$ is also assumed to remain fixed
throughout the evolution of the flux rope.  The major radius $\Major$ and
height $\Height$ above the photosphere are related by
\begin{equation}
\Major\equiv\frac{\Height^2+\Sf^2/4}{2\,\Height}\qquad\mbox{and}\qquad{H}>0.
\end{equation}
The length of the flux rope above the photosphere is
$L=2\,\pi\,\arclength\,\Major$ where
\begin{equation}
\arclength\equiv\left\lbrace\begin{array}{lr}
1-\varphi/\pi&\Height\ge\Sf/2,\\
\varphi/\pi&\Height<\Sf/2,
\end{array}\right.
\end{equation}
and $\varphi\equiv\arcsin\left(\Sf/2\,\Major\right)$
\cite[]{Chen1989,Krall2000}.  Note the variables that describe the plasma at
the apex of the flux rope $\left\langle{n_*}\right\rangle$,
$\left\langle{p_*}\right\rangle$, $B_{\poloidal{\aA}}$, $B_{\toroidal_*}$,
and $I_{\toroidal}$ will evolve with time and parameters of the interplanetary
medium $n_\mathrm{c}$, $B_\mathrm{c}$, and $V_\mathrm{SW}$ are implicit
functions of $H$. \cite[see][for a complete description of the interelations
  between variables and a model for the parameters of the interplanetary
  medium]{Chen1996} \cite[]{Krall2000}.\par
Flux-rope equilibria are determined from
\begin{equation}
f_{R*}=f_{a*}=0.
\end{equation}
\cite{Chen1994}, \cite{Chen1996}, \cite{Krall2000}, and \cite{Krall2005}
proposed local polar equilibrium profiles for $\partial_\localz\approx0$ and
large aspect ratio $\Major/a\gg1$
\begin{mathletters}
\label{eqn:chen:equilibrium}
\begin{equation}
B_\theta\left(r\right)=B_{\theta{a}}\,\left\lbrace\begin{array}{lr}
\displaystyle 3\,\frac{r}{a}\,\left(1-\frac{r^2}{a^2}+\frac{r^4}{3\,a^4}\right)& \displaystyle r<a,\\
\displaystyle\frac{a}{r}&\displaystyle r>a,
\end{array}\right.\label{eqn:chen:Bp}
\end{equation}
\begin{equation}
B_\toroidal\left(r\right)\simeq{B}_\localz\left(r\right)=3\,B_{\localz{a}}\,\left\lbrace\begin{array}{lr}
\displaystyle1-2\,\frac{r^2}{a^2}+\frac{r^4}{a^4}& \displaystyle r<a,\\
\displaystyle 0&\displaystyle r>a,
\end{array}\right.\label{eqn:chen:Bz}
\end{equation}
\end{mathletters}
compatible with the flux-rope model with a toroidal flux of
\begin{mathletters}
\begin{equation}
\Phi_\toroidal=\Phi_\localz\equiv\left\langle{B}_{\localz}\right\rangle\,\pi\,a^2=B_{\localz{a}}\,\pi\,a^2,\label{eqn:flux}
\end{equation}
 and carrying a bare toroidal current of
\begin{equation}
I_\toroidal=I_\localz={B}_{\localz{a}}\,{a}\,c/2.\label{eqn:tcurrent}
\end{equation}
\end{mathletters}
There is no return current which is consistent with the large-scale current
systems in active regions analyzed by \cite{Wheatland2000}.  Since
Equations~(\ref{eqn:chen:Bp}) and~(\ref{eqn:chen:Bz}) do not represent a
force-free equilibria $\J\cross\B=0$, the poloidal and toroidal magnetic
fields and corresponding currents are decoupled.\footnote{In equilibrium, the
  magnetic pressure is in detailed balance with gravity and kinetic pressure.}
Consequently, the cutoff of the toroidal field $B_\localz$ at $r=a$ is
arbitrary relative to the poloidal magnetic field. For example,
\begin{equation}
B_\localz\left(r\right)=B_{\localz{a}}\,e^{-r^2/a^2},
\end{equation}
is a valid toroidal field model and contains the identical
amount of flux as Equation~(\ref{eqn:chen:Bz}).\par
\begin{mathletters}
The toroidal energy per-unit-length of the flux rope is
\begin{equation}
\eden_\localz=\frac{1}{4}\,\int_0^a{dr}\,{r}\,B^2_\localz\left(r\right)\simeq\frac{9}{40}\,B_{\localz{a}}^2\,{a}^2,
\end{equation}
and the poloidal energy per unit length contained within a distance of
radius $r$ of the toroidal axis of the flux rope is
\begin{equation}
\eden_\poloidal\left(r\right)=\frac{1}{4}\,\int_0^r{dr'}\,{r'}\,B^2_\poloidal\left(r'\right)\simeq\frac{B_{\poloidal{a}}^2\,a^2}{480}\,\left[73+120\,\log\left(\rc/a\right)\right].
\end{equation}
\end{mathletters}
The local polar equilibrium profile
Equations~(\ref{eqn:chen:Bp}) and~(\ref{eqn:chen:Bz}) does not formally admit a
bounded poloidal energy per unit length along the flux rope\----a necessary
physical condition for admissibility\----because $B_\poloidal\sim{r}^{-1}$
and $\eden_\poloidal\left(r\right)\sim\log\left(r\right)$ for $r>a$. However,
for the closed circuit representative of the schematic flux rope shown in
Figure~\ref{fig:flux_injection}, the integral is cutoff at distances of
order of the dimension of the circuit \cite[see pp. 136-141 in][]{Landau1960}
which is roughly half the footpoint separation $\rc\simeq\Sf/2$.
The fraction of poloidal magnetic energy contained inside radius $r$ is then
\begin{equation}
\delta_{\poloidal}\left(r,a,\rc\right)\equiv\frac{\eden_\poloidal\left(r\right)}{\eden_\poloidal\left(\rc\right)}\simeq\frac{73+120\,\log\left(r/a\right)}{73+120\,\log\left(\rc/a\right)}.\label{eqn:delta}
\end{equation}
\cite{Chen2000a} have argued that a ``large fraction of the injected poloidal
energy is in the magnetic field outside the current channel,'' denoted by
$r=a$. Indeed, in the corona, roughly 1/3 of poloidal energy is contained
within the current channel $r\le{a}$. However, more than 2/3 of the poloidal
energy is contained within the region $r\le2\,a$ near the current channel. In
the photosphere, the amounts are reduced because the current channel is
narrower, but the poloidal energy is contained within the region $r\le2\,a$
near where the current channel remains significant varying from
$1/3\Rightarrow2/3$ depending on the strength of the toroidal field in the
photosphere. The comparison of the flux-injection hypothesis against
photospheric observations in Section~\ref{sec:observations} is limited to the
region $r\le2\,a$ where the poloidal energy transport is substantial and the
poloidal magnetic field in significant.\par
\begin{figure*}[!t]
\ifx\xfig\undefined
\centerline{\includegraphics[viewport= 195 467 426 738,width=3.5in,clip=]{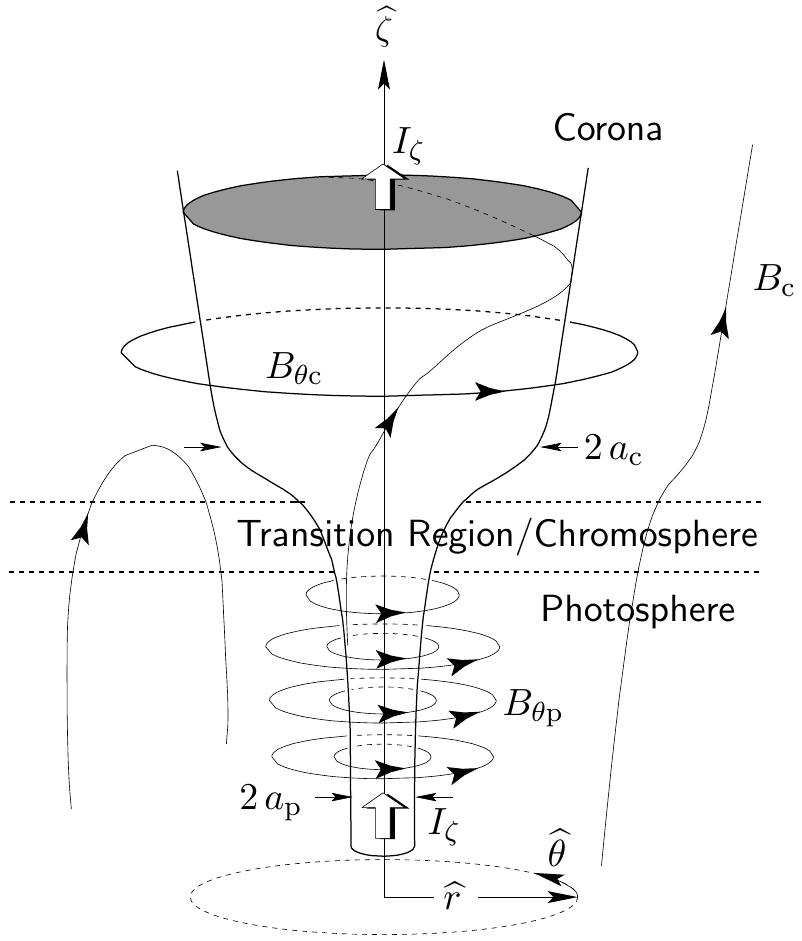}}
\else
\centerline{\input{leg.pstex_t}}
\fi
\caption{Schematic diagram of the flux-rope leg through the
  photosphere adapted from \cite{Chen2001}. The current channel
  carrying $I_\localz$, the top of which is shaded, and the 
  poloidal field $B_\poloidal$, and two community field lines
  extending into the corona $B_{\mathrm{c}}$ are shown. The subscripts ``c'' and ``p'' refer to coronal and photospheric values respectively.\label{fig:leg}}
\end{figure*}
Although the magnetic field model Equations~(\ref{eqn:chen:Bp})
and~(\ref{eqn:chen:Bz}) was originally proposed as an approximate local
description for large-aspect ratio coronal toroidal flux-rope magnetic fields,
it has also been used as an initial equilibrium for investigating photospheric
signatures in simulations of subphotospheric flux ropes
\cite[]{ChenHuba2005b,ChenHuba2005a,ChenHuba2006} and similar models have been
implemented for modeling the photosphere flux-rope dynamics
\cite[]{Sakai2000,Sakai2001}.  Figure~\ref{fig:leg} shows a schematic diagram
of the flux-rope leg through the photosphere adapted from \cite{Chen2001} in
the local polar $r$, $\poloidal$, $\localz$ coordinate system. The vertical
current channel carrying $I_\localz$, the top of which is shaded, and the
private poloidal field $B_\poloidal$, and two community field lines extending
into the corona $B_{\mathrm{c}}$ are shown. The subscripts ``c'' and ``p''
refer to values at the base of the corona and in the photosphere,
respectively.  Two general constraints may be applied quasi-adiabatically to
extend the coronal model into the photosphere: conservation of vertical flux
and vertical current \cite[]{Krall2005}
\begin{mathletters}
\begin{eqnarray}
B_{\localz{\mathrm{p}}}&=&B_{\localz{\mathrm{c}}}\,\frac{\ac^2}{\af^2},\label{eqn:krall1}\\
{B_{\poloidal{\mathrm{p}}}}&=&{B_{\poloidal{\mathrm{c}}}}\,\frac{\ac}{\af}.\label{eqn:krall2}
\end{eqnarray}
\end{mathletters} 
Combining these relationships with Equations~(\ref{eqn:chen:Bp}) and~(\ref{eqn:chen:Bz})
produces an approximate local model for the magnetic footpoints of the
flux rope in the photosphere with
$B_{\localz\mathrm{c}}=B_{\toroidal\mathrm{c}}$ at the base of the corona.
The details of the flux-tube expansion in the chromosphere is not explicitly
specified in this approximation.  However, the expansion implies a local radial
field $B_r\left(r\right)$ to balance the decrease of
$B_\localz\left(\localz\right)$ with height under the local constraint of
$\grad\cdot\B=0$.\par
\section{THE PHOTOSPHERIC FLUX BUDGETS OF THE FLUX-ROPE MODEL\label{sec:energy}}
The flux-rope model requires specification of two quantities in
addition to the geometrical aspects of the flux rope, namely the
toroidal and poloidal fluxes \cite[]{Chen1996}
\begin{mathletters}
\begin{eqnarray}
\Phi_\toroidal&=&B_{\toroidal\mathrm{c}}\,\pi\,\ac^2=\mathrm{constant},\label{eqn:toroidal_flux}\\
\Phi_\poloidal&=&c\,\mathcal{L}\,I_\toroidal,\label{eqn:poloidal_flux}
\end{eqnarray}
\end{mathletters}
where
\begin{equation}
\mathcal{L}\simeq\frac{4\,\pi\,\arclength\,\Major}{c^2}\,\left\lbrace\log\left(8\,\Major\right)-1+\frac{\xi}{2}-\frac{1}{\aA-\ac}\left[\aA\,\log\left(\aA\right)-\ac\,\log\left(\ac\right)\right]
\right\rbrace
\end{equation}
is the inductance \cite[]{Landau1960,Krall2000}. The toroidal 
energy is estimated from the arclength of the flux rope above the
photosphere.\footnote{The flux-rope minor current-channel radius
  varies linearly with arc length
  $a\left(\ell\right)=\ac+\left(\aA-\ac\right)\,\ell$ from the
  footpoint at $\ell=0$ to the apex at $\ell=1$.}
\begin{mathletters}
\begin{equation}
U_\toroidal\simeq\frac{9\,\pi\,\arclength}{20}\,\frac{\Major}{\aA}\,B_{\toroidal{\ac}}^2\,\ac^3=\frac{9\,\arclength}{20\,\pi}\,\frac{\Phi_{\toroidal}^2}{\ac}\,\frac{\Major}{\aA}\approx\mbox{constant},\label{eqn:toroidal_energy}\\
\end{equation}
whereas the poloidal energy is related to the inductance and poloidal flux
\begin{equation}
U_\poloidal=\frac{1}{2}\,\mathcal{L}\,I_\toroidal^2=\frac{1}{2\,c^2}\,\frac{\Phi_\poloidal^2}{\mathcal{L}}.\label{eqn:poloidal_energy}
\end{equation}
\end{mathletters}
The toroidal flux $\Phi_\toroidal=\Phi_\localz$ is conserved because the
toroidal field $B_{\toroidal{\ac}}=B_{\localz{\ac}}$ and current-channel
radius $\ac$ \textit{at the base of the corona} are held constant in time
during the eruption\----$B_{\toroidal{a}}$ and the minor radius $a$ may vary
along the flux rope subject to flux conservation.  The toroidal energy
$U_\toroidal$ is conserved because the flux rope erupts self-similarly
${\Major}/{\aA}\approx\mbox{constant}$. Consequently, the quantities that will
manifest dynamical changes in the photosphere during the eruption are the
poloidal power $dU_\poloidal/dt$ and the rate-of-change in poloidal flux $d\Phi_\poloidal/dt$ which are
related to the photospheric energy and helicity fluxes respectively. \par
The MHD induction equation derived by combining
Faraday's Law
\begin{equation}
\partial_t\B=-c\,\grad\cross\E,
\end{equation}
with Ohm's Law
\begin{equation}
\E=\frac{\J}{\sigma}-\frac{\vm}{c}\cross\B,
\end{equation}
to obtain
\begin{equation}
\partial_t\B=\grad\cross\left(\vm\cross\B\right)-c\,\grad\cross\left(\frac{\J}{\sigma}\right),\label{eqn:induction}
\end{equation}
where $\vm$ is the plasma velocity and $\sigma$ is a spatially variable
conductivity. 
The magnetic energy in the corona is formulated by
dotting the induction Equation~(\ref{eqn:induction}) with the magnetic field
$\B$ and using
 Ampere's Law without displacement currents
\begin{equation}
\grad\cross\B=\frac{4\,\pi}{c}\,\J,
\end{equation}
to derive Poynting theorem 
\begin{equation}
\frac{d U_{\mathrm{M}}}{d t}\equiv\frac{d}{dt}\int_{V_\mathrm{c}}\frac{dV}{8\,\pi}\,{B^2}=\frac{1}{4\,\pi}\,\oint_S{dS}\,\nhat\cdot\left[\B\cross\left(\vm\cross\B\right)+\frac{c}{\sigma}\,\left(\J\cross\B\right)\right]-\int_{V_\mathrm{c}}{dV}\,\left[\frac{\vm}{c}\cdot\left(\J\cross\B\right)+\frac{J^2}{\sigma}\right],\label{eqn:poynting}
\end{equation}
where the volume integrals $V_{\mathrm{c}}$ are over the corona, chromosphere,
and transition region, the surface integrals bounding the volume
$S\equiv{S}_\mathrm{p}+{S}_\mathrm{oc}$ are over the photosphere
${S}_\mathrm{p}$ and the outer corona ${S}_\mathrm{oc}$ at $R\gg{R}_\sun$, and
$\nhat$ is the surface normal pointing into the coronal volume (radially
outward at the photosphere ${S}_\mathrm{p}$ and radially inward at the outer
corona ${S}_\mathrm{oc}$). Assessing the energy budget of the region between
the photosphere and outer corona by tracking the Poynting flux through the
photosphere and the energy leaving the corona through eruptive phenomena
provides an estimate for the free energy available for producing flares and
CMEs \cite[see Figure~1 in][]{Kusano2002a}.  The first and second terms in
Equation~(\ref{eqn:poynting}) represent the $\E\cross\B$ Poynting flux through
the surfaces and the third and fourth terms represent the conversion of
magnetic energy to kinetic energy though work done by the $\J\cross\B$ force
on the plasma and Ohmic heating through resistivity respectively.  \par
The helicity for the flux rope may be written\footnote{See pp. 21,516 in
  \cite{Chen1996}. This expression is exact in the large aspect ratio limit
  $\Major/a\gg1$.} \cite[]{Chen1996,Chen2003}
\begin{equation}
\Helicity\equiv\int_{V_\mathrm{c}}{dV}\,\A\cdot\B\simeq\Phi_\toroidal\,\Phi_\poloidal.
\end{equation}
Comparisons between integrated photospheric helicity flux and the poloidal
flux injection profile for the flux-rope model can be made.  However, because
the field lines of the flux rope penetrate the photosphere, a gauge-invariant
\textit{relative} helicity must be used for estimating helicity fluxes through
the photosphere \cite[]{Berger1984}
\begin{equation}
\Delta{\Helicity}=\int_{V_\mathrm{c}}{dV}\,\left(\A\cdot\B-\Ap\cdot\Bp\right),\label{eqn:berger}
\end{equation}
where $V_\mathrm{c}$ corresponds to the volume above the photosphere and
$\Bp=\grad\cross\Ap$ are the reference fields which are chosen to match the
normal components of $\B$ and the tangential components of $\A$ respectively
at the surface:
\begin{mathletters}\label{eqn:helicity_bndry}
\begin{eqnarray}
\left.\left(\A-\Ap\right)\cross\nhat\right|_S&=&0,\label{eqn:helicity_bndry:a}\\
\left.\left(\B-\Bp\right)\cdot\nhat\right|_S&=&0.\label{eqn:helicity_bndry:b}
\end{eqnarray}
\end{mathletters}
These boundary conditions are sufficient for the equivalence of the relative
helicity defined by \cite{Berger1984} in Equation~(\ref{eqn:berger}) and
manifestly gauge invariant relative helicity defined by \cite{Finn1985}
\begin{equation}
\Delta{\Helicity}=\int_{V_\mathrm{c}}{dV}\,\left(\A+\Ap\right)\cdot\left(\B-\Bp\right),\label{eqn:finn}
\end{equation}
because
\begin{eqnarray}
\int_{V_\mathrm{c}}{dV}\,\left(\Ap\cdot\B-\A\cdot\Bp\right)&=&\int_{V_\mathrm{c}}{dV}\,\left(\Ap\cdot\grad\cross\A-\A\cdot\grad\cross\Ap\right),\nonumber\\
&=&\int_{V_\mathrm{c}}{dV}\,\grad\cdot\left(\A\cross\Ap\right),\nonumber\\
&=&-\int_{S}{dS}\,\nhat\cdot\left(\A\cross\Ap\right),\nonumber\\
&=&0,\label{eqn:berger-finn}
\end{eqnarray}
with Equation~(\ref{eqn:helicity_bndry}).\par
A judicious choice for the reference field is a potential field
\begin{mathletters}
\begin{equation}
\Bp=\grad\cross\Ap=\grad\Psi_\mathrm{R},\label{eqn:potential}
\end{equation}
in the Coulomb gauge
\begin{equation}
\grad\cdot\Ap=0,\label{eqn:coulomb}
\end{equation}
with the additional boundary condition
\begin{equation}
\left.\nhat\cdot\Ap\right|_S=0.\label{eqn:potential_bndry}
\end{equation}
\end{mathletters}
For these conditions, this reference field has zero
helicity
\begin{equation}
\Helicity_\mathrm{R}=\int_{V_\mathrm{c}}{dV}\,\Ap\cdot\Bp=\int_{V_\mathrm{c}}{dV}\,\Ap\cdot\grad\Psi_\mathrm{R}=\oint_{S}{dS}\,\nhat\cdot\Ap\Psi_\mathrm{R}-\int_{V_\mathrm{c}}{dV}\,\Psi_\mathrm{R}\,\grad\cdot\Ap=0.
\end{equation}\par
Using Equation~(\ref{eqn:berger}) or~(\ref{eqn:finn}) with
Equations~(\ref{eqn:potential}) and (\ref{eqn:coulomb}) and boundary
conditions (\ref{eqn:helicity_bndry}) and (\ref{eqn:potential_bndry}), the
Poynting theorem for the magnetic helicity into the corona then takes a
particularly simple form
\begin{mathletters}
\begin{eqnarray}
\frac{d\Delta{\Helicity}}{dt}&=&\frac{d}{dt}\,\int_{V_\mathrm{c}}{dV}\,\A\cdot\B,\\
&=&2\,\oint_S{dS}\,\nhat\cdot\left[\Ap\cross\left(\vm\cross\B\right)-\frac{c}{\sigma}\,\Ap\cross\J\right]-2\,c\,\int_{V_\mathrm{c}}{dV}\,\frac{\J\cdot\B}{\sigma}.\label{eqn:helicity_flux}
\end{eqnarray}
\end{mathletters}
The first and second terms in Equation~(\ref{eqn:helicity_flux}) represent the
helicity flux through the photosphere and the third term represents helicity
dissipation in the coronal volume.  \par
Although the photosphere is not ideal, with a magnetic Reynolds number of
$R_{\mathrm{M}}={U\,L}/{\eta}\sim10^5\--10^6$ where $U$ is the typical
velocity, $L$ is the typical gradient scale, and
$\eta=c^2/\left(4\,\pi\,\sigma\right)$ is the magnetic diffusivity, the ideal
approximation has been demonstrated to be adequate for inferring plasma
velocities from magnetic field dynamics in convection zone
simulations\footnote{See \cite{Abbett2004}.} with
magnetic Reynolds numbers as low as $R_{\mathrm{M}}\sim10^3$
\cite[]{Welsch2007,Schuck2008d}. The ideal MHD induction equation becomes
\begin{equation}
\partial_\TIME\B=\grad\cross\left(\vm\cross\B\right).
\end{equation}
Within the flux injection paradigm, consistent with
Equation~(\ref{eqn:toroidal_flux}), the toroidal magnetic field in the
photosphere does not change during the eruption \cite[]{Chen1997a}
\begin{equation}
\partial_\TIME{B}_\localz=\zhat\cdot\grad\cross\left(\vm\cross\B\right)\approx0.\label{eqn:toroidal}
\end{equation}
This relationship also represents an \textit{observational constraint} on the
vertical magnetic field of the flux rope in the photosphere because large
changes in line-of-sight magnetograms near disk center have not been observed
during eruptions. This constraint on the vertical magnetic field implies 
\begin{mathletters}
\label{eqn:velocities}
\begin{equation}
\vm\cross\B\approx\grad_h\psi+\left(v_r\,B_\theta-B_r\,v_\theta\right)\,\zhat,\label{eqn:22a}
\end{equation}
where the subscript ``h'' refers to the horizontal $\left(r,\theta\right)$
coordinates of the local polar coordinate system and $\psi$ is the
electrostatic potential.  Solving Equation~(\ref{eqn:22a}) for $\vm$ produces
\begin{equation}
\vm=\frac{\zhat\cross\B\,\left(\B\cdot\grad_h\psi\right)}{B_\localz\,B^2}-\frac{\grad_h\psi\cross\B}{B^2}+v_\parallel\,\B/\left|\B\right|,\label{eqn:vm}
\end{equation}
\end{mathletters}
where $v_\parallel$ is the field-aligned plasma velocity.  For a locally
cylindrical flux rope with $\partial_\theta=0$ and $B_r=0$ consistent
with Equation~(\ref{eqn:chen:equilibrium}), the evolution of the radial and poloidal
fields at the photosphere are determined by
\begin{mathletters}
\begin{eqnarray}
\partial_\TIME B_r&=&0,\\
\partial_\TIME B_\theta&=&\partial_{r\localz}\psi-B_\localz^{-1}\,\left(\partial_\localz{B_\localz}\right)\,\left(\partial_r\psi\right),
\label{eqn:poloidal}\\
\partial_\TIME B_\localz&=&0.
\end{eqnarray}
\end{mathletters}
\cite{Chen1996} and \cite{Chen2010} have noted that the poloidal magnetic
field at the base of the corona may only increase by 20\%\--50\% during the
first 30 minutes of the eruptions and \cite{Krall2001} have shown an event
that requires no perceptible increase in the poloidal magnetic field \cite[see
  Figure~8(a) in][]{Krall2001}. To the extent that $\partial_\localz\approx0$,
these results are consistent with MHD. \par
\subsection{The Ideal Poynting Fluxes in the Photosphere}
From Equations~(\ref{eqn:poynting}), (\ref{eqn:helicity_flux}),
and~(\ref{eqn:vm}), the ideal MHD energy flux through the photosphere is
\begin{eqnarray}
u_\localz&=&\frac{1}{4\,\pi}\,\left(v_\localz\,\B_h-B_\localz\,\vm_h\right)\cdot\B_h,\nonumber\\
&=&-\frac{1}{4\,\pi}\,\left(\zhat\cross\grad_h\psi\right)\cdot\B_h,\label{eqn:ideal:poynting}
\end{eqnarray}
and the helicity flux is
\begin{eqnarray}
k_\localz&=&2\,\left(v_\localz\,\B_h-B_\localz\,\vm_h\right)\cdot\Ap,\nonumber\\
&=&2\,\zhat\cdot\left(\Ap\cross\grad_h\psi\right).\label{eqn:ideal:helicity}
\end{eqnarray}
Equations~(\ref{eqn:velocities}), (\ref{eqn:poloidal}),
(\ref{eqn:ideal:poynting}), and~(\ref{eqn:ideal:helicity}) are general with
respect to the constraint~(\ref{eqn:toroidal}); the fluxes through the
photosphere are completely specified by the electric potential $\psi$ and the
magnetic field.  The flux-injection hypothesis has always appealed to some
``unspecified sub-photospheric process'' for initiating and driving the CME
\cite[\citeauthor{Krall2000} \citeyear{Krall2000}; see
  also][]{Chen1989,Chen1993,Chen1996,Chen1997a,Chen1997b,Chen2000a,Chen2001,Chen2003,Krall2001,Chen2010}. Nonetheless,
in the ideal MHD limit, energy is transported through the photosphere by the
term $\widehat{\localz}\cdot\B\cross\left(\vm\cross\B\right)$ and helicity is
transported by
$\widehat{\localz}\cdot\Ap\cross\left(\vm\cross\B\right)$. These terms
integrated over the footpoints in the photosphere (including the region
outside the current channel) must balance the poloidal power requirements
$dU_\poloidal/dt$ and helicity requirements
$d\Helicity/dt\simeq\Phi_\toroidal\,d\Phi_\poloidal/dt$ of a CME described by
the flux-rope model. Temporally, the increase in the fluxes at the footpoints
should precede the eruptions by at least $\delta t\simeq4$~minutes to account
for the transport of magnetic field along the flux rope to the apex of the CME
at the coronal Alfv\'en speed\footnote{The Alfv\'{e}n speed is based on
  flux-rope initial conditions from \cite[]{Chen1996} with
  $\left\langle{n}\right\rangle\simeq7.5\times10^7\,\mbox{cm}^{-3}$ and
  $B\simeq6$~G. Note that this estimate is an order of magnitude larger than
  estimates corresponding to the same height range by \cite{Regnier2008}.}
$V_{\mathrm{A}}\simeq1.5\times10^8\,\mbox{cm}\,\mbox{s}^{-1}$.  \par
Assuming azimuthal symmetry $\partial_\theta=0$, the poloidal power injected
through the flux-rope footpoints can be written in terms of
Equation~(\ref{eqn:poloidal_energy}) or~(\ref{eqn:ideal:poynting})
\begin{equation}
\frac{dU_\poloidal}{dt}\simeq\left(\frac{2}{\Phi_\poloidal}\,\frac{\Phi_\poloidal}{dt}-\frac{1}{\mathcal{L}}\,\frac{d\mathcal{L}}{dt}\right)\,\frac{\Phi_\poloidal^2}{2\,\mathcal{L}\,c^2}\simeq-2\times\frac{1}{2}\,\int_{0}^{\rc}{dr}\,{r}\,{B}_\theta\,\partial_r\psi=-\int_{0}^{\rc}{dr}\,{r}\,{B}_\theta\,\partial_r\psi,\label{eqn:power}
\end{equation}
and because the toroidal flux is constant, the rate of change of helicity is
\begin{equation}
\frac{d\Helicity}{dt}\simeq\Phi_\toroidal\,\frac{d\Phi_\poloidal}{dt}\simeq-2\times4\,\pi\int_0^{\rc}{dr}\,r\,A_{\mathrm{R}\poloidal}\,\partial_r\psi,\label{eqn:helicity}
\end{equation}
where the fluxes are assumed to be equipartitioned between the two
footpoints. Equations~(\ref{eqn:power}) and~(\ref{eqn:helicity}) represent
\textit{independent} constraints between the flux-rope dynamics and the energy
and helicity fluxes in the photosphere.\par
Balancing the photospheric Poynting fluxes against the poloidal energy and
helicity budgets of CME trajectories fit by the flux-rope model requires
choosing a photospheric model for the magnetic field and a velocity profile or
$\partial_r\psi$. For the former, the simple photospheric extension of the
flux-rope model described in Section~\ref{sec:model} is used.  For the latter, the
\textit{optimal} velocities for transporting energy and helicity across the
photosphere consistent with the respective overall budgets of the flux-rope
model are estimated via constrained variational calculus (see
Appendices~\ref{app:power} and~~\ref{app:helicity}). \textit{Optimal}, in this
context, means the \textit{minimum} root-mean-squared (rms) photospheric
velocities corresponding to the minimum photospheric kinetic energy (for
constant density).\par
There are two physical mechanisms for transporting energy and helicity
through the photosphere and into the corona:
\begin{enumerate}
\item Twisting the magnetic field in the photosphere through poloidal
  motion. Under this mechanism and the magnetic field model in
  Equation~(\ref{eqn:chen:equilibrium}), all the energy and helicity must be
  transported through the photosphere and into the corona within the
  current channel where $B_\localz\neq0$. Minimizing
  $\int{dr}\,{r}\,v_\poloidal^2$ with $v_\localz=0$, the rms poloidal
  photospheric velocity inside $r\le{\af}$ constrained by the energy budget is
  derived in Appendix~\ref{app:power}
\begin{eqnref}{eqn:vpoloidal}
\left\langle{v}_\poloidal^2\right\rangle_{\af}^{1/2}=\left|\frac{dU_\poloidal}{dt}\right|\,\frac{4\,\sqrt{70}}{\sqrt{437\,\left|B_{\poloidal{\mathrm{c}}}^2\,B_{\localz{\mathrm{c}}}\right|}\,\ac^2}\,\frac{1}{\sqrt{\left|B_{\localz{\mathrm{p}}}\right|}}\,\propto\frac{1}{\sqrt{\left|B_{\localz{\mathrm{p}}}\right|}},
\end{eqnref}
and constrained by the helicity budget derived in Appendix~\ref{app:helicity}
\begin{eqnref}{eqn:helicity:vpoloidal}
\left\langle{v}_\poloidal^2\right\rangle_{\af}^{1/2}=\left|\frac{d\Phi_\poloidal}{dt}\right|\,\sqrt{\frac{70}{437}}\,\frac{1}{\sqrt{\left|B_{\localz\mathrm{c}}\,B_{\localz\mathrm{p}}\right|}}\,
\frac{1}{\ac}.
\end{eqnref}
The twisting of footpoints does not change the net current carried by the
flux rope (see discussion on pp.~\pageref{error:coupling} and
Appendix~\ref{app:Krall}).
\item Flux injection involving the emergence of poloidal flux transported
  through the photosphere by vertical motion. Under this mechanism and the
  magnetic field model in Equation~(\ref{eqn:chen:equilibrium}), the energy
  and helicity transport are not limited to the current channel, but
  significant transport occurs within about $r\lesssim2\,\af$. Minimizing
  $\int{dr}\,{r}\,v_\localz^2$ with $v_\poloidal=0$, the rms vertical
  photospheric velocity inside $r\le{2\,\af}$ constrained by the energy budget
  is  derived in Appendix~\ref{app:power}
\begin{eqnref}{eqn:vtoroidal}
\left\langle{v_\localz^2}\right\rangle^{1/2}_{2\,\af}=\left|\frac{dU_\poloidal}{dt}\right|\,\frac{\sqrt{12516735}\,\left|B_{\localz\mathrm{p}}\right|\,\rc^2}{2\,B_{\poloidal\mathrm{c}}^2\,\ac^2\,\left|2998\,B_{\localz\mathrm{p}}\rc^2-1155\,B_{\localz\mathrm{c}}\,\ac^2\right|}.
\end{eqnref}
Note that Equation~(\ref{eqn:vtoroidal}) approaches a constant asymptotically if
either $\rc\rightarrow\infty$ or $B_{\localz\mathrm{p}}\rightarrow\infty$,
while holding the coronal flux constant $B_{\localz\mathrm{c}}\,\ac^2$
\begin{equation}
\left\langle{v_\localz^2}\right\rangle^{1/2}_{2\,\af}\ge\lim_{\rc\rightarrow\infty}\left\langle{v_\localz^2}\right\rangle^{1/2}_{2\,\af}=\left|\frac{dU_\poloidal}{dt}\right|\,\frac{\sqrt{12516735}}{5996\,B_{\poloidal\mathrm{c}}^2\,\ac^2},
\end{equation}
and this value is entirely determined by the current-channel radius and the
poloidal magnetic field \textit{at the base of the corona}.  For efficient energy
transport, a significant fraction of the energy transport for flux emergence
will occur \textit{in and near} the current channel. The corresponding rms
vertical photospheric velocity inside $r\le{2\,\af}$ constrained by the
helicity budget is derived in Appendix~\ref{app:helicity}
\begin{eqnref}{eqn:helicity:vtoroidal}
\left\langle{v_\localz^2}\right\rangle^{1/2}_{2\,\af}=\left|\frac{d\Phi_\poloidal}{dt}\right|\,\frac{\sqrt{12516735}\,\left|B_{\localz\mathrm{p}}\right|\,\rc^2}{8\,B_{\poloidal\mathrm{c}}\,\ac\,\left|2998\,B_{\localz\mathrm{p}}\rc^2-1155\,B_{\localz\mathrm{c}}\,\ac^2\right|}.
\end{eqnref}
 For comparison, minimizing
 $\int{dr}\,{r}\,\left(\partial_r{v}_\localz\right)^2$ with
 $v_\poloidal=0$ produces the minimum shear estimate (constant
 velocity $r<\rc$) which has rms vertical photospheric (everywhere) of
\begin{mathletters}
\begin{equation}
\left\langle{v}_\localz^2\right\rangle^{1/2}_\mathrm{p}=\left|\frac{dU_\poloidal}{dt}\right|\,\frac{1}{B_{\poloidal\mathrm{c}}^2\,\ac^2}\,\frac{120}{73+120\,\log\left(B_{\localz\mathrm{p}}\,\rc/B_{\localz\mathrm{c}}\,\ac\right)},
\end{equation}
for the energy constraint and
\begin{equation}
\left\langle{v}_\localz^2\right\rangle^{1/2}_\mathrm{p}=\left|\frac{d\Phi_\poloidal}{dt}\right|\,\frac{1}{B_{\poloidal\mathrm{c}}\,\ac}\,
\frac{30}{73+120\,\log\left(B_{\localz\mathrm{p}}\,\rc/B_{\localz\mathrm{c}}\,\ac\right)},
\end{equation}
for the helicity constraint.
\end{mathletters}
\item Both limits above require wasted
  flows along field lines $v_\parallel\neq0$. However by minimizing
  $\int{dr}\,{r}\,\left(v_\poloidal^2+v_{\smash{\localz}}^2\right)$ with
  $v_\parallel=0$, the most efficient rms total velocity inside
  $r\le{2\,\af}$ constrained by the energy budget is derived in Appendix~\ref{app:power}
\begin{eqnref}{eqn:vall}
\left\langle{v^2}\right\rangle^{1/2}_{2\,\af}=\left|\frac{dU_\poloidal}{dt}\right|\,\frac{2\,B_{\localz\mathrm{p}}\,\rc^2\,\sqrt{2310}\,\sqrt{21674\,B_{\poloidal\mathrm{c}}^2 + 14421\,B_{\localz\mathrm{c}}\,B_{\localz\mathrm{p}}}}
 {\ac^2\,\left|B_{\poloidal\mathrm{c}}\,B_{\localz\mathrm{p}}\,(23984\,B_{\poloidal\mathrm{c}}^2 + 14421\,B_{\localz\mathrm{c}}\,B_{\localz\mathrm{p}})\,\rc^2-9240\,\ac^4\,B_{\poloidal\mathrm{c}}^3\,B_{\localz\mathrm{c}}\right|},
\end{eqnref}
and the most efficient rms total velocity inside $r\le{2\,\af}$ constrained by
the helicity budget is derived in Appendix~\ref{app:helicity}
\begin{eqnref}{eqn:vall:helicity}
\left\langle{v^2}\right\rangle^{1/2}_{2\,\af}=\left|\frac{d\Phi_\poloidal}{dt}\right|\,\frac{B_{\localz\mathrm{p}}\,\rc^2\,\sqrt{2310}\,\sqrt{21674\,B_{\poloidal\mathrm{c}}^2 + 14421\,B_{\localz\mathrm{c}}\,B_{\localz\mathrm{p}}}}
 {2\,\ac\,\left|B_{\localz\mathrm{p}}\,(23984\,B_{\poloidal\mathrm{c}}^2 + 14421\,B_{\localz\mathrm{c}}\,B_{\localz\mathrm{p}})\,\rc^2-9240\,\ac^4\,B_{\poloidal\mathrm{c}}^2\,B_{\localz\mathrm{c}}\right|},
\end{eqnref}
where $v_\parallel=0$ everywhere.
\end{enumerate}
These three driving scenarios are used to estimate the expected rms velocities
near the current channel of CME flux-rope footpoints.
\subsection{Velocity Estimates Constrained by CME Trajectories}
\begin{figure*}[!t]
\centerline{\hskip0.25in\includegraphics[viewport=10 0 512 360,width=4in,clip=]{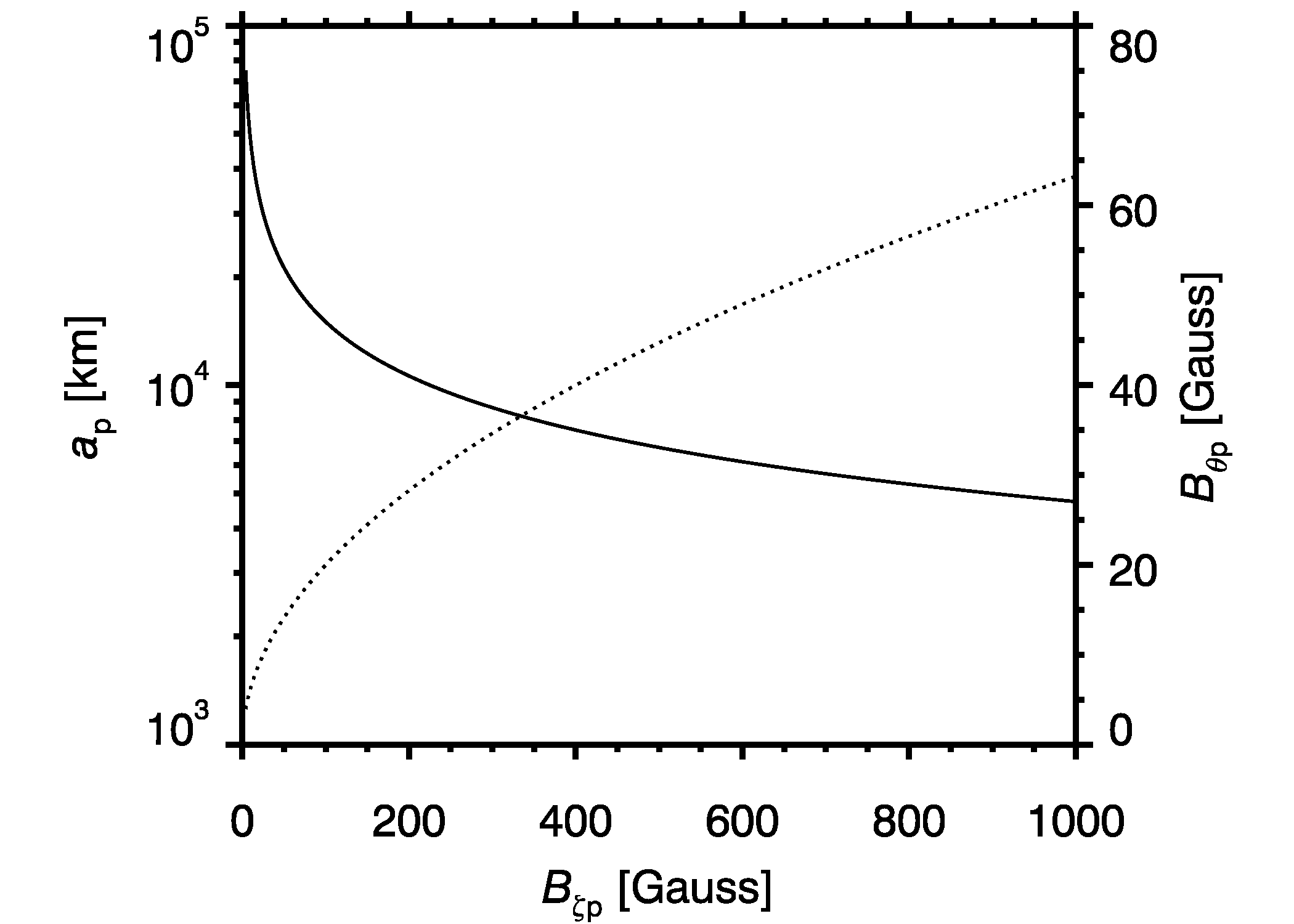}\hskip-0.25in\includegraphics[width=4in]{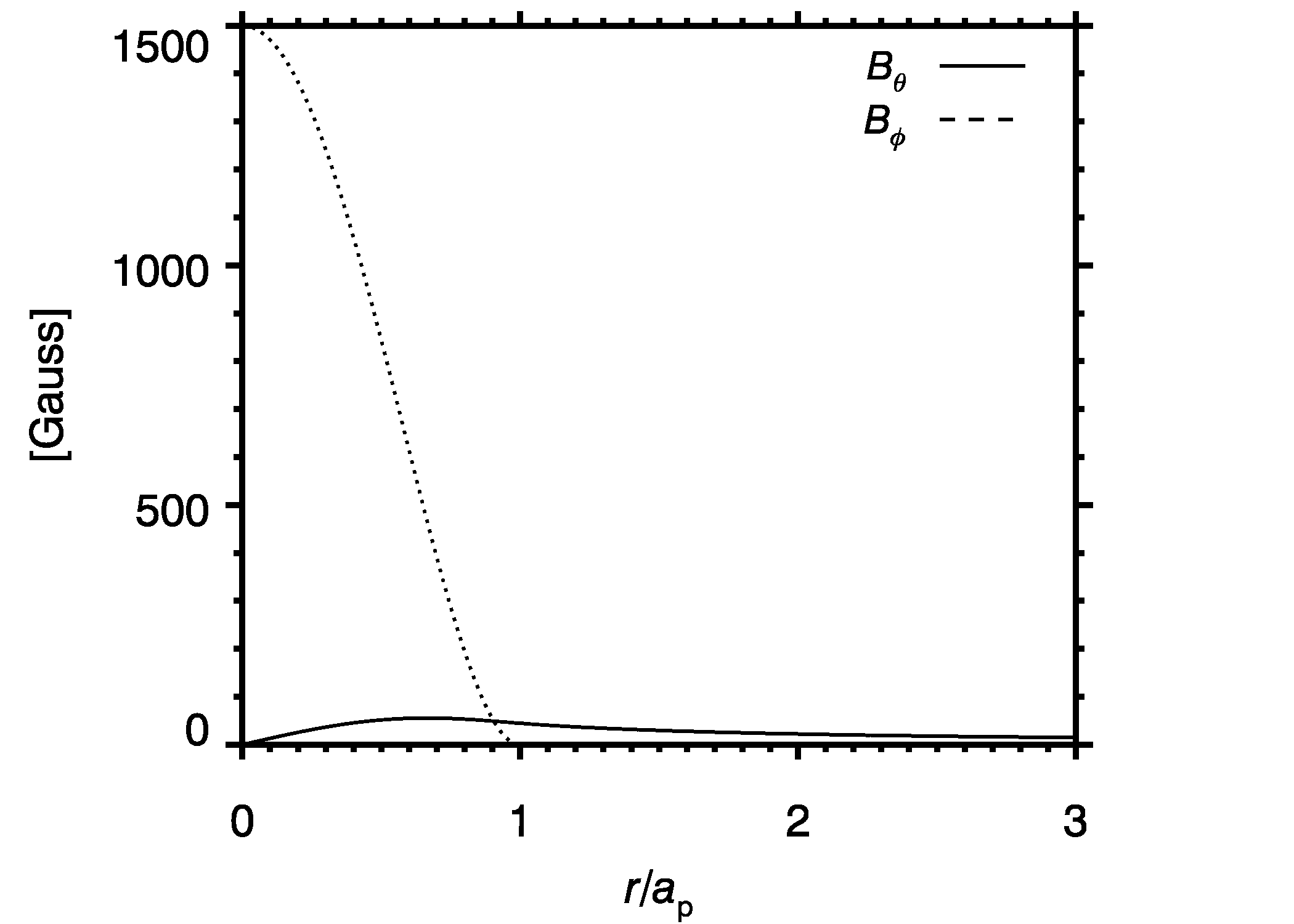}}
\caption{left: photospheric current-channel radius $\af$ (solid line)
  and poloidal magnetic field $B_{\poloidal\mathrm{p}}$ (dashed line)
  as a function of the average vertical magnetic field in the current channel
  $B_{\localz\mathrm{p}}$ assuming coronal parameters consistent with
  the 2000 September 12 CME trajectory: $\ac=\AC$~cm,
  $B_{\poloidal\mathrm{c}}=B_{\localz\mathrm{c}}\simeq4$~G. Right:
  photospheric magnetic field profiles for
  $B_{\localz\mathrm{p}}=\BTp$~G\label{fig:a2000}}
\end{figure*}
The flux-rope model has been fitted to the CME trajectory observed on 2000
September 12 \cite[]{Chen2006,Chen2010}. \cite{Chen2006} and \cite{Chen2010}
find that the observed dynamics is consistent with a peak poloidal
flux injection rate of
$d\Phi_\poloidal/dt=1.4\times10^{19}\,\mathrm{Mx}\,\mathrm{s}^{-1}$ and an
energy injection of $2\--4\times10^{32}\,\mathrm{erg}$ over 40 minutes (James
Chen, personal communication 2008 November). The average power requirement of
the flux-rope model
$\left\langle{dU_\poloidal/dt}\right\rangle\simeq1.3\times10^{29}\,\mbox{erg}\,\mbox{s}^{-1}$
is the \textit{lower limit} for the power budget of the CME over 40
minutes\----the instantaneous power budget is likely to exceed this value
$dU_\poloidal/dt\ge\left\langle{dU_\poloidal/dt}\right\rangle$.  The
separation distance between the filament footpoints and the flux-rope
footpoints were estimated to be $\sf\simeq3.5\times10^{10}$~cm and
$\Sf\simeq5\times10^{10}$~cm respectively which leads to a coronal
current-channel radius $\ac=\left(\Sf-\sf\right)/2\simeq\AC$~cm. The magnetic
field at the base of the corona is estimated to be
$B_{\poloidal\mathrm{c}}=B_{\localz\mathrm{c}}\simeq4$~G
\cite[]{Chen2010}. Figure~\ref{fig:a2000} shows the photospheric current-channel radius
$\af$ (solid line) and poloidal magnetic field $B_{\poloidal\mathrm{p}}$
(dashed line) as a function of the average vertical magnetic field in the
current channel determined from Equations~(\ref{eqn:krall1})
and~(\ref{eqn:krall2}).\par
\begin{figure*}[!t]
\centerline{\hskip0.25in\includegraphics[width=4in]{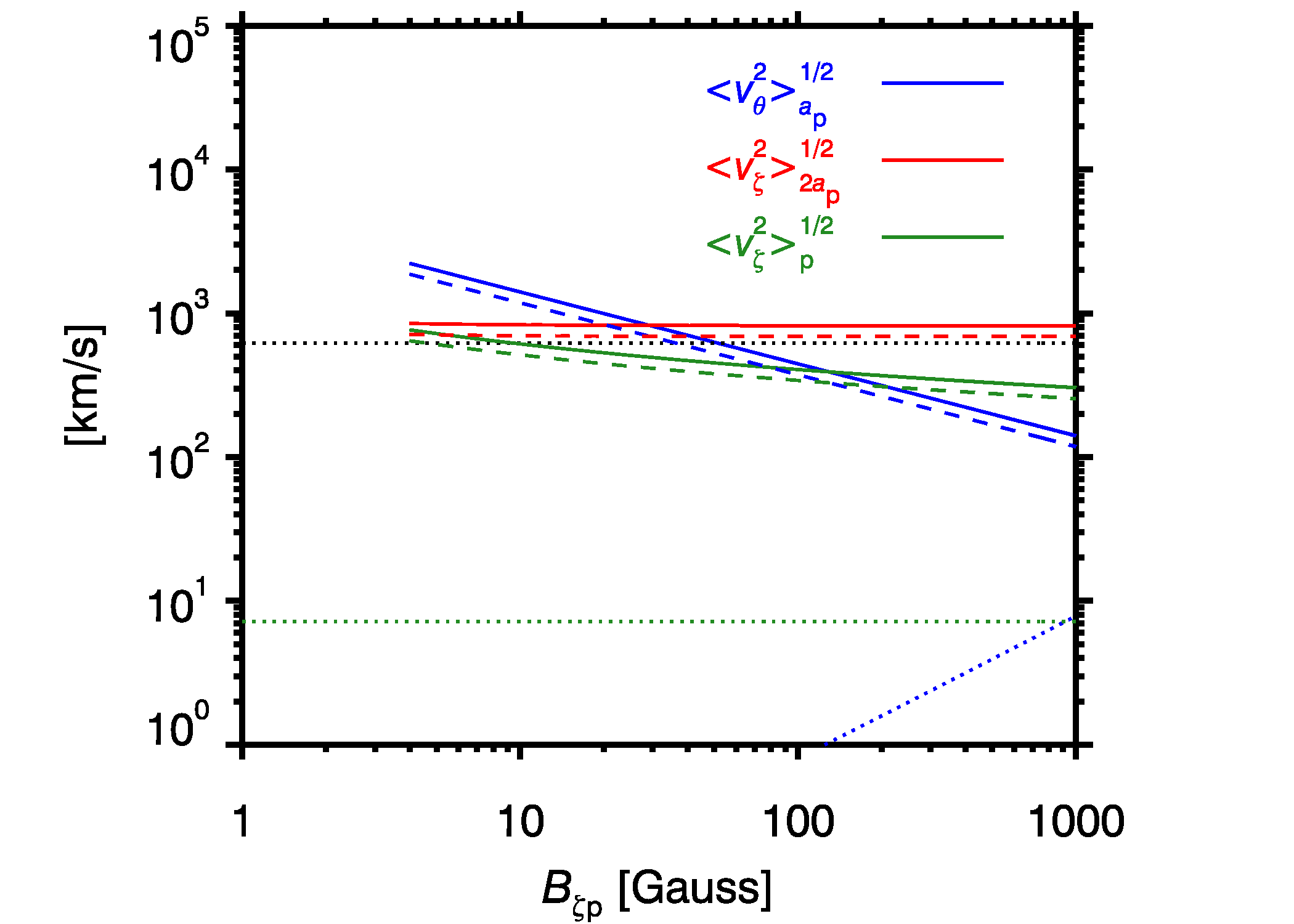}\hskip-0.5in\includegraphics[width=4in]{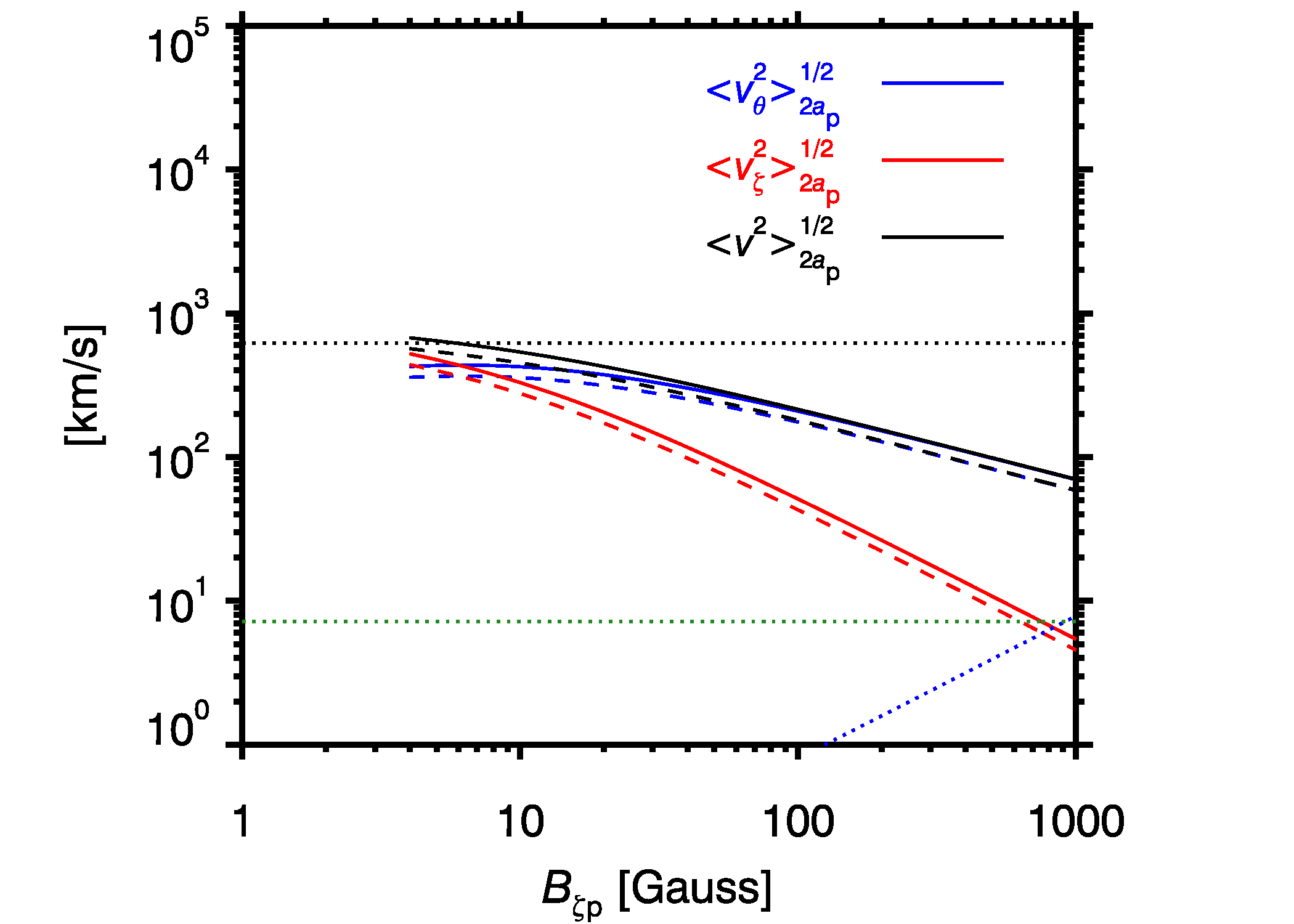}}
\caption{Optimal velocity estimates for the 2000 September 12 CME trajectory
  modeled by \cite{Chen2006} and \cite{Chen2010} as a function of the average
  vertical magnetic field in the current channel.  The solid lines correspond
  to rms velocities estimated from the energy budget and the dashed lines
  correspond to rms velocities estimated from the helicity budget. Left: the
  blue and red lines are the optimal twisting
  $\left\langle{v}_{\poloidal}^2\right\rangle^{1/2}_{\af}$ and emergence
  $\left\langle{v}_{\smash{\localz}}^2\right\rangle^{1/2}_{2\,\af}$ velocities
  at the flux-rope footpoint inside $r\le\af$ and $r\le2\,\af$
  respectively. The green lines indicate the emergence velocities
  corresponding to the minimum shear (constant velocity, $r<\rc$)
  solution. Right: the blue, red and black lines correspond to the poloidal
  $\left\langle{v}_{\poloidal}^2\right\rangle^{1/2}_{2\,\af}$ and vertical
  $\left\langle{v}_{\smash{\localz}}^2\right\rangle^{1/2}_{2\,\af}$ components
  of the optimal total velocity
  $\left\langle{v}^2\right\rangle^{1/2}_{2\,\af}$ respectively at the
  flux-rope footpoint inside $r\le2\,\af$. (both) The dotted black lines are
  the solar escape velocity $v_\sun=\vsun\,\mbox{km}\,\mbox{s}^{-1}$, the
  dotted green lines are the photospheric sound speed
  $C_{\mathrm{s}}\simeq\Cs$, and the dotted blue lines are the average
  Alfv\'en speed $\left\langle{V}_\mathrm{A}\right\rangle_{2\,\af}$.
\label{fig:V2000}}
\end{figure*}
Figure~\ref{fig:V2000} shows velocity estimates for the 2000 September 12 CME
trajectory modeled by \cite{Chen2006} and \cite{Chen2010} as a function of the
average vertical magnetic field in the current channel using a cutoff scale
$\rc=\Sf/2=2.5\times10^{10}$~cm.  The magnitude of the vertical magnetic field
can be interpreted as height with
$B_{\localz\mathrm{p}}=B_{\localz\mathrm{c}}=4$~G corresponding to the base of
the corona and $B_{\localz\mathrm{p}}\simeq\BTp$~G corresponding to the
photosphere.  The solid lines correspond to rms velocities estimated from the
energy budget and the dashed lines correspond to rms velocities estimated from
the helicity budget.  In the left panel, the blue and red lines are the
optimal twisting $\left\langle{v}_{\poloidal}^2\right\rangle^{1/2}_{\af}$ and
emergence $\left\langle{v}_{\smash{\localz}}^2\right\rangle^{1/2}_{2\,\af}$
velocities at the flux-rope footpoint inside $r\le\af$ and $r\le2\,\af$
respectively, corresponding to the minimum velocity solutions in
Appendices~\ref{app:power} and~\ref{app:helicity}. The green lines indicate
the emergence velocity corresponding to the minimum shear (constant velocity
$r<\rc$) solution. In the right panel, the blue, red and black lines
correspond to the poloidal
$\left\langle{v}_{\poloidal}^2\right\rangle^{1/2}_{2\,\af}$ and vertical
$\left\langle{v}_{\smash{\localz}}^2\right\rangle^{1/2}_{2\,\af}$ components
of the optimal total velocity $\left\langle{v}^2\right\rangle^{1/2}_{2\,\af}$
respectively with $v_\parallel=0$ at the flux-rope footpoint inside
$r\le2\,\af$. In both panels, the dotted black line is the solar escape
velocity $v_\sun=\vsun\,\mbox{km}\,\mbox{s}^{-1}$, the dotted green line is
the photospheric sound speed
$C_{\mathrm{s}}\equiv\sqrt{\gamma\,p_{\mathrm{p}}/\rho_{\mathrm{p}}}\simeq\Cs$
with $\gamma=5/3$, mass density $\rho_{\mathrm{p}}\simeq\Rhoph$, and pressure
$p_{\mathrm{p}}\simeq\Pph$ from VAL-C model for the quiet sun
\cite[]{Vernazza1981} interpolated to the $\tau=1$ height of $\Hp$ for
\ion{Ni}{1} 6767.8~\AA~line imaged by MDI \cite[]{Jones1989,Bruls1993}. The
dotted blue line is the average Alfv\'en speed
$\left\langle{V}_{\mathrm{A}}\right\rangle_{2\,\af}\equiv\left\langle{B}\right\rangle_{2\,\af}/\sqrt{4\,\pi\,\rho_{\mathrm{p}}}$
inside $r\le2\,\af$. The velocities based on the energy and helicity budgets
are in close agreement and this agreement is nontrivial as
Equations~(\ref{eqn:power}) and~(\ref{eqn:helicity}) indicate. Exact agreement
for the magnetic field profile Equations~(\ref{eqn:chen:Bp}) and
(\ref{eqn:chen:Bz}) corresponds to
\begin{equation}
\frac{dU_\poloidal}{dt}=\frac{B_{\poloidal\mathrm{c}}\,\ac}{4}\,\frac{d\Phi_\poloidal}{dt}.
\end{equation} \par
\begin{figure*}[!t]
\centerline{\hskip0.5in\includegraphics[width=4in]{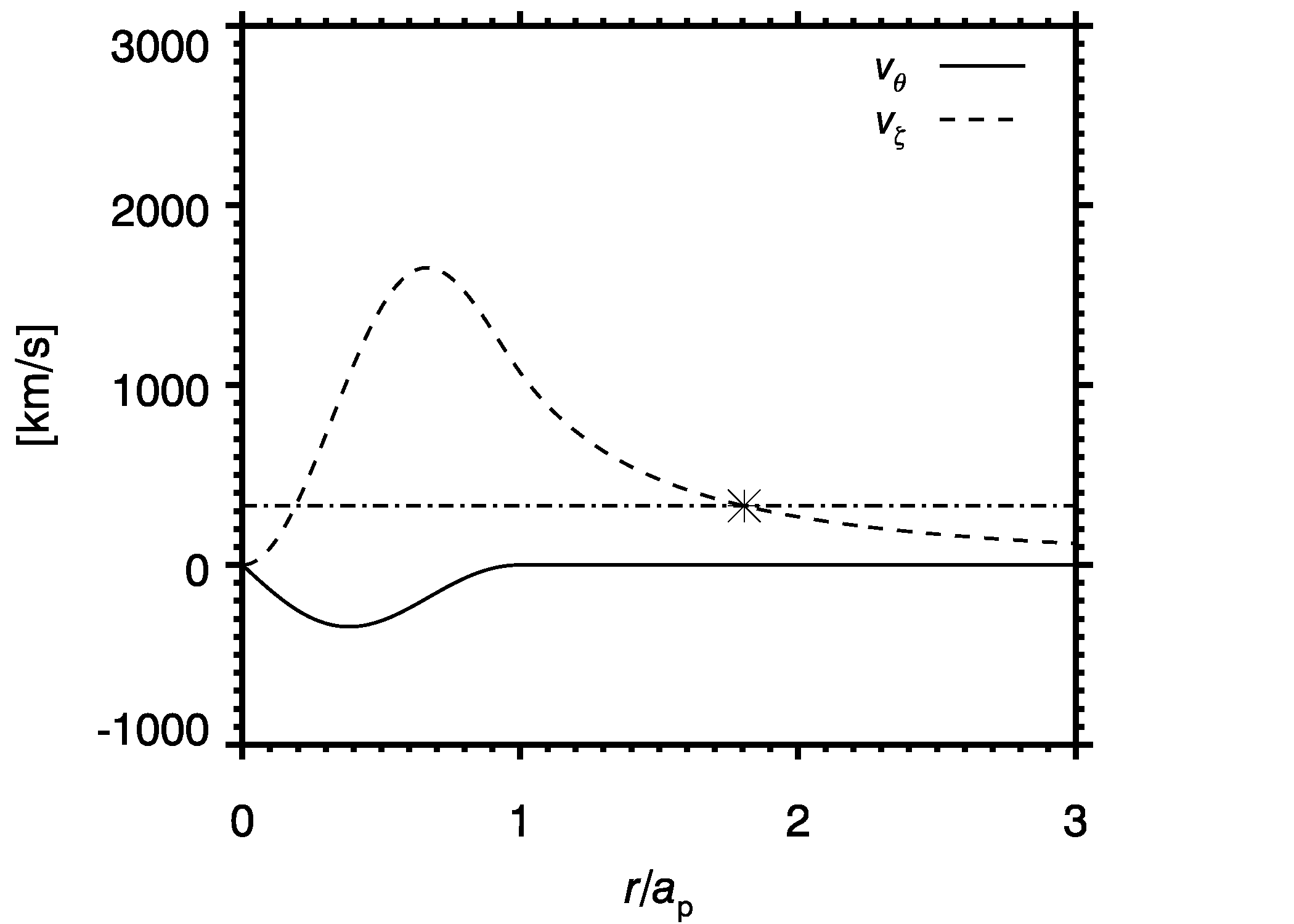}\hskip-0.5in\includegraphics[width=4in]{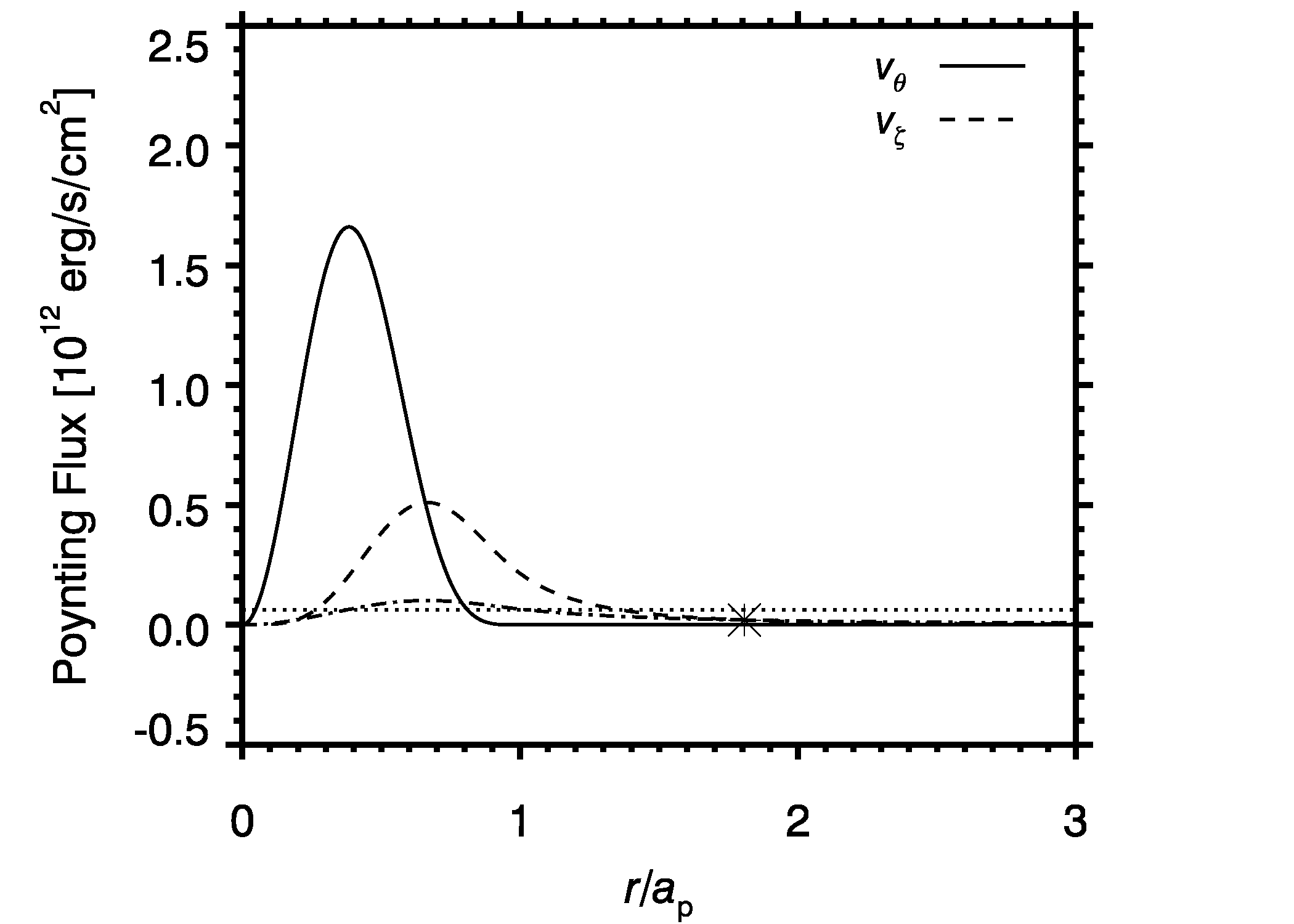}}
\caption{Example optimal velocity and Poynting flux profiles in the
  photosphere for the 2000 September 12 CME assuming
  $B_{\poloidal\mathrm{c}}=B_{\localz\mathrm{c}}\simeq4$~G and
  $B_{\localz\mathrm{p}}=\BTp$~G. The dot-dashed lines correspond to the
  minimum shear (constant velocity $r<\rc$) solutions for the vertical velocity. The
  crossing point between the optimal emergence and constant velocity solutions
  occurs at the asterisks. The radiation emittance at the solar surface
  $\mathcal{F}\simeq6.317\times10^{10}\,\mathrm{erg}\,\mathrm{s}^{-1}\,\mathrm{cm}^{-2}$
  is indicated by the horizontal dotted line.\label{fig:VP2000}}
\end{figure*}
Figure~\ref{fig:VP2000} shows example optimal velocity and Poynting flux
profiles\footnote{The optimal solutions for the energy flux and helicity flux
  are proportional with $B_\poloidal$ in the former replaced with
  $A_\poloidal$ in the latter and
  $B_\poloidal/A_\poloidal=2\,B_{\poloidal\mathrm{p}}/\af\,B_{\localz\mathrm{p}}=2\,B_{\poloidal\mathrm{c}}/\ac\,B_{\localz\mathrm{c}}\simeq2/\af$,
  The optimal velocity and helicity flux profiles are proportional to those
  shown in Figure~\ref{fig:VP2000}.} for the 2000 September 12 CME assuming
$B_{\poloidal\mathrm{c}}=B_{\localz\mathrm{c}}\simeq4$~G and
$B_{\localz\mathrm{p}}=\BTp$~G. The dot-dashed lines correspond to the minimum
shear (constant velocity $r<\rc$) solution for the vertical velocity. The
crossing point between the optimal emergence and constant velocity solutions
occurs at the asterisks.  The optimal emergence velocity profile $v_\localz$,
constrained by the energy budget, exceeds $1500\,\mbox{km}\,\mbox{s}^{-1}$
inside the current channel $r\lesssim\af$ and exceeds
$100\,\mbox{km}\,\mbox{s}^{-1}$ over most of the range $0<r\lesssim3\,\af$.
The Poynting flux should be compared with the radiation emittance at the solar
surface
$\mathcal{F}\simeq6.317\times10^{10}\,\mathrm{erg}\,\mathrm{s}^{-1}\,\mathrm{cm}^{-2}$
indicated by the horizontal dotted line in the right panel of
Figure~\ref{fig:VP2000}. All three velocity profiles produce Poynting fluxes
which exceed the radiation emittance at the solar surface and the fluxes of
white light flare kernels
$1\--2\times10^{10}\,\mathrm{erg}\,\mathrm{s}^{-1}\,\mathrm{cm}^{-2}$
\cite[]{Neidig1989}. The absence of strong photospheric signatures associated
with CMEs suggests that flux injection cannot be responsible for the CME
eruption.  \par
\subsection{Discussion}
\begin{figure*}[!t]
\centerline{\includegraphics[width=5in]{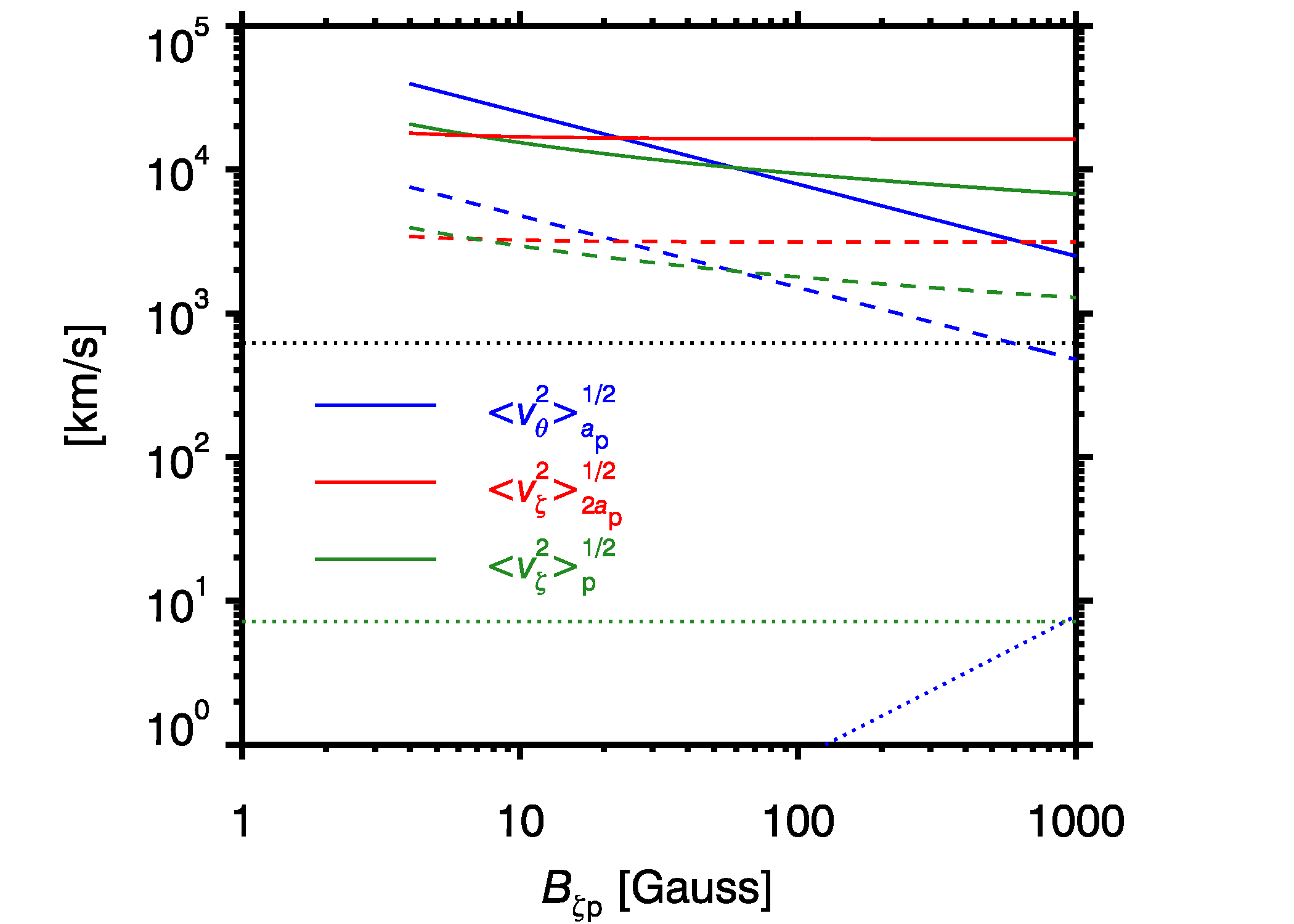}}
\caption{Optimal velocity estimates for the 2003 October 28, CME/ICME
  trajectory modeled by \cite{Krall2006a} in the same format as
  Figure~\ref{fig:V2000}.\label{fig:V2003}}
\end{figure*}
I emphasize that the 2000 September 12 event is not an extreme
CME. \cite{Krall2006a} describe the 2003 October 28 CME, first
observed in LASCO C3 images at 11:30~UT, that requires
$\Delta{U}_\poloidal\simeq2\times10^{33}\,\mathrm{erg}$ and
$\Delta\Phi_\poloidal\simeq6\times10^{22}~\mathrm{G}\,\mathrm{cm}^{-2}$ over
$\Delta t\simeq18$~minutes with $a_{\mathrm{c}}=8.1\times10^9\,\mathrm{cm}$,
$\Sf=3\times10^{10}$~cm and
$B_{\poloidal\mathrm{c}}=B_{\localz\mathrm{c}}=3.2$~G. The timescale of
approximately $15$~minutes and energy requirements $2\times10^{32}$~erg are
in close agreement with the estimates used by \cite{Manchester2008} to
simulate the initiation and propagation of this eruption.
Figure~\ref{fig:V2003} shows optimal velocity estimates for the 2003 October
28 CME/ICME trajectory modeled by \cite{Krall2006a} in the same format as
Figure~\ref{fig:V2000}. This event requires extreme photospheric velocities in
the range $v\simeq2500\--16000\,\mbox{km}\,\mbox{s}^{-1}$ to satisfy the energy budget and
$v\simeq500\--3000\,\mbox{km}\,\mbox{s}^{-1}$ to satisfy the helicity budget.\par
Returning to the 2000 September 12 event, Figures~\ref{fig:V2000}
and~\ref{fig:VP2000} indicate that significant photospheric signatures of
flux injection should be detectable in or near the current-channel radius
$r\lesssim2\,\af$.  For the left panel in Figure~\ref{fig:V2000},
corresponding to ideal footpoint twisting \textit{or} flux injection
(emergence), all of the rms velocities exceed the local characteristic speeds
of an MHD plasma.  The optimal rms emergence velocities
$\left\langle{v}_{\smash{\localz}}^2\right\rangle^{1/2}_{2\,\af}$
\textit{exceed the solar escape velocity}
$v_\sun=\vsun\,\mbox{km}\,\mbox{s}^{-1}$ inside of $r\lesssim2\,\af$. Thus,
gravitational forces cannot restrain the photospheric material in this
region. Even for emergence velocities of $100\,\mbox{km}\,\mbox{s}^{-1}$, the
photospheric material would require 12 minutes to return to the surface by
gravity alone and this is roughly the timescale of the 2003 October 28,
CME/ICME eruption. Consequently, a characteristic of flux injection should be
\textit{hypersonic} upflows concomitant with and preceding the
eruption. Optimal photospheric velocities imply a \textit{sustained} mass
transport rate of $6\times10^{20}\,\mbox{g}\,\mbox{s}^{-1}$ for just the
ring-shaped regions with $v_\localz\ge\vsun$ maintained over roughly
40~minutes and spatial scales of $\pi\,\ac^2\simeq2\times10^{20}\,\mbox{cm}^2$
in the corona (see Figure~\ref{fig:sun} for the scale of the twice the
current channel in the corona). These photospheric flows would eject a net
mass of $10^{24}$~g of which is 9~orders of magnitude larger than the typical
CME mass $10^{15}$~g as estimated from LASCO images. Ballistic photospheric
mass outflows of this magnitude should be straight-forward to detect using EIT
and LASCO C1 and C2 coronagraphs\----mass outflows of this magnitude are
not observed by these instruments.
It is worth repeating that the rms optimal emergence velocities in the
photosphere are largely independent of $B_{\localz\af}$ and the cutoff scale
$\rc$ and $\left\langle{v}_{\smash{\localz}}\right\rangle^{1/2}_{2\,\af}$ is
asymptotically determined by the values at the base of the corona
$B_{\localz\ac}$ and $\ac$.
The left panel of Figure~\ref{fig:V2000} appears to indicate that the
minimum shear velocity (green) is more efficient than either footpoint
twisting (blue) or flux injection (red) since it exhibits lower velocities
inside $r\le2\,\af$. However the left panel of Figure~\ref{fig:VP2000} shows
that this is simply because the profiles of the optimal twisting and emergence
velocities reduce to essentially zero beyond $r=\af$ and $r\gtrsim1.75\,\af$
respectively whereas the minimum shear velocity remains constant out to the
cutoff scale $\rc$. For the right panel in
Figure~\ref{fig:V2000}, corresponding to combined footpoint twisting
\textit{and} flux injection (emergence), flux emergence is only efficient when
$B_\poloidal\simeq B_\localz$. For $B_\localz\gg{B}_\poloidal$ footpoint
twisting is more efficient for transferring energy and helicity from the
photosphere to the corona. \par
\begin{figure*}[!t]
\centerline{\hskip0.25in\includegraphics[width=4in]{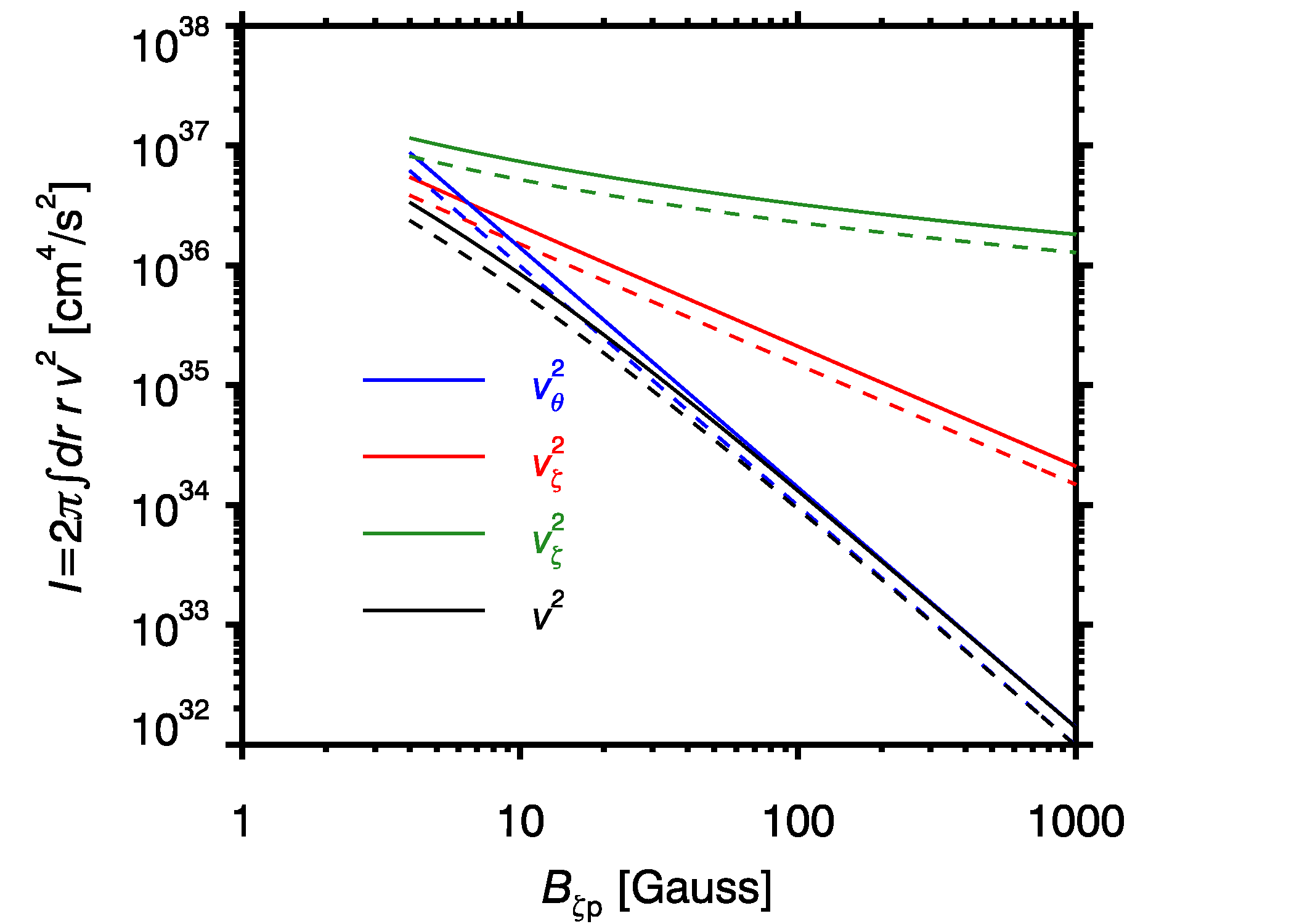}\hskip-0.25in\includegraphics[width=4in]{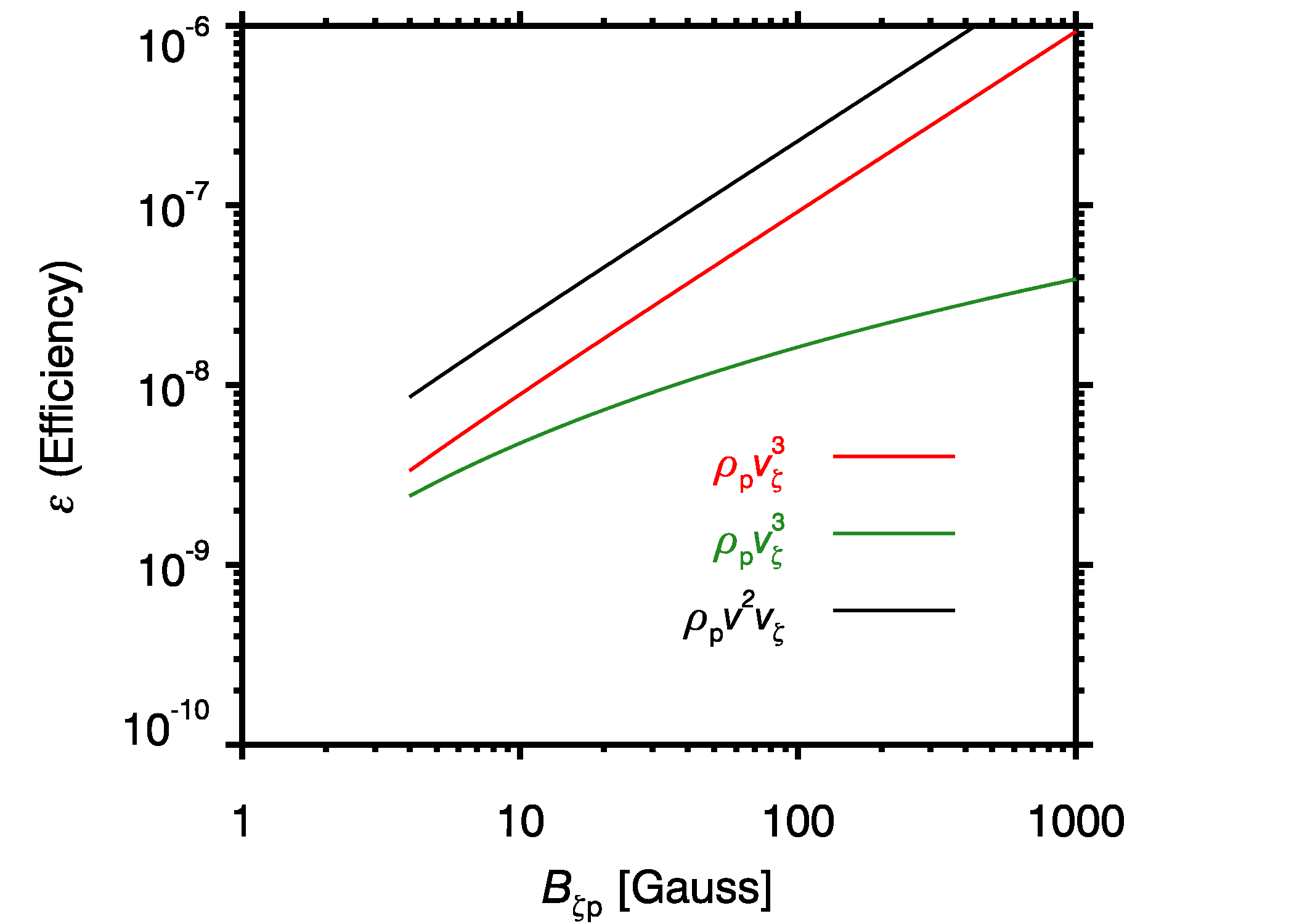}}
\caption{Efficiencies of the driving scenarios for the 2000 September 12
  CME. Left: density normalized kinetic energy $I=\mathcal{I}/\rho_\mathrm{p}$
  in a horizontal slice and right: ratio of the kinetic energy transport
  across the photosphere to the magnetic power requirements of the CME under
  the flux-injection hypothesis. The solid lines correspond to rms velocities
  estimated from the energy budget and the dashed lines correspond to rms
  velocities estimated from the helicity budget. Blue, red, green, and black
  lines correspond to optimal twisting, optimal emergence, minimum shear
  (constant velocity $r<\rc)$, and optimal combined twisting and emergence
  velocity profiles respectively (blue line is not shown in right
  panel).\label{fig:efficiencies}}
\end{figure*}
The efficiencies of the processes may be ranked either with the
magnitude of the integral
\begin{mathletters}
\begin{equation}
\mathcal{I}=2\,\pi\,\rho_{\mathrm{p}}\,\int_0^{\rc}{dr}\,r\,v^2,\label{eqn:I}
\end{equation}
or
\begin{equation}
\epsilon\equiv\frac{\left|dU_\poloidal/dt\right|}{\left|dU_\poloidal/dt\right|+2\,\pi\,{\rho_\mathrm{p}}\,\int_0^{\rc}{dr}\,r\,v^2\,v_\localz},\label{eqn:efficiencies}
\end{equation}
\end{mathletters}
where $F=\pi\,{\rho_\mathrm{p}}\,\int_0^{\rc}{dr}\,r\,v^2\,v_\localz$
is the kinetic energy flux through the photosphere integrated over one
footpoints.  The former metric is simply the photospheric kinetic
energy which is used to optimize the velocity profiles, whereas the
latter metric is the ratio of the power requirements of the CME to the
total energy transported across the photosphere under the flux-injection
hypothesis.\par
Figure~\ref{fig:efficiencies} shows the efficiencies of the driving scenarios
for the 2000 September 12 CME based on $\mathcal{I}$, the kinetic energy in a
horizontal slice (Equation~(\ref{eqn:I})) and $\epsilon$, the ratio of the
magnetic power requirements of the CME to the total energy transported across
the photosphere (Equation~\ref{eqn:efficiencies}) under the flux-injection
hypothesis in the left and right panels respectively. The solid lines
correspond to rms velocities estimated from the energy budget and the dashed
lines correspond to rms velocities estimated from the helicity budget. Blue,
red, green, and black lines correspond to optimal twisting, optimal emergence,
minimum shear, and optimal combined twisting and emergence velocities
respectively\----the blue line is not shown in the right panel because the
efficiency is $\epsilon=1$. Based on either criterion~(\ref{eqn:I})
or~(\ref{eqn:efficiencies}) the minimum shear (constant velocity $r<rc$) is
the \textit{least} efficient velocity profile (shown in green) for
transporting magnetic energy across the photosphere on the timescale of the
eruption. However, the most efficient velocity is different for $\mathcal{I}$
and $\epsilon$. Using the photospheric kinetic energy $\mathcal{I}$, the
combined twisting and emergence is the most efficient velocity profile (shown
as the black line in the left panel). Using the ratio of the magnetic power
requirements of the CME to the total energy transported across the photosphere
$\epsilon$, footpoint twisting is the most efficient velocity profile
$\left(\epsilon=1\right)$ because footpoint twisting does not transport mass
across the photosphere\----the velocity field is tangent to the surface.\par
The left panel Figure~\ref{fig:efficiencies} shows that, for small vertical
magnetic fields $B_{\localz\mathrm{p}}\lesssim10$~G, the kinetic energy
$\mathcal{I}$ for pure emergence is slightly less than for pure footpoint
twisting. For $B_{\localz\mathrm{p}}\gtrsim10$~G, the kinetic energy for
footpoint twisting less than for pure emergence. Consequently, the twisting
motions dominate the kinetic energy of the combined twisting and emergence
velocity profile since twisting motions are more efficient for
transporting energy and helicity into the corona. The combined twisting and
emergence velocity profile might be more efficient than the right panel of
Figure~\ref{fig:efficiencies} indicates. The rms vertical velocities for this
solution in Figure~\ref{fig:V2000} do not exceed the escape velocity and
significant photospheric material could \textit{eventually} return to the
surface via strong downflows along magnetic fields.\par
\section{COMPARISON AGAINST PHOTOSPHERIC OBSERVATIONS\label{sec:observations}}
The vertical photospheric plasma velocities in and near the current channel of
the CME footpoints must be large to satisfy the flux-injection
hypothesis. Such extreme velocities probably would have been detected in
previous studies of CME eruptions, but not no such observations have been
reported in the literature. Nonetheless, comparing expected values with
observations is a necessary final step to establish the likelihood that a
theory is compatible with nature. High spatial resolution high-cadence
($\sim1$~minute) line profiles would be ideal for examining the footpoints of
CMEs during eruptions. This suggests that limited field-of-view line-profile
data would be the best candidate data set \cite[see][for observations of
  coronal outflows during the gradual phase of a flare]{Harra2007,Imada2007}.
However, CME footpoints are always identified post-facto and are usually
outside the main flux concentration of the active region that is often the
focus of high-resolution campaigns. Thus, full-disk data are required to
ensure that the dynamics of both CME footpoints are captured. Full-disk
line-profile data are presently scarce. One possible candidate is the Naval
Research Laboratory \textit{Skylab} \ion{He}{2} $304~\AA$ spectroheliograms
with $2\arcsec$ spatial resolution, but these observations are usually at low
temporal cadence relative to flare/CME dynamics, in the wrong wavelength range
for photospheric observations, and are difficult to disambiguate\----the
spatial and wavelength information is convolved.  Finally, for direct
comparison with the flux-rope model, photospheric observations concomitant
with published results are desirable
\cite[]{Chen1997a,Chen2000a,Wood1999,Krall2001,Chen2006,Krall2006a,Chen2010}\----and
this event must correspond to a front-side CME to ensure that both footpoints
are visible in the photosphere.\par
After surveying the CMEs modeled with the flux-rope model, the 2000 September
12 CME \cite[]{Chen2006,Chen2010} was determined to be the best candidate
because: (1) the event was front side and associated with a M1.0 flare, (2)
the filament footpoint locations identified in \cite[]{Chen2006} were distinct
from the flare ribbons simplifying the interpretation of the observed emission
lines, and (3) the dynamics of the filament was captured by EIT, LASCO C2 and
C3 coronagraphs, and KSO H$\alpha$-observations.  The MDI instrument was in
Flarewatch mode on this day, but there is a gap in high-cadence coverage
during the flare between 07:00~UT and 15:30~UT. However, the solar
oscillations investigation (SOI) aboard \textit{SOHO} provides continuous
monitoring of Doppler velocities of low- to intermediate degree $l$
\cite[]{Scherrer1995}. The medium-$l$ data are the result of Gaussian spatial
filtering to mitigate spatial aliasing followed by a reduction in resolution
from $1024\times1024$~pixels at $2\arcsec\,\mathrm{pixel}^{-1}$ to
$192\times192$ at roughly $10\arcsec\,\mathrm{pixel}^{-1}$. This is
supplemented with level 1.8.2 $2\arcsec$ 5-minute averaged 96-minute cadence
magnetograms with the most recent magnetic field inter-calibrations
\cite[]{Tran2005,Ulrich2009}.  \par
\begin{figure*}[!t]
\centerline{\includegraphics[width=5in]{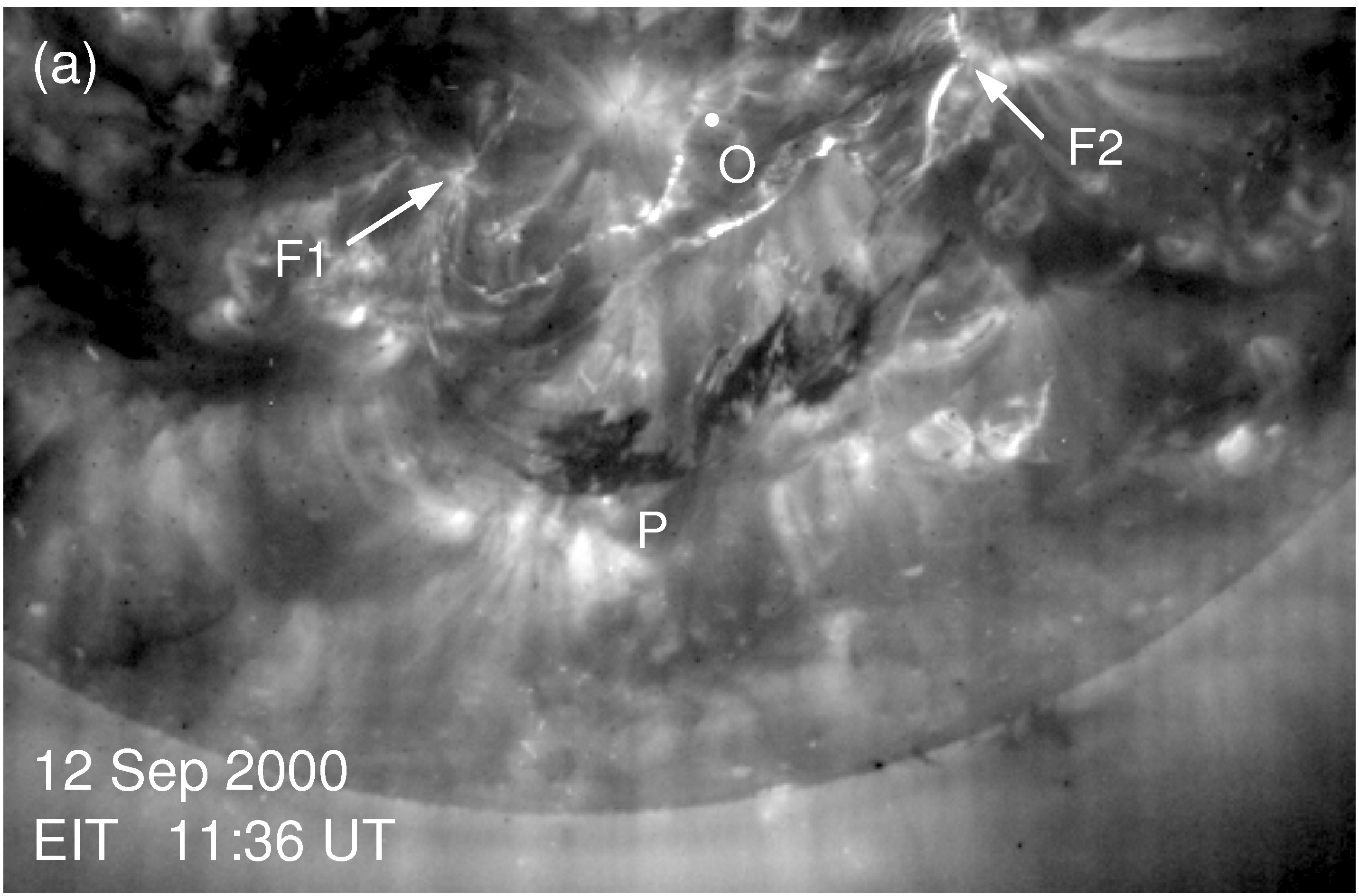}}
\caption{Source region of eruption observed by EIT (\ion{Fe}{12} 195 \AA) at
  11:35 UT. The erupting prominence, denoted by ``P'' appears as a dark
  absorption feature. The prominence footpoints are indicated ``F1'' and
  ``F2'' and their midpoint is designated ``O.'' The time 11:36 in the figure
  indicates the \textit{uncorrected start time}. After Figure 2(a) in
  \cite{Chen2006}.\label{fig:EIT}}
\end{figure*}
\cite{Chen2006} identified the locations F1 and F2 of the prominence
footpoints in the EIT (\ion{Fe}{12} 195 \AA) image shown in Figure~\ref{fig:EIT}. 
\begin{figure*}
\centerline{\includegraphics[viewport=17 40 438 305 ,clip=,width=5in]{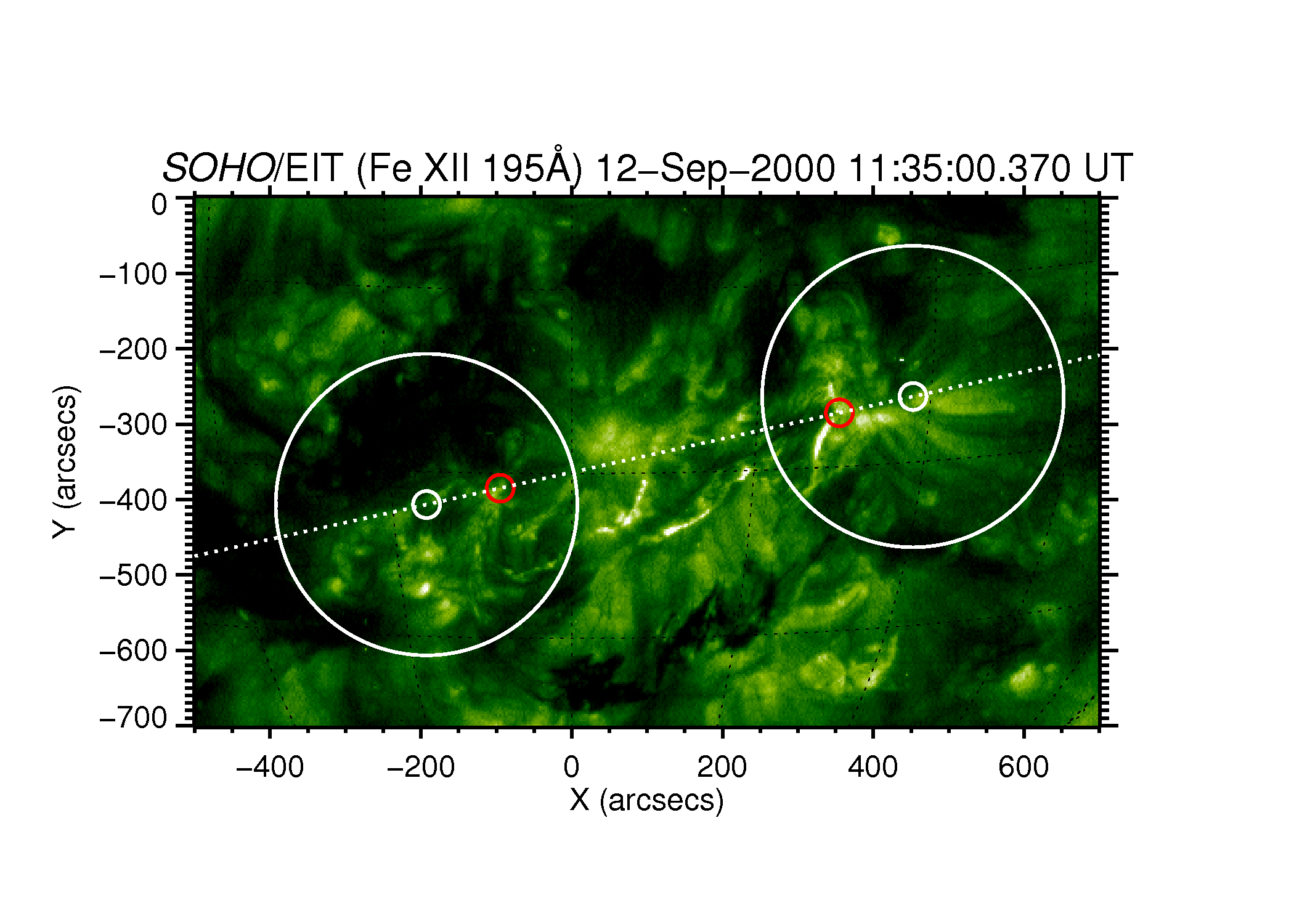}}
\centerline{\includegraphics[clip=,width=5in]{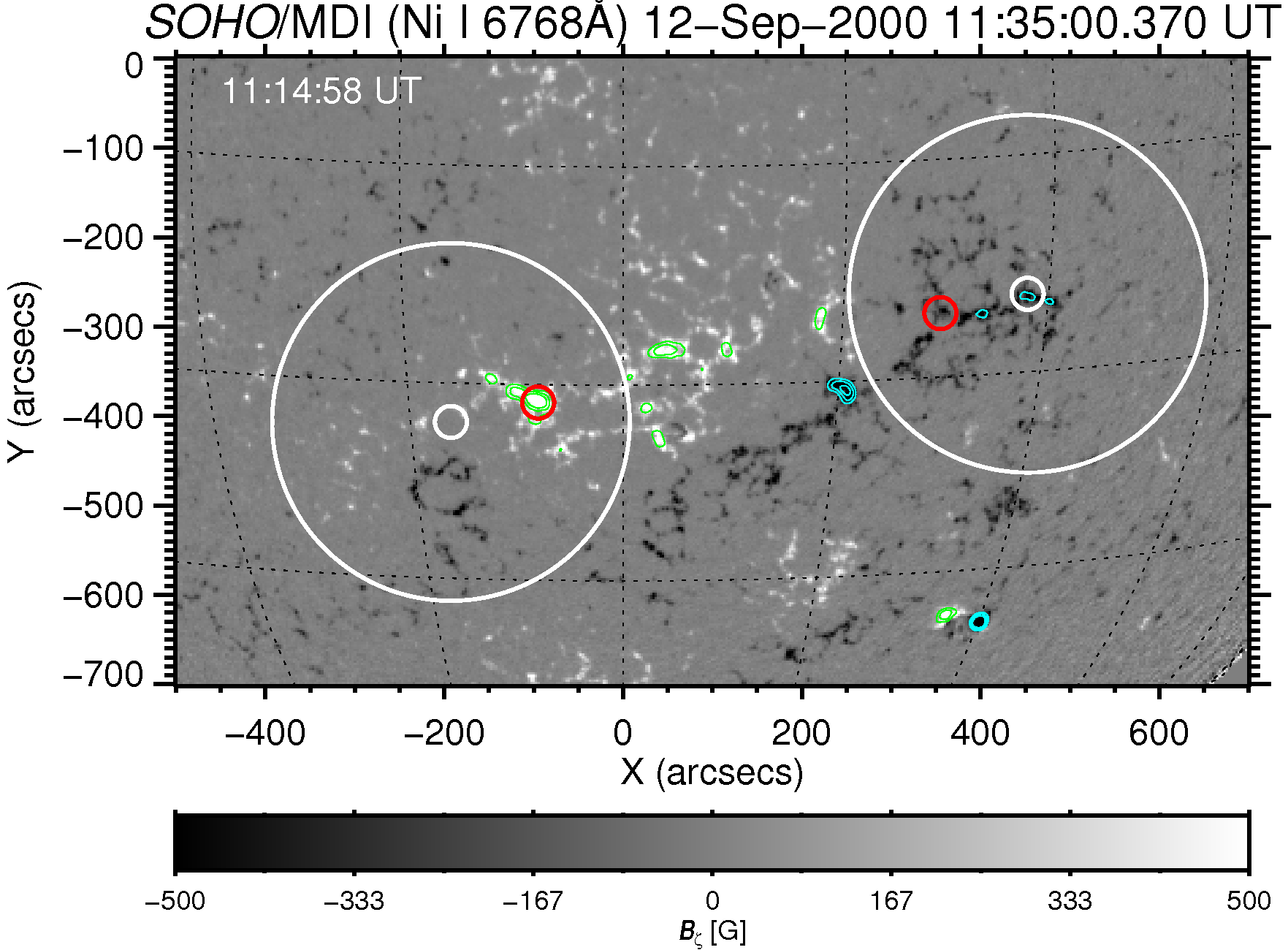}}
\caption{Nominal geometry of the flux-rope model for the base of the corona
  and photosphere. Top: EIT (\ion{Fe}{12} 195 \AA) at 11:35 UT.  Bottom:
  MDI (\ion{Ni}{1} 6767.8~\AA)
  at 11:15~UT de-rotated to 11:35 UT to match the time of the EIT image. The
  red circles correspond to the filament footpoints in
  Figure~\ref{fig:EIT}. The small and large white circles correspond to the
  extent of twice the current-channel radius in the photosphere and corona
  respectively assuming $B_{\localz\mathrm{p}}\simeq\BTp$~G and
  $B_{\localz\mathrm{c}}\simeq\BTc$~G. Cyan and green contours correspond to
  average magnetic fields of (-500,-400,-300) and (300,400,500)~GE
  respectively on a scale size of
  $1.5\times10^{18}\,\mbox{cm}^2$. \label{fig:sun}}
\end{figure*}
Using these locations, the nominal geometry of the flux-rope model for the
base of the corona and photosphere is diagramed in Figure~\ref{fig:sun}.  The
top shows the EIT (\ion{Fe}{12} 195 \AA) image at 11:35 UT and the bottom shows the MDI
(\ion{Ni}{1} 6767.8~\AA) at 11:15~UT differentially de-rotated to 11:35 UT to match the
time of the EIT image using Dominic Zarro's mapping package.\footnote{The
  SolarSoft mapping software is located at
  \url{http://www.lmsal.com/solarsoft/gen/idl/maps/}.}  The filament and
nominal current-channel footpoints are distinct from the flare ribbons in the
EIT image. The dark curved adsorbtion feature in the EIT image spanning
-175$\arcsec$ to 400$\arcsec$ in $X$ and -375$\arcsec$ to -675$\arcsec$ in $Y$
is the filament.  \par
The MDI level 1.8.2
magnetograms (BLDVER18=60100) incorporate the latest sensitivities from
inter-calibrations with the Mount Wilson Observatory
\cite[]{Tran2005,Ulrich2009}. The line-of sight magnetic field was corrected for
geometrical effects with a factor of $\mu^{-1}$ assuming that the
field is purely vertical where 
\begin{equation}
\mu=\frac{\Rsun-\Dsun\,\sqrt{1-\varrho_1^2/\rsun^2}}{\sqrt{\Rsun^2-2\,\Rsun\,\Dsun\,\sqrt{1-\varrho_1^2/\rsun^2}+\Dsun^2}}\approx\sqrt{1-\varrho_1^2/\rsun^2},
\end{equation}
where the impact parameter $\varrho_1$ (radians) is measured from disk center,
$\rsun$ is the radius of the Sun in the telescope (radians or arcsecs), and
$\Rsun/\Dsun$ is the ratio of the radius of the Sun to the distance between
the observer and Sun center. 
\cite{Chen2006} argue that the filament footpoints
shown in red and the nominal current-channel footpoints are not co-located,
but are related by 
\begin{equation}
\Sf=\sf+2\,\ac.\label{eqn:geom}
\end{equation}
This idealized geometry is reflected in Figure~\ref{fig:sun} where the red
circles correspond to the filament footpoints in Figure~\ref{fig:EIT} and the
large and small white circles correspond to the extent of \underline{twice}
the current-channel radius in the corona and photosphere respectively. The
left large white circle is associated with mainly positive magnetic field and
the right large white circle is associated with mainly negative magnetic
field.\par
\begin{figure*}[!t]
\ifx\xfig\undefined
\centerline{\includegraphics[viewport=126 462 486 735,clip=,width=5in]{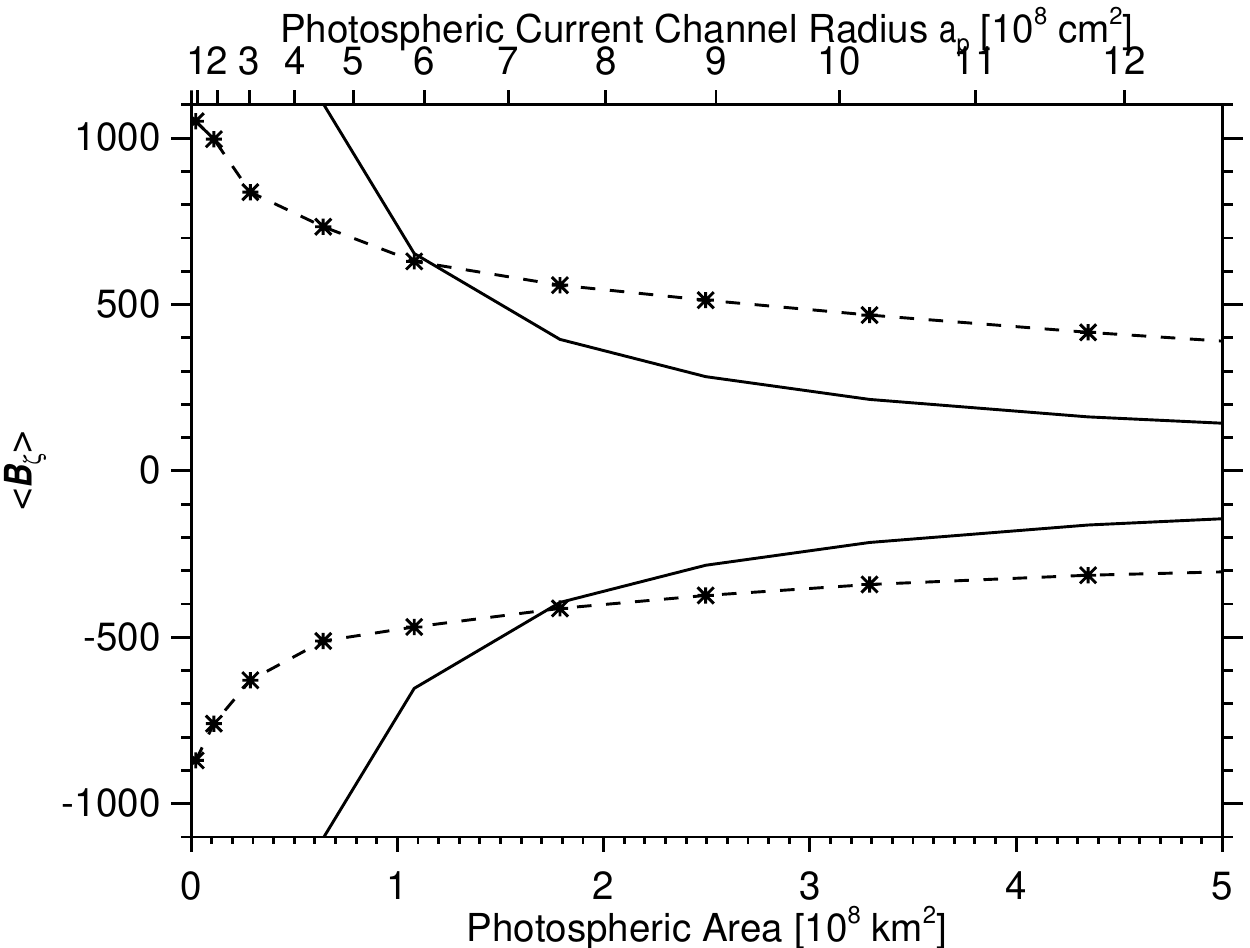}}
\else
\centerline{\includegraphics[viewport=126 462 486 735,clip=,width=5in]{idl/eps/mag_footpoints}}
\fi
\caption{Estimate for the vertical magnetic field (flux-rope toroidal magnetic
  field) in the photosphere.  Asterisks connected by dashed lines are the
  largest magnitude MDI values adjusted for line-of-sight projection and
  averaged over circular masks with increasing areas within the large white
  circles of Figure~\ref{fig:sun}.  Solid lines are
  $B_{\localz{\mathrm{p}}}=B_{\phi{\mathrm{c}}}\,\ac^2/\af^2$ estimated from
  the flux-rope model with $B_{\phi{\mathrm{c}}}=\BTc$~G and $\ac=\AC$~cm. The
  top and bottom lines correspond to the left and right footpoints of the
  flux rope. See the text for discussion.\label{fig:mag_footpoints}}
\end{figure*}
The current-channel radius in the photosphere should be determined by
enforcing consistency between the flux-rope model photospheric extrapolation
and the MDI magnetogram though flux
conservation (Equation~(\ref{eqn:flux})). The current-channel radius at the base
of the corona is $B_{\localz\mathrm{c}}\simeq4$~G with a radius $\ac=\AC$~cm
\cite[]{Chen2006}.  If all the toroidal flux in the coronal current channel
mapped to a single MDI pixel at disk center the resulting magnetic field would
exceed 30~kG. Such extreme values of magnetic field have never been observed
in the photosphere.  Consequently, the photospheric current channel will be
larger than one MDI pixel except perhaps very near the limb. To estimate the
current-channel size in the photosphere while minimizing discrepancies between
the photospheric observations and the flux-rope model, circular masks of
increasing scale size were computed. These masks were convolved with the MDI
values adjusted for line-of-sight projection in regions bounded by both the
left and right large white circles in Figure~\ref{fig:sun}. At each scale, the
maximum magnitude magnetic fields were selected from the output of the
convolution and plotted as asterisks connected by dashed lines in
Figure~\ref{fig:mag_footpoints}; these values correspond to the maximum
magnitude average magnetic field at that scale size contained within the
regions bounded by the large white circles in Figure~\ref{fig:sun}.  The solid
lines are $B_{\localz{\mathrm{p}}}=B_{\phi{\mathrm{c}}}\,\ac^2/\af^2$
estimated from the flux-rope model with $B_{\phi{\mathrm{c}}}=\BTc$~G and
$\ac=\AC$~cm. The top and bottom lines correspond to the left and right
footpoints of the flux rope.  The photospheric extrapolation of flux-rope
model agrees with the MDI magnetogram where the solid and dashed lines cross.
The photospheric area most consistent with the constraints of the flux-rope
model is $1\--2\times10^8\,\mbox{km}^2$ corresponding to a current-channel
radius of $\af\simeq\AP\,\mbox{cm}$ and an average photospheric toroidal field
of $\left\langle{B}_{\localz{\mathrm{p}}}\right\rangle\simeq\BTp$~G.  \par
The geometrical relationship~(\ref{eqn:geom}) proposed in \cite{Chen2006}
isn't entirely consistent with the photospheric observations in
Figure~\ref{fig:sun}.  For the left footpoint, the main concentration of
positive flux is associated with the filament footpoint not the nominal
current channel. Indeed the magnetic field in the nominal current channel
ranges from -24 to +24~G. For the right footpoint the magnetic field
pixel ranges in the filament footpoint and the nominal current channel
are similar ranging from $-665$ to $-42$~G and $-600$ to $-62$~G
respectively.  However, the average magnetic field magnitude in the right
nominal filament footpoint does not exceed 300~G whereas the magnitude in the
nominal current channel is somewhat higher as indicated by the cyan contour
corresponding to -300~G contained within the small white circle.
\begin{figure*}[!t]
\centerline{\includegraphics[width=5in]{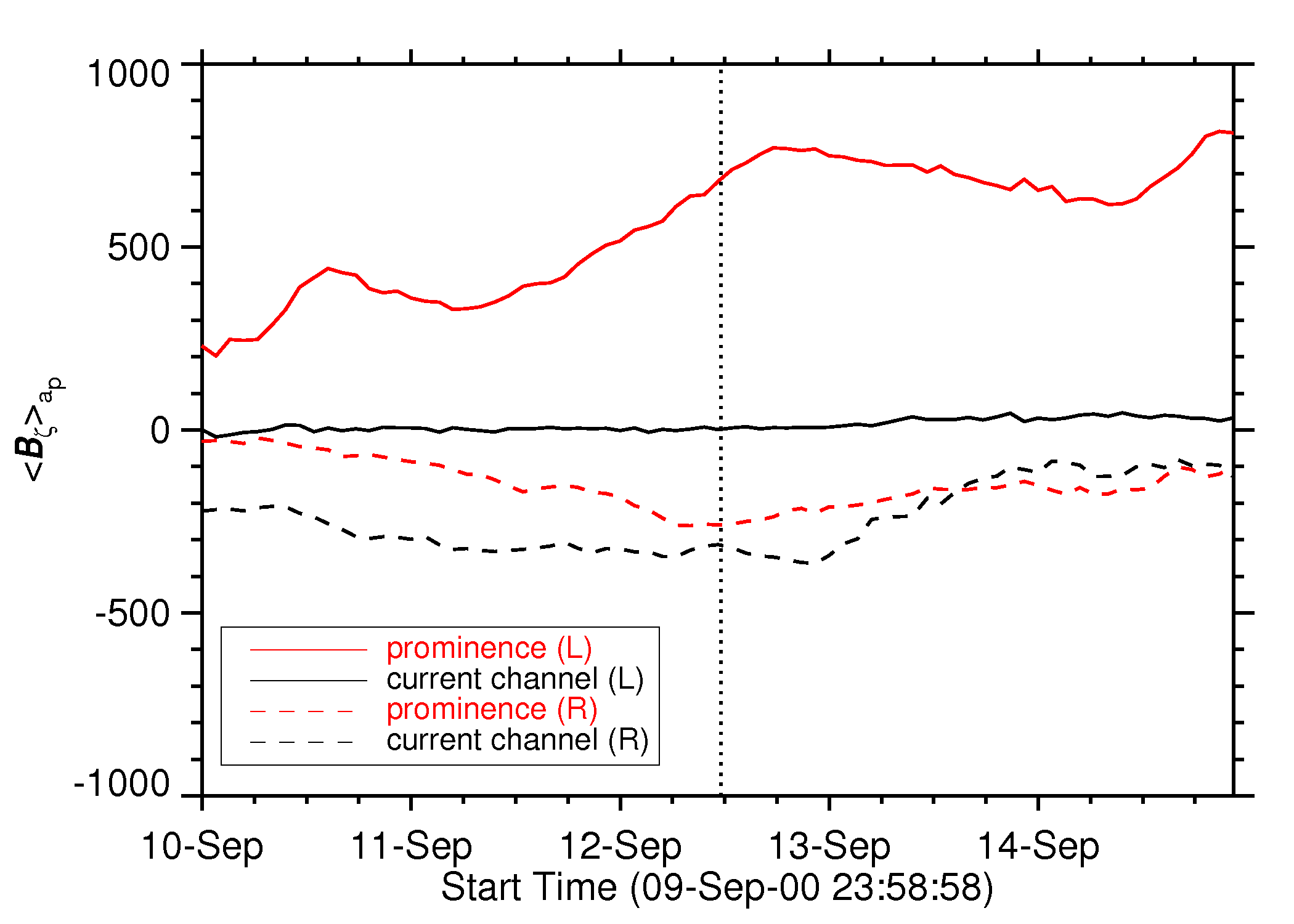}}
\caption{Temporal development of the average vertical magnetic field
  within $r\le\af\simeq\AP$~cm for the left (L) and right (R) filament and
  current-channel footpoints shown in Figure~\ref{fig:sun}. The vertical
  dashed line corresponds to the rise of the M-class flare.\label{fig:avgmag}}
\end{figure*}
Figure~\ref{fig:avgmag} shows the temporal development of the average vertical
magnetic field within $r\le\af\simeq\AP$~cm for the left (L) and right (R)
filament and current-channel footpoints shown in Figure~\ref{fig:sun}.  The
vertical dashed line corresponds to the rise of the M-class flare. Around the
time of the flare, the vertical flux is strengthening at the left
prominence footpoint at a rate of $8\times10^{15}\,\mbox{Mx}\,\mbox{s}^{-1}$,
but the flux in the nominal left current channel remains near zero over the
whole five day period. The average fluxes at the right prominence and
current-channel footpoints are flat around the eruption time. The
flux-imbalance between the left and right footpoint indicates that the
connectivity between these photospheric magnetic features is not trivial.\par
While gravity certainly effects the height of the heavy prominence material
potentially producing a difference between the position of the prominence
material and the center of the current channel at the apex of the flux rope,
there is no stated reason in \cite{Chen2006} for why the flux-rope
current-channel and prominence footpoints aren't co-located in the
photosphere.\footnote{Although, the radius of the current channel at the
  footpoint is defined at the base of the corona the footpoints separation is
  fixed by the ``dense subphotospheric plasma'' \cite[see pp.~457
    in][]{Chen1989} or the ``massive photospheric density'' \cite[see
    pp.~2320]{Chen1993} \cite[see also][]{Krall2000,Chen2008}. This implies
  that $\Sf$ and $\sf$ should be interpreted as photospheric footpoint
  separations \cite[see also][]{ChenAGU2004}.} The footpoint separation $\Sf$
is a critical parameter because the height $H_\mathrm{max}$ of the maximum
acceleration for the CME scales with $\Sf$ in the flux-rope model
$\Sf/2\leq{H}_\mathrm{max}\lesssim3/2\,\Sf$ \cite[]{Chen2006,Chen2007}. If
$\Sf$ is not consistent with observations, then the flux-rope model cannot be
correct.\par
\subsection{Doppler Data Preparation}
The MDI vector-weighted Dopplergrams were analyzed following the procedures
outlined in \cite{Snodgrass1984} and \cite{Hathaway1992}. First, the motion of
the observer was removed from each Dopplergram using
\begin{equation}
V_{\mathit{SOHO}}\left(\varrho_1,\thetaCCW\right)=V_R\,\left(1-\varrho_1^2/2\right)+V_\mathrm{W}\,\varrho_1\,\sin\,\thetaCCW-V_\mathrm{N}\,\varrho_1\,\cos\thetaCCW,
\end{equation}
where the impact parameter $\varrho_1$ (radians) is measured from disk center and
the position angle $\thetaCCW$ (radians) measured counterclockwise from solar
north are the heliocentric radial coordinates and $V_R$, $V_\mathrm{W}$, and
$V_\mathrm{N}$ are the \textit{SOHO} satellite velocities radial outward,
westward parallel to equator, and northward along the rotation axis,
respectively, using the appropriate keywords provided with the MDI data.\par
Second, the Dopplergrams are co-registered and time-averaged. The time-averaged
Dopplergram was fitted with orthonormal disk functions to eliminate cross-talk between
coefficients \cite[]{Snodgrass1984}.
\begin{mathletters}
\begin{equation}
\overline{V}\left(B_0,\varrho,\Lat,\Phi\right)=\omega\left(\Lat\right)\,R_\sun\,\sin\Phi\,\cos
B_0+V_{\mathrm{LS}}\left(\varrho\right)+V_{\mathrm{MF}}\left(\Lat\right)+H\,\sin\Lat,\label{eqn:Vbar}
\end{equation}
where differential rotation profile 
\begin{equation}
\omega\left(\Lat\right)=\sum_{n=0}^2A_{2n}\,\mathrm{T}^1_{2n}\left(\sin\Lat\right)
\end{equation}
is expanded in a truncated series of even orthonormal Gegenbauer polynomials
\cite[]{Morse1953a}, the limbshift function is represented by 
\begin{equation}
V_{\mathrm{LS}}\left(\varrho\right)=\sum_{n=0}^4L_n\,\mathcal{L}_n\left(1-\cos\varrho\right),
\end{equation}
with \cite[see pp. 174\--175 in][]{Smart1977}
\begin{equation}
\varrho=-\varrho_1+\arcsin\left[\Dsun\,\sin\left(\varrho_1\right)/\Rsun\right],
\end{equation}
and the meridional flow represented by 
\begin{equation}
V_{\mathrm{MF}}\left(\Lat\right)=\sum_{n=1}^{2}M_n\,\mathcal{M}_m\left(\sin\Lat\right)\,S\left(\Lat\right)\,\left(\cos\Phi\,\sin\Lat\,\cos{B_0}-\cos\Phi\,\sin{B_0}\right),
\end{equation}
\end{mathletters}
are expanded in truncated series of $\left(1-\cos\varrho\right)$ and Fourier
series of latitude respectively where the function classes
$\mathcal{L}_n\left(x\right)$ and $\mathcal{M}_n\left(x\right)$ were
orthonormalized by the Gram\--Schmidt procedure on the interval
$\left(0,1\right)$ by \cite{Snodgrass1984}. The latitude, longitude, and
solar-B angle are denoted $\Lat$, $\Lon$ and $B_0$ respectively and
$\varrho=\arcsin\left(\varrho_1/\rsun\right)$ is the angle measured from the
center of the Sun between the sub-Earth point and a point on the surface of
the Sun and $S\left(\Lat\right)=+1$ for $\Lat>0$ and $S\left(\Lat\right)=-1$
for $\Lat<0$.
\newcommand{\dopplercaption}{Fit-determined Coefficients with
  $\chi^2=$\chiall~for a Reduced Set of Disk-orthonormalized Functions
  Determined from \limbhours~hr of MDI Vector Weighted Dopplergams on 2003
  September 12. Estimates $\widehat{\overline{\mu}}$ and population standard
  deviations $\widehat{\sigma}_{\overline{\mu}}$ from Table~2 in
  \cite{Snodgrass1984} for data taken between 1967 January 1 and 1984 March 5
  at the Mount Wilson Solar Observatory.}
\newcommand{\chiall}{3898.0}
\newcommand{\dofall}{25455}
\newcommand\Bot{\rule[-1.5ex]{0pt}{0pt}}
\newcommand\Top{\rule{0pt}{3.2ex}}
\begin{deluxetable*}{ccccccccccccc}
\tabletypesize{\footnotesize}
\newcommand{\limbhours}{24}
\tablecaption{\dopplercaption\label{tab:limb_all}}
\tablehead{
\colhead{}&\multicolumn{3}{c}{$\mu\,\mbox{Rads}\,\mbox{s}^{-1}$}&
\colhead{\phantom{h}}&\multicolumn{8}{c}{$\mbox{m}\,\mbox{s}^{-1}$}\\
\cline{2-4}\cline{6-13}
\colhead{\Top}&
\colhead{          $A_0$}  &
\colhead{          $A_2$}  &
\colhead{          $A_4$}  &
&
\colhead{          $L_0$}  &
\colhead{          $L_1$}  &
\colhead{          $L_2$}  &
\colhead{          $L_3$}  &
\colhead{          $L_4$}  &
\colhead{          $M_1$}  &
\colhead{          $M_2$}  &
\colhead{$H$}}
\startdata
$\widehat{\theta}$&
 3.1870  &
-0.1610  &
-0.0216  &
&
131  &
174  &
 88  &
 10  &
-3.1  &
-17.4  &
0.6  &
29.1 \\
$\widehat{\sigma}_\theta$&
                                         0.0021  &
                                    \phs 0.0027  &
                                    \phs 0.0027  &
&
                                    \phn 19  &
                                    \phn 26  &
                                         21  &
                                         11  &
                                    \phs 3.5  &
                                \phs\phn  3.0  &
                                        0.4  &
                                    \phn 5.0 \\
\noalign{\vskip0.05in}
$\widehat{t}_\theta$&
 1496.81  &
  -59.19  &
   -8.05  &
&
    6.89  &
    6.64  &
    4.19  &
    0.92  &
   -0.88  &
   -5.76  &
    1.59  &
    5.87 \\
\%\Bot&
0.00  &
0.00  &
0.00  &
&
 0.0  &
 0.0  &
 0.0  &
35.8  &
37.7  &
 0.0  &
11.1  &
 0.0 \\
\hline
$\widehat{\overline{\mu}}$\Top&
 3.1556  &
-0.1610  &
-0.0312  &
&
127  &
160  &
 88  &
 14  &
 2.0  &
  6.2  &
0.1  &
 8.1 \\
$\widehat{\sigma}_{\overline{\mu}}$\Bot&
                                         0.1366  &
                                    \phs 0.0238  &
                                    \phs 0.0136  &
&
                                        252  &
                                    \phn 29  &
                                         24  &
                                         12  &
                                         5.4  &
                                         36.0  &
                                        8.6  &
                                        77.1 \\
\hline
$\widehat{t}_{\theta\overline{\mu}}$\Top&
   0.23  &
   0.00  &
   0.69  &
&
   0.02  &
   0.34  &
  -0.00  &
  -0.25  &
  -0.79  &
  -0.65  &
   0.06  &
   0.27 \\
\%&
  85.63  &
  99.99  &
  61.49  &
&
  98.98  &
  78.94  &
  99.92  &
  84.30  &
  57.42  &
  63.10  &
  96.16  &
  83.13 \\
\enddata
\end{deluxetable*}
 
The function~(\ref{eqn:Vbar}) was fitted to the time-averaged Dopplergram with
the standard deviations estimated from the variance of each pixel from its
respective time average. The results, with a weighted $\chi^2=\chiall$ and
\dofall~degrees of freedom, are summarized in Table~\ref{tab:limb_all}. The
first four rows of entries represent the best estimate of each parameter
$\widehat{\theta}$, their standard deviations $\widehat{\sigma}$, their
$t$-scores $\widehat{t}\equiv\widehat{\theta}/\widehat{\sigma}$, and their
two-sided significance probability in percent that the coefficients would be
larger by chance\----the smaller the probabilities, the more significant the
coefficients. The fifth and sixth rows represent the average parameter
estimates $\widehat{\overline{\mu}}$ and population standard deviations
$\widehat{\sigma}_{\overline{\mu}}$ from Table~2 in \cite{Snodgrass1984} for
the data taken between 1967 January~1 and 1984 March~5 at the Mount Wilson
Solar Observatory (WSO). The seventh and eighth rows represent the t-scores
$\widehat{t}_{\theta\overline{\mu}}=\left(\widehat{\theta}-\widehat{\overline{\mu}}\right)/\sqrt{\widehat{\sigma}_\theta^2+\widehat{\sigma}_{\overline{\mu}}^2}$
and their two-sided significance probability in percent that the best
estimates $\widehat{\theta}$ and the population averages
$\widehat{\overline{\mu}}$ would differ more by chance\----the larger the
probabilities, the better the two results agree. All of the best estimates are
within one rms standard deviation of the population averages
reported by \cite{Snodgrass1984}.\par
\begin{figure*}[!t]
\begin{center}
\hbox{\includegraphics[width=2.5in]{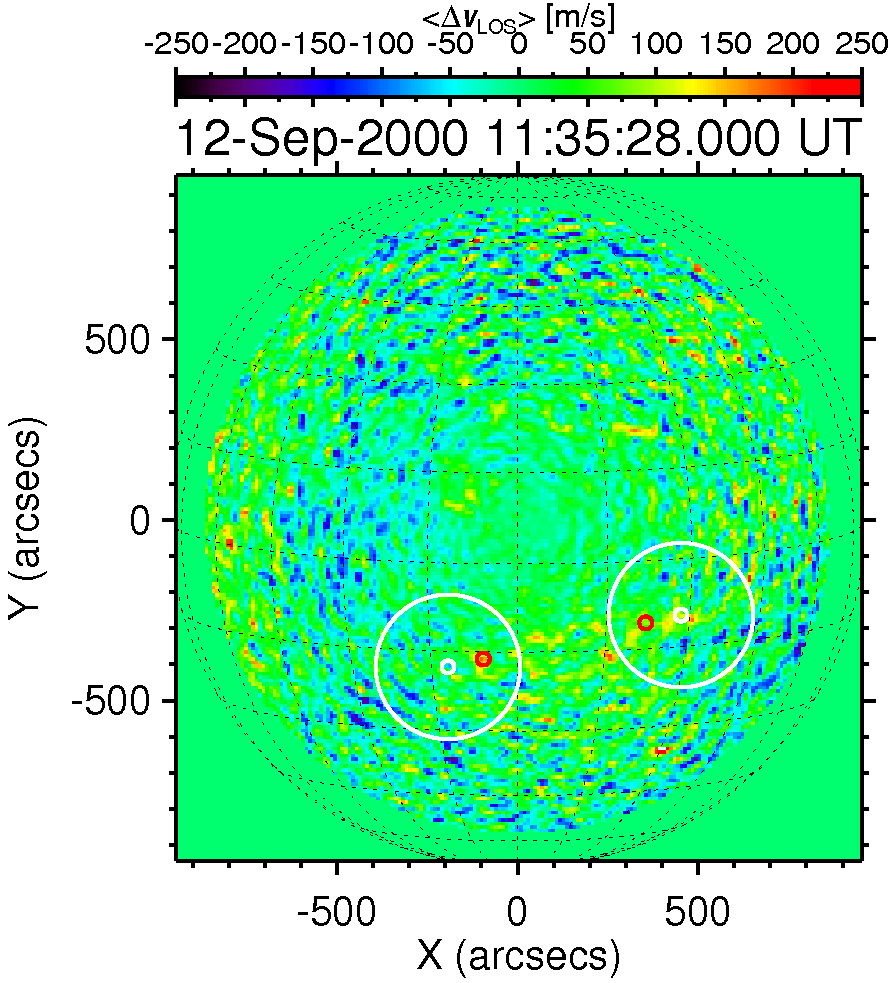}
\includegraphics[width=3.5in]{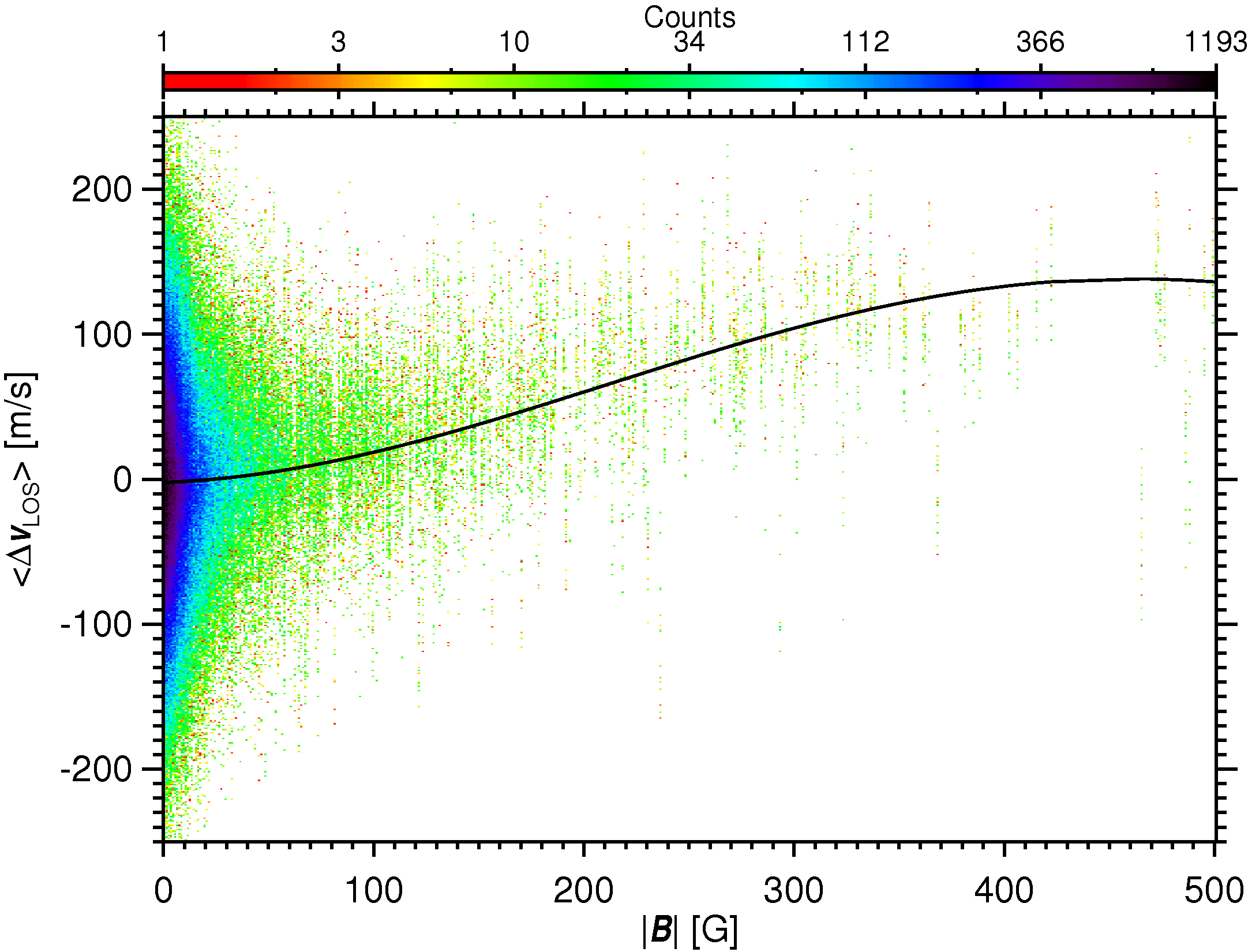}}
\includegraphics[width=6in]{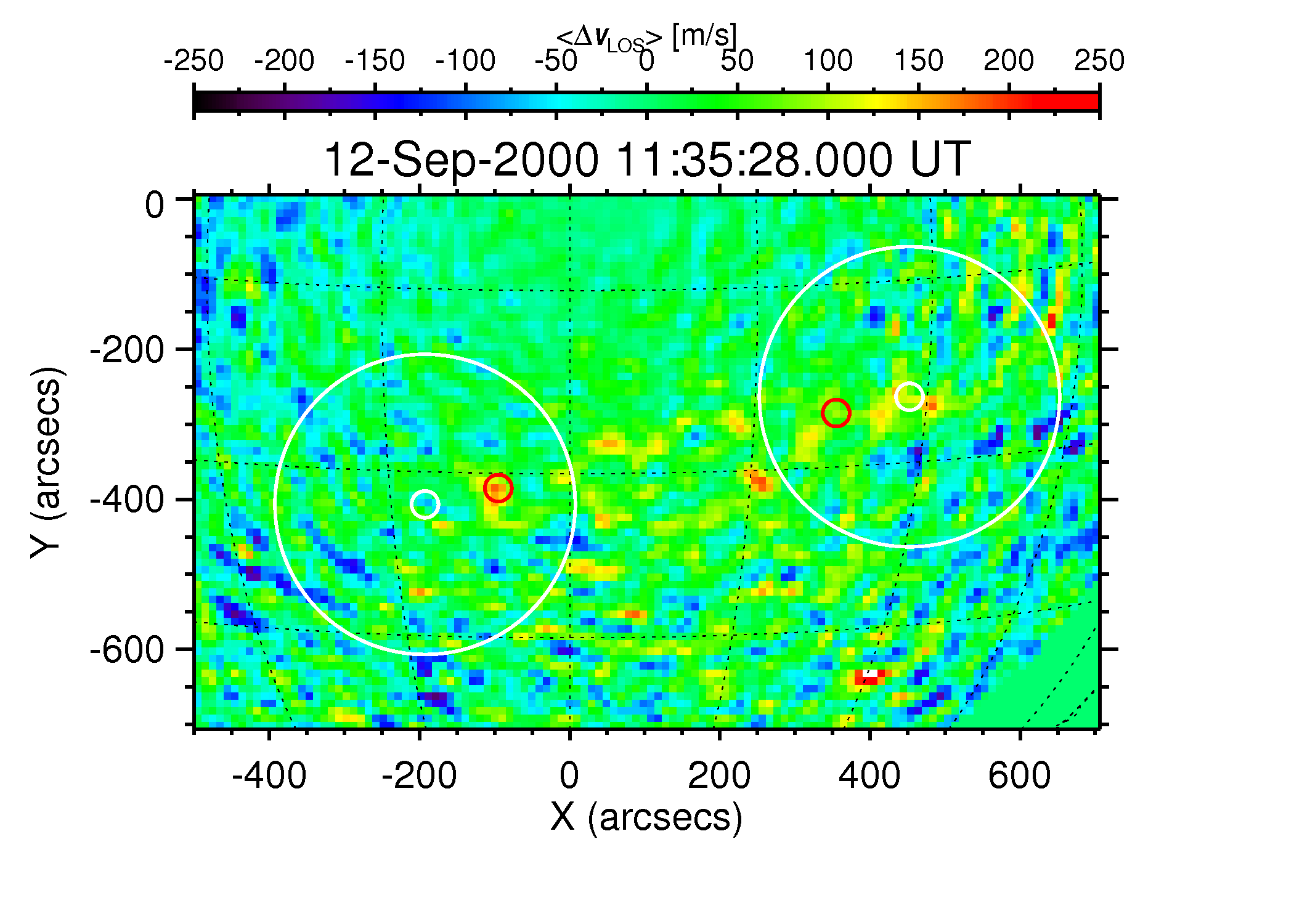}
\end{center}
\vskip-0.5in
\caption{Top left: residual full disk Doppler velocities from the
  Medium-$l$ program averaged over 24 hr. Top right: histogram showing the
  correlation between the magnitude of the magnetic field and the
  time-averaged residual Doppler velocity. Bottom: magnified view of the
  time-averaged residual Doppler velocities in the region containing the
  filament. The red circles correspond to the filament footpoints and the
  large and small white circles correspond to the extent of twice the
  current-channel radius in the corona and photosphere,
  respectively.\label{fig:bias}}
\end{figure*}
The best estimates were used to construct a model line-of-sight velocity for
the solar disk which was subtracted from each co-registered Dopplergram and
then the Dopplergrams were de-rotated to coincide with the time of EIT image in
Figure~\ref{fig:sun} to produce the residual velocity
$\Delta{v}_{\mathrm{LOS}}$. The time average of the residual Doppler images is
shown on the top left of Figure~\ref{fig:bias}.  The red circles correspond to
the filament footpoints and the large and small white circles correspond to
the extent of twice the current-channel radius in the corona and photosphere
respectively. \par
Unfortunately the MDI Dopplergrams are not absolutely calibrated. The
limb-shift coefficient $L_0=131\pm19\,\mbox{m}\,\mbox{s}^{-1}$ represents a
constant offset for the Dopplergram series. Since the coefficient is in
excellent agreement with measurements by \cite[]{Snodgrass1984},
cross-calibration against WSO with a laboratory source could permit absolute
calibration of MDI Doppler velocities. The well known pseudo redshift in
active regions contributes the ambiguity. The top right panel of
Figure~\ref{fig:bias} shows a two-dimensional histogram pairing the magnitude
of the magnetic field and the Doppler velocity. The magnetic data were
computed by spatially filtering the 96-minute MDI magnetograms (also
de-rotated to coincide with the time of the EIT image) with a two-dimensional
Gaussian kernel consistent with the Medium-$l$ program \cite[$a=4$
  in][]{Kosovichev1996} followed by subsampling to the resolution of the
Medium-$l$ program. Each Dopplergram was compared to the magnetogram that
minimized differences in observation times. This panel demonstrates that the
Doppler velocities in magnetic regions are systematically redshifted with
respect to the quiet sun with $B=0$. The histogram with $\left|B\right|<1$G is
Gaussian with an offset of $-2.6\pm0.2\,\mbox{m}\,\mbox{s}^{-1}$ and a
standard deviation of $\sigma=65.7\pm0.1\,\mbox{m}\,\mbox{s}^{-1}$.  Fitting a
fifth degree polynomial to the scatterplot of
$\left\langle\Delta{v}_{\mathrm{LOS}}\right\rangle$ versus $B$ produces the
line in the right panel described by
\begin{mathletters}
\begin{equation}
\left\langle{v}_{\mathrm{LOS}}\right\rangle\left(B\right)=\sum_{n=0}^5\,q_n\,B^n,
\end{equation}
where
\begin{eqnarray}
q_0&=&-2.2\pm0.1\,\mbox{m}\,\mbox{s}^{-1},\\
q_1&=&\left( 5.3\pm0.4\right)\times10^{-2}\,\mbox{m}\,\mbox{s}^{-1}\,\mbox{G},\\
q_2&=&\left( 1.75\pm0.02\right)\times10^{-3}\,\mbox{m}\,\mbox{s}^{-1}\,\mbox{G}^{-2},\\
q_3&=&\left(-1.56\pm0.08\right)\times10^{-6}\,\mbox{m}\,\mbox{s}^{-1}\,\mbox{G}^{-3},\\
q_4&=&\left(-4.53\pm0.09\right)\times10^{-9}\,\mbox{m}\,\mbox{s}^{-1}\,\mbox{G}^{-4},\\
q_5&=&\left( 4.9\pm0.2\right)\times10^{-12}\,\mbox{m}\,\mbox{s}^{-1}\,\mbox{G}^{-5}.
\end{eqnarray}

\end{mathletters}
However, this apparent pseudo redshift in magnetic regions is actually
produced by the blueshifting of the quiet-sun regions caused by convective
motion and the brightness velocity correlation in the convection
cells.\footnote{P. Scherrer and B. Welsch, personal communication 2009
  September; see also \cite{Dravins1982} and \cite{Bumba1995}.} The convective
blueshift is suppressed in magnetic regions causing them to appear
redshifted relative to the quiet sun. Consequently, the Doppler shifts may
only be discussed relative to the quiet-sun motions which dominate the
coefficient $L_0$ and serve as the zero-point Doppler velocity of
$\Delta{v}_{\mathrm{LOS}}$.\par
Third, the $p$-mode oscillations are removed from the residuals by temporal
filtering.  \cite{Hathaway2000} employ a Gaussian weighted average of 31
Dopplergrams to reduced the $p$-mode signal in the 2\---4 mHz frequency band
where the weights are given by
\begin{mathletters}
\begin{equation}
W_i=\frac{w_i}{\sum_{i=-15}^{15}{w_i}},
\end{equation}
with 
\begin{equation}
w_i=\exp\left(-\frac{\Delta
  t_i}{2\,\alpha^2}\right)-\exp\left(-\frac{\beta^2}{2\,\alpha^2}\right)\,\left(1+\frac{\beta^2-{\Delta t_i^2}}{2\,\alpha^2}\right)
\end{equation}
\end{mathletters}
and $\Delta t_i\equiv{t_i}-t_0$ is the time difference of the central Dopplergram
with $\alpha=8$~minutes and $\beta=16$~minutes.\par
Alternatively, nonparametric uniform B-splines
$\model_{\horder}\left(\veta|t\right)=\sum_{k=1}^{\Nd}\,\neta_k\,B_{k,2\,\horder-1}\left(t\right)$
may be employed to filter temporally  the data. B-splines are solutions to the
optimization problem
\begin{equation}
\min\left|\sum_{\dindex=1}^\Nd\,\frac{\left[\model_{\horder}\left(\veta|t_\dindex\right)-\reality_\dindex\right]^2}{\delta\reality_\dindex^2}+\parm\,\int\limits_{-\infty}^{\infty}{dt}\left[\frac{d^\horder\model_{\horder}\left(\veta|t\right)}{dt^\horder}\right]^2\right|,\label{eqn:var}
\end{equation}
where the summation is the weighted $\widehat{\chi}_{\horder}^2$, $\delta
\reality_i^2$ are the uncertainties, $\Nd$ is number of data or knots,
$\horder$ is the spline half-order, and $\veta\equiv\veta\left(\parm\right)$
is a vector of B-spline parameters and a function of the global regularization
parameter $\parm\ge0$ \cite[]{Woltring1986}.  The variational
principle~(\ref{eqn:var}) may be justified by Bayesian arguments when the data
$\reality_\dindex$ is associated with certain priors \cite[see][and references
  therein]{Craven1979}.  The parameter $\parm$ controls the balance between
smoothness measured by the $\horder$th derivative and fidelity measured by the
variance with $\parm\rightarrow0$ corresponding to spline interpolation with
$\widehat{\chi}_{\horder}^2=0$ and $\parm\rightarrow\infty$ corresponding to a
least squares polynomial of order $\Np_{\horder}=\horder$, degree
$\degree=\horder-1$.  The boundary conditions at $t=t_1$ and $t_\Nd$ are
determined by ${d^\horder\model_{\horder}}/{dt^\horder}=0$, where
$\horder=1,2,\ldots$, consistent with a well-posed solution to the variational
principle.\par
B-spline basis functions have many attractive properties for
filtering. For example, they do not require uniform sampling, they
adapt to local structure, and the effect of outliers is localized
because the basis functions are defined on compact support. For the
special case of uniformly sampled, uniformly weighted, periodic data,
the spline smoother behaves as two cascaded $m$th order Butterworth
filters without phase distortion; thus B-splines may be viewed as a
generalized low pass filter adapted for \textit{nonuniform weighting}, \textit{uneven
sampling}, and \textit{non-periodic boundary conditions} \cite[]{Craven1979}.
Butterworth filters are an approximation to an ideal filter because they
have maximally flat frequency response in the passband and are monotonically
decreasing roll-off in the stop-band \cite[]{Butterworth1930}.\par
\begin{figure*}[!t]
\centerline{\includegraphics[viewport=18 6 495
    340,height=2in,clip=]{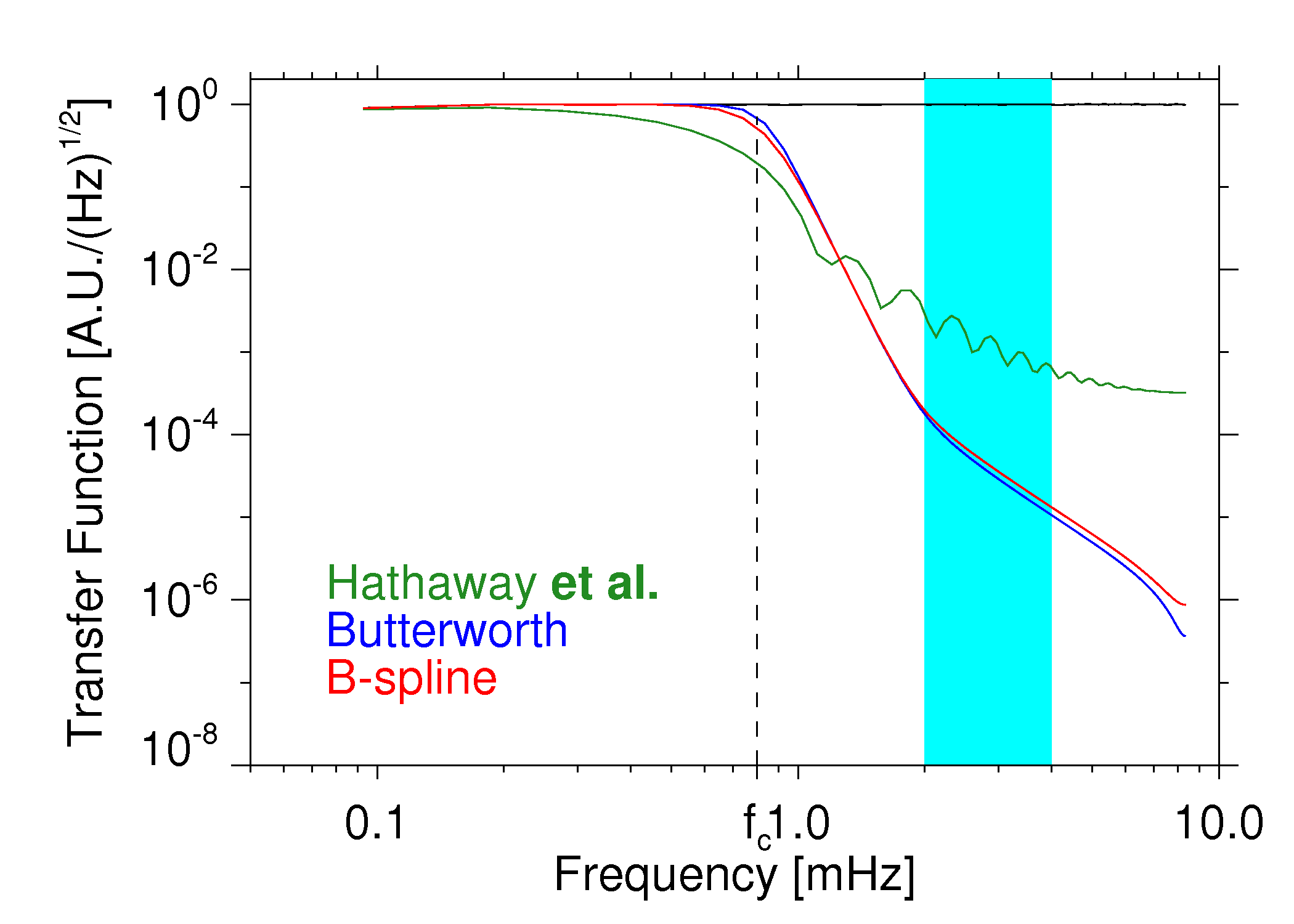}\hskip0.2in\includegraphics[height=2in,
    viewport=-5 0 452 360,clip=]{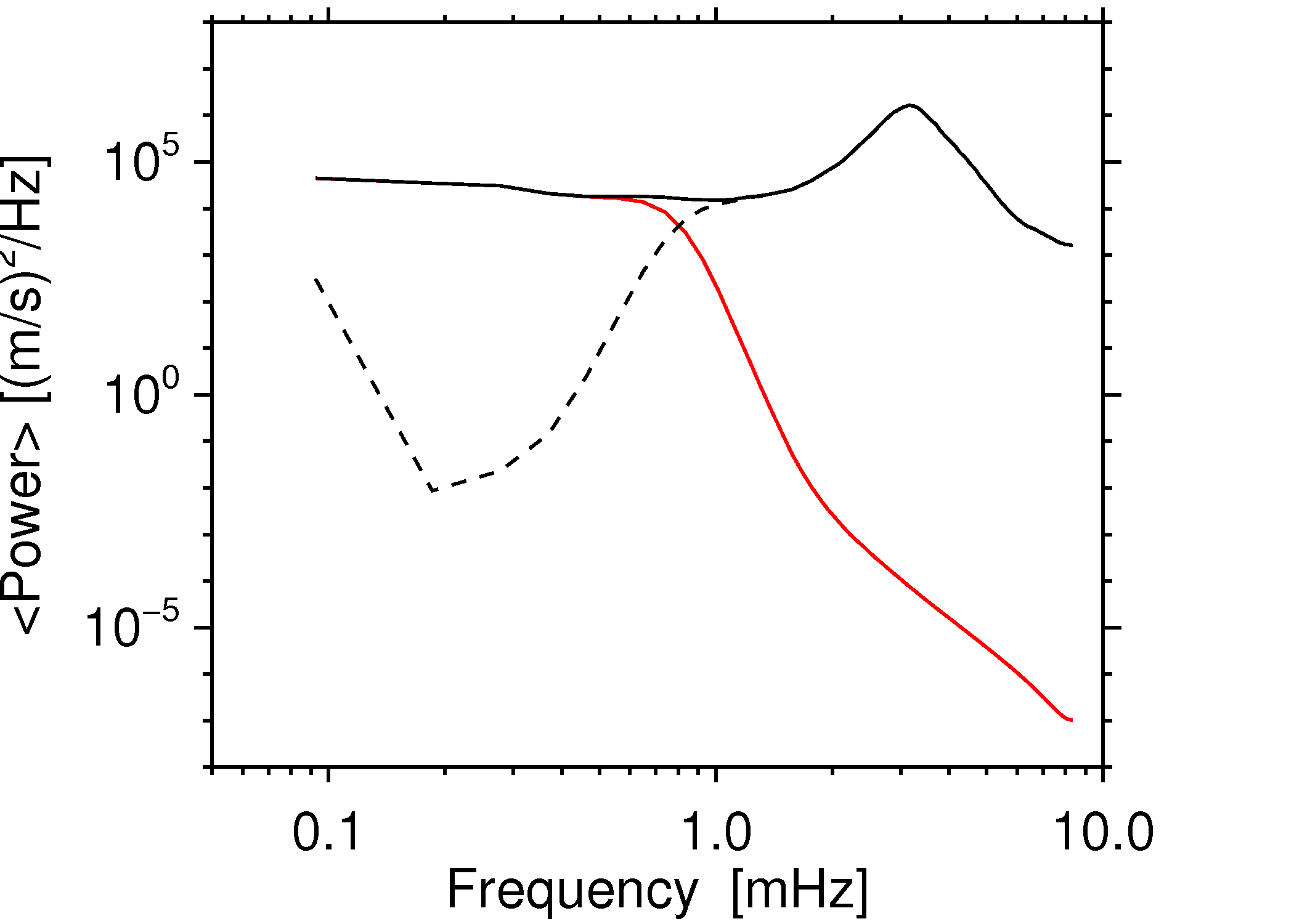}}
\caption{Left: the white noise filter responses for the Gaussian weighted
  average filter from \cite{Hathaway2000} in green, and two $m=5$th order
  cascaded Butterworth filter with $f_\mathrm{c}=8$~mHz in blue and the
  B-spline filter in green. The filter responses were constructed with
  \NR~realizations of a 24~hr white noise time series with $\Delta
  t=1$~minute whose average spectrum is shown in black. The 2\---4~mHz regime
  is shaded in light blue. Right: average power spectra of co-registered and
  de-rotated Doppler time-series from the inner 10\% of the solar disk where
  the solid black line denotes the unfiltered data exhibiting an enhanced
  $p$-mode spectrum in the 2\--5 mHz, the solid red line is for the smoothed
  results after B-spline filtering, and the dashed black line is for the
  residual data (unfiltered minus smoothed). Both left and right: the spectra
  were computed using 3~hr segments with 50\% overlap. \label{fig:butter}}
\end{figure*}
The left panel of Figure~\ref{fig:butter} shows the filter responses for the
Gaussian weighted average filter from \cite{Hathaway2000} in green, and the
two $m=5$ order cascaded Butterworth filters with $f_\mathrm{c}=8$~mHz in blue
and the B-spline filter in red with $\tau=1/f_\mathrm{c}=1250$~s and
\begin{equation}
\lambda=\left[\left(\frac{\tau_\mathrm{c}}{2\,\pi}\right)^{2\,m}-{\left(\frac{2\,\Delta t}{2\,\pi}\right)^{2\,m}}\right]/{\Delta t}=1.62\times10^{21}\,\mathrm{s}^9.
\end{equation}
The filter responses were constructed with \NR~realizations of a 24~hour white
noise time series with $\Delta t=1$~minute whose average spectrum is shown in
black. The 2\---4~mHz regime is shaded in light-blue. The B-splines exhibit a
flatter power spectrum in the pass-band and a steeper roll-off in the
stop-band than the Gaussian filter. Furthermore, the B-spline filter does not
exhibit any ripples in the stop-band. Finally, the B-spline filter can
accommodate occasional missing data or uneven sampling and simultaneously
interpolate to a different temporal grid as part of the filtering
process. Indeed, the B-spline filters were used to replace missing data in the
central regions of four MDI Dopplergrams.  The right panel of
Figure~\ref{fig:butter} shows the average power spectra of co-registered and
de-rotated Doppler time series from the inner 10\% of the solar disk where the
solid black line denotes the unfiltered data exhibiting an enhanced $p$-mode
spectrum in the 2-5 mHz, the solid red line is for the smoothed results after
B-spline filtering, and the dashed black line is for the residual data
(unfiltered minus smoothed). The peak power of the $p$-mode oscillation at
$f\simeq3.15$~mHz is reduced by a factor of $10^{-10}$.\par
\subsection{Photospheric Doppler Signatures During the CME}
\begin{figure*}[!t]
\begin{center}
\includegraphics[width=4.5in]{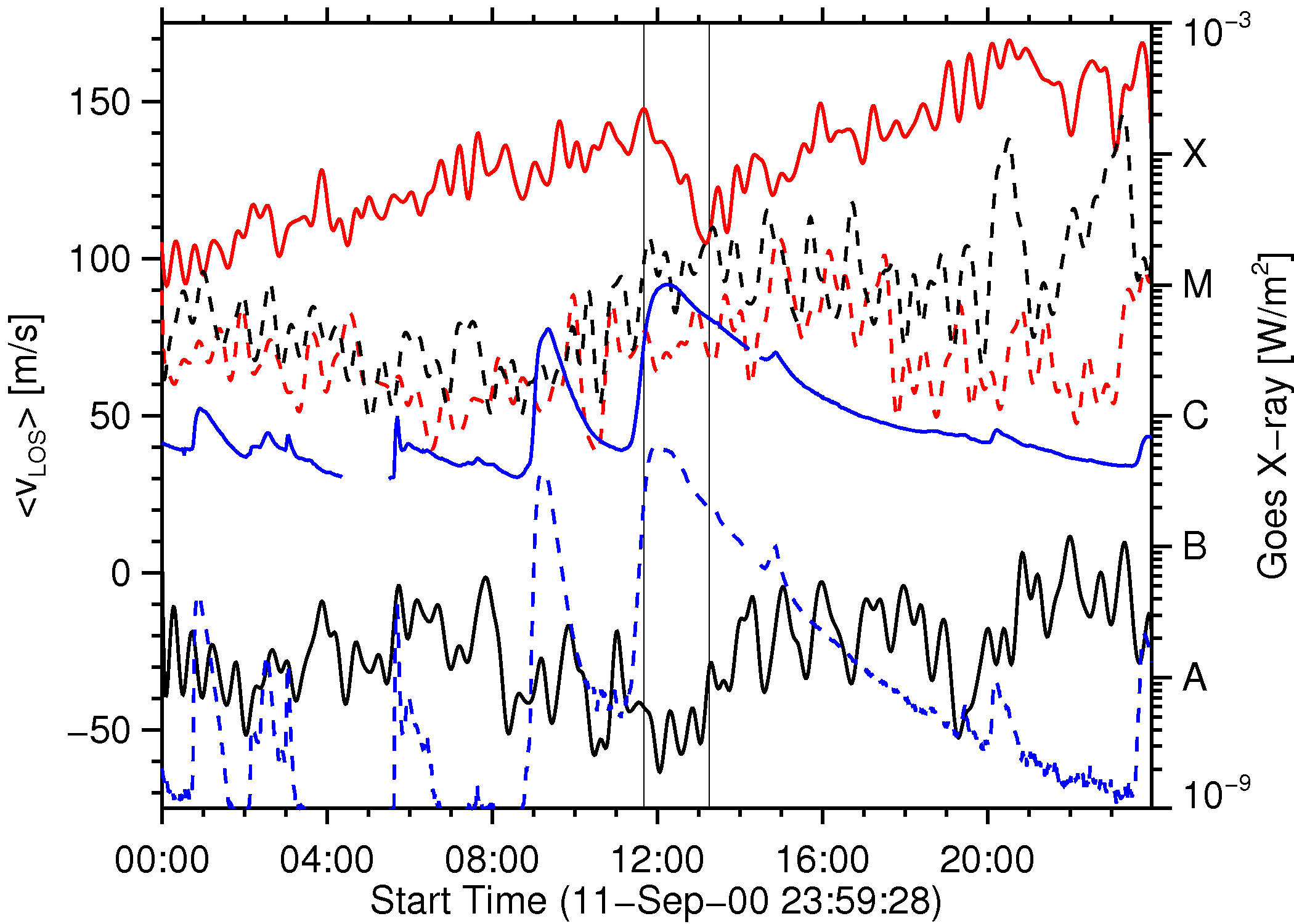}\\ \vskip0.25in
\includegraphics[width=4.5in]{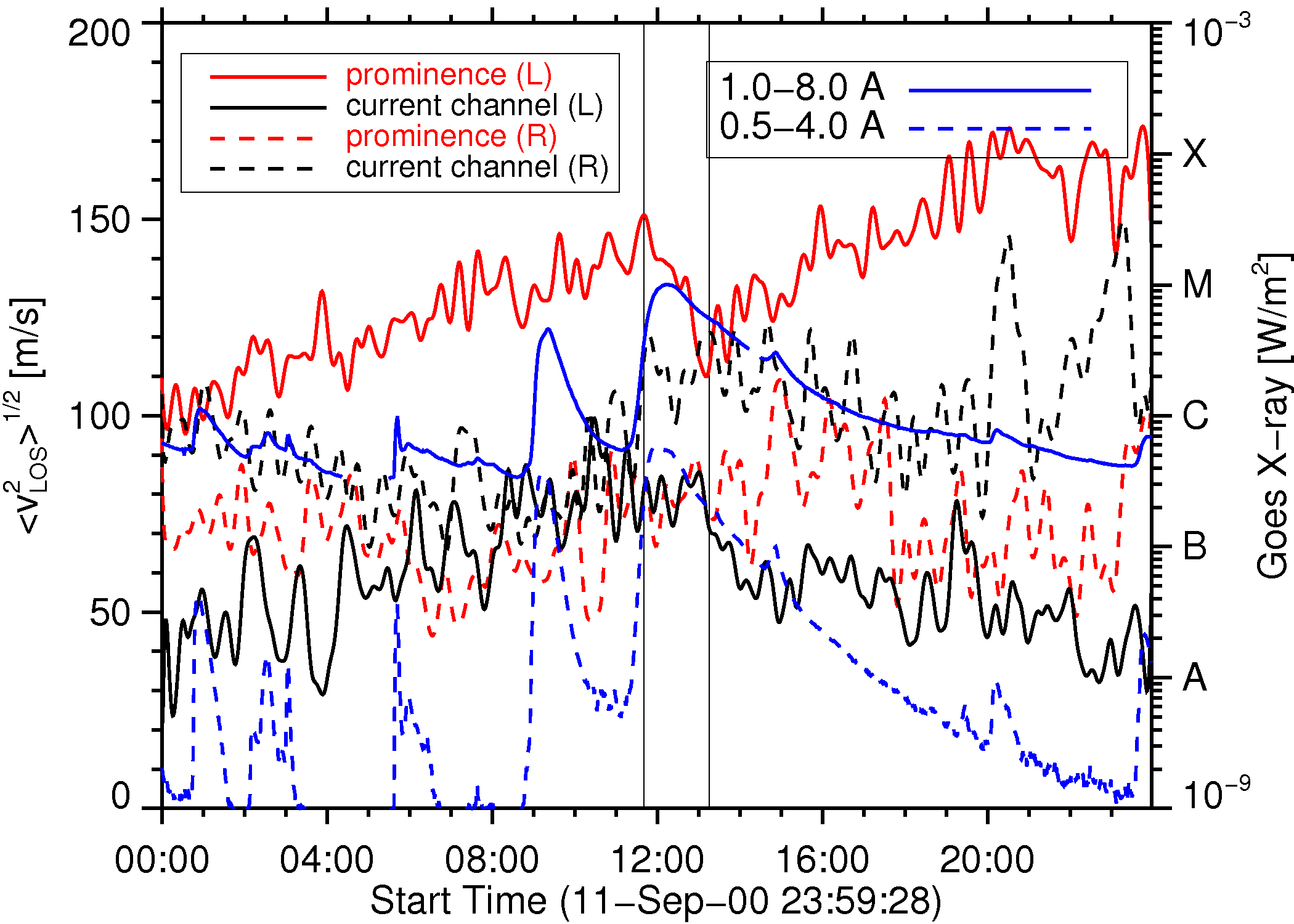}
\end{center}
\caption{Line-of-sight velocity (top) and the rms line-of sight velocity
  (bottom) averaged over the current channel of radius $\af=\AP$~cm for the
  filament channel (red) and current channels (black) on the left (solid) and
  right (dashed) and the soft X-ray flux observed by GOES 8
  (blue).\label{fig:footpoints}}
\end{figure*}
Figure~\ref{fig:footpoints} shows the line-of-sight velocity (top) and the rms
line-of sight velocity (bottom) averaged over the current channel of radius
$\af=\AP$~cm for the filament channel (red) and current channels (black) on
the left (solid) and right (dashed) side of Figure~\ref{fig:sun} and the soft
X-ray flux observed by GOES 8 (blue). The M-class flare associated with the
2000 September 12 CME occurs shortly after 11:30~UT. As discussed in the
previous section, all velocities are measured relative to the quiet sun which
serves as the zero-point.  The top plot shows that the average velocity in the
left and right prominence channel and in the right current channel is
redshifted with respect to the quiet sun.  The left current channel show some
evidence of some blueshift (upflows) with respect to the zero-point in the
range of -60 to +$10\,\mbox{m}\,\mbox{s}^{-1}$. These velocities are within
the standard deviation of $\sigma=65.7\pm0.1\,\mbox{m}\,\mbox{s}^{-1}$
reported for the quiet sun $\left|B\right|<1$~G bins in top right panel
Figure~\ref{fig:bias}. That velocities are quite similar to quiet sun
conditions is not surprising because this current-channel footpoint region
contains weak magnetic field ranging from -24 to 24~G. The weak field has
already been discussed as an inconsistency between the photospheric
observations and the simplified flux-rope geometry which should have strong
magnetic field in the current channel. Although, the absolute Doppler
velocities are not known, they are bounded. The absolute errors must be much
less that the passband of the filter $\approx7\,\mbox{km}\,\mbox{s}^{-1}$ and
are likely less than $100\,\mbox{m}\,\mbox{s}^{-1}$ because the average quiet
sun is not moving toward the Earth at several~$\mbox{km}\,\mbox{s}^{-1}$ given
the agreement for the coefficient $L_0$ for the MDI Doppler measurements and
\cite[]{Snodgrass1984}. None of the observed Doppler velocities approach
magnitudes of $10^3\,\mbox{km}\,\mbox{s}^{-1}$ which would be consistent with
the analysis in Section~\ref{sec:energy} \---- these upflows would be too weak
to transport the poloidal flux into the corona necessary to drive the CME
eruption within the context of the flux-rope model. These data falsify the
flux injection mechanism as a driver for this CME. However, an interesting
feature is the sharp decrease of roughly $50\,\mbox{m}\,\mbox{s}^{-1}$ peak to
minimum in the redshift of the left prominence channel (solid red) beginning
just after the GOES X-ray flux begins to rise and lasting 1~hr and 35~minutes
bounded by the two vertical lines. The flow velocity then appears to recover
to its initial value over the next $2\frac{1}{2}$ hr. Through this dynamic
change of Doppler velocity in the filament channel, the magnitude remains
redshifted $100\,\mbox{m}\,\mbox{s}^{-1}$ relative to the quiet sun.  To
determine if dynamic change in the Dopper velocity of filament footpoints is a
common characteristic of filament eruptions producing CMEs will require
evaluating more events.\par
\section{COMPARISON WITH PREVIOUS WORK AND CONCLUSIONS\label{sec:conclusions}}
The flux-injection hypothesis requires a large energy transport of
$2\times10^{33}$~erg across the photosphere on timescales of 600\--1200~s. If
this hypothesis is correct, then the absence of any significant photospheric
signature of this transport is certainly surprising given that the power
requirements $\simeq3\times10^{30}\,\mbox{erg}\,\mbox{s}^{-1}$ exceed that of
the typical solar flare
$\lesssim10^{29}\,\mbox{erg}\,\mbox{s}^{-1}$. \cite{Sudol2005} and
\cite{Fletcher&Hudson2008} have observed permanent changes in longitudinal
magnetograms concomitant with solar flares, and a more comprehensive follow-on
study by \cite{Petrie2009} has demonstrated that these changes are associated
with every flare. However, these changes \textit{always} lag the rise of the
flare and thus would lag the peak in the CME acceleration which is strongly
correlated with the flare rise time
\cite[]{Shanmugaraju2003,Zhang2001,Zhang2004}. Furthermore, the changes in the
longitudinal magnetograms are largest for events near the limb suggesting that
they are caused by a rapid change in the angle of the magnetic field through
the photosphere rather than a true change in photospheric magnetic field
strength.\par
The present study has failed to find evidence of the magnetized photospheric
plasma velocities required for the flux-injection hypothesis to satisfy the
CME energy budget on timescales of the eruption.  \cite{Chen2000a},
\cite{Chen2001}, \cite{Krall2001}, \cite{Chen2003} and \cite{Chen2010} have
attempted to address the criticisms of the flux-injection hypothesis based on
energy arguments.  \cite{Chen2000a} cite one-dimensional simulations by
\cite{Huba1996} of a step function horizontal magnetic field of tens of Gauss
injected into the low chromosphere. The unspecified current system generating
and maintaining the magnetic piston is assumed to lie below the simulation
boundary in the photosphere\----this is a different situation than the
diagrams in Figures~\ref{fig:flux_injection} and \ref{fig:leg}.  The
simulation does not include the photosphere (J. Huba, personal communication
2008 December) and the plasma and neutrals only attain vertical velocities of
$v_\localz\simeq$tens of $\mbox{m}\,\mbox{s}^{-1}$ in the low
chromosphere. \cite{Chen2000a} infer that these simulations imply photospheric
upflow velocities of \textit{merely meters per second}. Can these results be
used to satisfy the power requirements of the flux-rope model?  Assuming a
constant horizontal magnetic field $B_h\simeq30$~G over a large area
$\pi\,\left(\Sf/2\right)^2=2\times10^{21}\,\mathrm{cm}^2$ surrounding each
current channel and a vertical plasma velocity of
$v_\localz=10\,\mbox{m}\,\mbox{s}^{-1}$, the power supplied by this
one-dimensional simulation over two footpoints would be
$dU_\poloidal/dt\simeq{v}_\localz\,B_h^2\,\left(\Sf/2\right)^2/2=8\times10^{26}\,\mbox{erg}\,\mbox{s}^{-1}$
which is \textit{still} more than 2 orders of magnitude less than the power
requirements of the 2000 September 12 CME \cite[]{Chen2006,Chen2010} which was
associated with an M1.0 flare and more than 3 orders of magnitude less than
the power requirements of the 2003 October 28 CME which was associated with an
X17 flare \cite[]{Krall2006a}.\par
\cite{Krall2000} consider the constraints of driving the flux-rope model by
footpoint twisting. They conclude that footpoint twisting is inefficient and
cannot reproduce the characteristics of the CME trajectories event with large
photospheric poloidal velocities of $10\,\mbox{km}\,\mbox{s}^{-1}$. However,
this estimate corresponds to velocities at the base of the corona based on
their values of current-channel radius, and poloidal and toroidal magnetic
field \cite[see Eqs~(31-33) and Tables~1 and~2 in][]{Krall2000}. They coupled
footpoint twisting to the flux-rope model by relating \textit{net} current and
poloidal magnetic field at the edge of the current channel to the number of
twists at the base of the corona.  However, footpoint twisting merely
rearranges the current distribution in the current channel leaving the net
current constant. Footpoint twisting should be coupled to the flux-rope model
through the change in internal inductance as outlined in
Appendix~\ref{app:Krall} not by modifying the net current and poloidal
magnetic field at the edge of the current channel.\label{error:coupling} \par
\cite{Krall2001} argue that the flux-injection hypothesis implies radial
plasma velocities of the order of $v_r\simeq1\,\mbox{km}\,\mbox{s}^{-1}$ at
the footpoints of the flux rope implied by the inductive toroidal electric
field $E_\toroidal\simeq-{c^{-1}}\,\partial_t A_\toroidal$. These velocities
are already substantial and should be well within the observational
capabilities of MDI aboard \textit{SOHO} and ground based
observatories. However, this electric field is not relevant to estimating the
Poynting flux because
$\E_\toroidal\cross\B\simeq\left(B_\poloidal/c\right)\,\partial_t
A_\toroidal\,\widehat{r}$ is roughly tangent to the photosphere\----the
vertical Poynting flux is zero (see Figure~\ref{fig:leg} for the geometry).
Similarly the vertical helicity flux $\E\cross\A_R$ is also zero
for the toroidal electric field.\par
\cite{Chen2003} have argued that the photospheric velocities should be highly
nonuniform with coherence scales of less than $\sim10^5$~km consistent with a
high $\beta\gg1$ plasma outside the current channel although no theoretical
estimate of this coherence length has been provided. High-resolution
convection simulations exhibit rms fluctuations in the $\tau=1$ surface of
$\approx30$~km which is much less than the local photospheric pressure scale
height,\footnote{B. Abbett, personnel communication 2009 July.} casting doubt upon
these assertions. Although the region far from the current channel is
$\beta\gg1$, the region containing the toroidal field and toroidal currents is
$\beta\simeq1$.  Section~\ref{sec:energy} has demonstrated that a significant
amount of poloidal energy is injected in the region very near and interior to
the current channel $r\le\,2\,\af$. Finally, the spatial scale of the
$10\arcsec\,\mbox{pixel}^{-1}$ for MDI corresponds to $\simeq7500$~km.  The
photospheric dynamics detected at this scale concomitant with the eruption at
the footpoints of the 2000 September 12 CME are not sufficient to drive the
eruption.\par
Emerging horizontal magnetic fields will carry mass into the chromosphere. For
velocities much less than the escape velocity
$v_\sun=\vsun\,\mbox{km}\,\mbox{s}^{-1}$ this mass will flow downward as the
magnetic field at the footpoints of the forming loops become more vertically
inclined. Indeed, \cite{Ishikawa2008} have observed a strongly redshifted
$+5\,\mbox{km}\,\mbox{s}^{-1}$ Stokes $V$ profile (downflows) at one end of an
emerging horizontal magnetic feature. However, the turbulence argument of
\cite{Chen2003} does not seem to appreciate the very large and continuous
\textit{average} velocities$\simeq$hundreds of $\mbox{km}\,\mbox{s}^{-1}$
necessary to satisfy the power budget of the flux-rope model. In the limit
$v_{\poloidal\mathrm{p}}=0$, the \textit{average} vertical velocities in the
current channel exceed the escape velocity $v_\sun$ and the mass could only be
restrained by magnetic forces. The expected velocities also exceed the Afv\'en
$V_\mathrm{A}$ and sound speed $C_\mathrm{s}$ in an ideal MHD plasma.
Furthermore, regions of downflow embedded in the upflow would imply
\textit{larger} upflow velocities to enhance vertical transport to account for
the smaller effective upflow area and to balance any energy transported out of
the corona through the photosphere in the downflow regions (unless downflow
regions are assumed special and contain no horizontal field). \par
In a similar vein \cite{Krall2001} and \cite{Chen2003} have argued that
filling factor $f$\----the percentage of a pixel that contains magnetized
atmosphere\----in the photosphere is unknown and the inherent
nonuniformity of the magnetic field in the photosphere explains the lack of
observational evidence for flux injection.  However, the filling factor can in
principle be estimated by modern magnetographs by enforcing consistency
between the Stokes $I$ profile and the $Q$ (linear), $U$ (linear), and $V$
(circular polarization) profiles \cite[]{Keppens1996}. Indeed, the filling
factor will be a standard data product of the next generation vector
magnetograms produced by \textit{SDO}/HMI. The non-magnetic component only
contributes to the Stokes $I$ profile, whereas the Doppler velocities are
usually determined from the $V$ profile providing a relatively unambiguous
estimate for the velocity of the \textit{magnetized} atmosphere. Although
Stokes $V$ spectra may exhibit features of both upflows and downflows
simultaneously \cite[]{BellotRubio2001}, the observations in
Section~\ref{sec:observations} require precise cancellation between the
upflows and downflows to produce no evidence of significant dynamics in the
flux-rope footpoint regions necessary for consistency with the flux-injection
hypothesis. This perfect cancellation seems improbable.\par
The large photospheric velocities implied by the flux-rope model power budget
would shift the spectral lines used to estimate the magnetic field and Doppler
velocity out of the pass-band of many modern telescopes.  For example, the MDI
instrument records filtergrams around the \ion{Ni}{1} spectral line
\ion{Ni}{1} 6767.8~\AA~with a 94~m\AA~bandpass. Under normal operation,
filtergrams are made at five tuning positions separated by 75~m\AA~spanning
377~m\AA~\cite[]{Scherrer1995}.  The maximum velocity that could be measured
by the MDI instrument is
$v=c\,\Delta\lambda/\lambda=0.150/6767.8\approx7\,\mbox{km}\,\mbox{s}^{-1}$. The
absence of significant changes in the magnetic field indicate that a large
fraction of the magnetized atmosphere \textit{is not} moving with velocities
that exceed $5\,\mbox{km}\,\mbox{s}^{-1}$.\par
A simple photospheric magnetic field model for flux-rope footpoints has been
developed by extrapolating the flux-rope magnetic field model of
\cite{Chen1994}, \cite{Chen1996}, \cite{Krall2000}, and \cite{Krall2005} into
the photosphere with conservation of toroidal flux and toroidal current.  This
is equivalent to magnetic field models implemented in simulation studies of
photospheric flux injection
\cite[]{ChenHuba2005b,ChenHuba2005a,ChenHuba2006}. This magnetic field model
has been used to estimate the minimum photospheric velocities necessary to
satisfy the power budget of the flux-rope model for CMEs fitted by
\cite{Chen2006}, \cite{Krall2006a}, and \cite[]{Chen2010}. The flux-rope power
budget requires large average poloidal or vertical velocities of the order of
\textit{hundreds to thousands of $\mbox{km}\,\mbox{s}^{-1}$} over large
photospheric areas of $10^8\,\mathrm{km}^2$ to transport the necessary
poloidal magnetic field into the corona on a timescale of the eruption.  While
Chen, Krall, and Kunkel might argue that the photospheric magnetic field
implemented in this study is oversimplified, there is agreement that enhanced
photospheric activity should be detected in the region near the
footpoints. Indeed, \cite{Krall2001} affirm ``with certainty we can state that
flux injection as discussed above should be accompanied by increased
photospheric flow activity over a large spatial area, near the footpoints, for
a period of hours during and following a CME eruption.''  To address this,
Doppler and magnetic field observations at the footpoints of the 2000
September 12 CME have been analyzed.  No significant dynamics at the flux-rope
current-channel footpoints concomitant with the CME eruption have been
detected.  The flux-injection hypothesis is incompatible with these
observations.\par
Although, the flux-injection hypothesis has been falsified, the flux-rope
model \cite[]{Chen1989,Chen1996} could remain a useful theoretical tool for
modeling and interpreting CME dynamics because there are other hypotheses for
forming or increasing the poloidal flux of a flux rope. For example, shearing
\cite[]{Mikic1994,Antiochos1999,Amari2000}, converging flows
\cite[]{Forbes1995}, or nearby emerging flux \cite[]{ChenShibata2000} may
convert coronal arcade field into flux-rope fields via rapid
reconnection.\footnote{Falsification of one hypothesis does not imply
  verification of another.} Consequently, the flux-injection hypothesis is
incorrect, but the flux-rope model could correctly describe the dynamics of an
erupting CME.  Finally, the flux-rope model has brought to the forefront, the
paradigm that CMEs are current carrying coherent magnetic structures
consistent with the three-part morphology observed by LASCO coronagraphs and
predicts the scaling law that height of the CME at maximum acceleration scales
with the footpoint separation distance $\Sf$ \cite[]{Chen2006}.\par
While the focus of this investigation has been the flux-injection hypothesis
\cite[]{Chen1997a,Chen2000a,Wood1999,Krall2001,Chen2003,Chen2006,Krall2006a,Chen2010},
the observational component of this study places important constraints on any
hypothesis that relies on the photosphere for the power
$10^{29}-10^{30}\,\mbox{erg}\,\mbox{s}^{-1}$ driving a CME. These hypotheses
are likely incompatible with the present investigation, because the small
velocities observed in the photospheric cannot supply the necessary energy on
the timescale of the main acceleration phase of the CME. In contrast, the
storage-release mechanism, where the energy is transported across the
photosphere over a period of hours or days and then released rapidly through
reconnection, is compatible with the present study. Under storage release, the
power supplied by reconnection is limited by the reconnection rate, in part,
determined by the velocity of the reconnection point as it unzips the
overlying arcade.\par
\acknowledgements
The author thanks the reviewer for his/her constructive criticism that greatly
clarified and improved the paper.  The author gratefully acknowledges
insightful conversations with Jonathan Krall, Joseph Huba, Mark Linton,
K. D. Leka, Todd Hoekesema, Werner Poetzi, and Uri Feldman. The author thanks
the \textit{SOHO}/SOI team for providing the Dopplergrams and magnetograms.
\appendix
\section{THE MINIMUM PHOTOSPHERIC VELOCITIES CONSISTENT WITH THE POWER BUDGET\label{app:power}}
The minimum photospheric velocities consistent with the power requirements of
the flux-injection hypothesis may be found from a constrained variational principle where 
\begin{equation}
\mathcal{I}_1=\int_0^{\rc}{dr}{r}\,v_\poloidal^2,\label{eqn:I1}
\end{equation}
\begin{equation}
\mathcal{I}_2=\int_0^{\rc}{dr}{r}\,v_\localz^2,\label{eqn:I2}
\end{equation}
or
\begin{equation}
\mathcal{I}_3=\int_0^{\rc}{dr}{r}\,\left(v_\poloidal^2+v_\localz^2\right),\label{eqn:I3}
\end{equation}
is minimized subject to the integral constraint
\begin{equation}
\constraint\equiv\frac{dU_\poloidal}{dt}=-\int_{0}^{\rc}{dr}{r}{B}_\theta\left(r\right)\,\psi'\left(r\right),\label{eqn:constraint}
\end{equation}
where $\psi'\left(r\right)=\partial_r\psi$.
This leads to the functional
\begin{equation}
\mathcal{H}\equiv\left[\psi'\left(r\right)\right]^2\,{f}^2\left(r\right)-\lambda\,B_\theta\left(r\right)\,\psi'\left(r\right),\label{eqn:functional}
\end{equation}
and the Euler equation
\begin{equation}
\frac{\partial\mathcal{H}}{\partial\psi^\prime}=\kappa=\mathrm{constant},\label{eqn:dH}
\end{equation}
where $\lambda$ is a Lagrange multiplier.
There are two limiting cases and a third general case to consider:\\
\textbf{Case 1:} $v_\parallel\neq0$, $v_\localz=0$,
and $f=B_\localz^{-1}$.\\
Solving for $\psi'$
\begin{mathletters}
\begin{eqnarray}
\psi'&=&\frac{B_\localz^2}{2}\,\left(\kappa_1+\lambda\,B_\poloidal\right),\\
v_\poloidal&=&\frac{B_\localz}{2}\,\left(\kappa_1+\lambda\,B_\poloidal\right).
\end{eqnarray}
\end{mathletters}
Physical considerations require  $v_\poloidal\rightarrow0$ when
$r\rightarrow0$ which is equivalent to $B_\poloidal\rightarrow0$. This implies
$\kappa_1\equiv0$ and
\begin{mathletters}
\begin{eqnarray}
\psi'&=&\frac{\lambda}{2}\,B_\poloidal\,B_\localz^2,\label{eqn:psi1}\\
v_\poloidal&=&\frac{\lambda}{2}\,B_\poloidal\,B_\localz.\label{eqn:vtheta}
\end{eqnarray}
\end{mathletters}
Substituting Equation~(\ref{eqn:psi1}) into Equation~(\ref{eqn:constraint})
determines the Lagrange multiplier
\begin{equation}
\lambda\equiv-\frac{1120\,\constraint}{437\,B_{\localz{\mathrm{p}}}^2\,B_{\poloidal{\mathrm{p}}}^2\,\af^2}.\label{eqn:lambda1}
\end{equation}
Substituting Equations~(\ref{eqn:lambda1}) and Equation~(\ref{eqn:vtheta})
into~(\ref{eqn:I1}), integrating from $r=0\Longrightarrow\af$ and using
Equations~(\ref{eqn:krall1}) and~(\ref{eqn:krall2}) the rms poloidal plasma
velocity inside $r\le\af$ is
\begin{equation}
\left\langle{v}_\poloidal^2\right\rangle_{\af}^{1/2}=\left|\frac{dU_\poloidal}{dt}\right|\,\frac{4\,\sqrt{70}}{\sqrt{437\,\left|B_{\poloidal{\mathrm{c}}}^2\,B_{\localz{\mathrm{c}}}\right|}\,\ac^2}\,\frac{1}{\sqrt{\left|B_{\localz\mathrm{p}}\right|}}.\label{eqn:vpoloidal}
\end{equation}
\textbf{Case 2:} $v_\parallel\neq0$, $v_\poloidal=0$, and $f=B_\poloidal^{-1}$ \\
Solving for $\psi'$
\begin{mathletters}
\begin{eqnarray}
\psi'&=&\frac{B_\poloidal^2}{2}\,\left(\kappa_2+\lambda\,B_\poloidal\right),\label{eqn:psi2}\\
v_\localz&=&-\frac{B_\poloidal}{2}\,\left(\kappa_2+\lambda\,B_\poloidal\right).\label{eqn:vz}
\end{eqnarray}
\end{mathletters}
The Lagrange multiplier is determined by substituting
Equation~(\ref{eqn:psi2}) into Equation~(\ref{eqn:constraint})
\begin{equation}
\lambda\equiv-\frac{6\,\rc\,\left( 85085\,B_{\poloidal{\mathrm{p}}}^3\,\af^3\,
        \kappa_2 - 170170\,\constraint\,\rc - 
       144048\,B_{\poloidal{\mathrm{p}}}^3\,\af^2\,\rc\,\kappa_2
       \right) }{221\,B_{\poloidal{\mathrm{p}}}^4\,\af^2\,
     \left( 1155\,\af^2 - 2998\,{\rc}^2 \right) }.\label{eqn:lambda2}
\end{equation}
Substituting Equations~(\ref{eqn:lambda2}) and~(\ref{eqn:vz}) into
Equation~(\ref{eqn:I2}), differentiating with respect to $\kappa_2$, and
solving determines the value of $\kappa_2=0$ corresponding to the minimum mean
squared velocity over the region $r=0\longrightarrow\rc$.  Integrating
Equation~(\ref{eqn:I2}) from $r=0\rightarrow2\,\af$ with
Equations~(\ref{eqn:krall1}) and~(\ref{eqn:krall2}) produces the rms vertical
velocity inside $r\le2\,\af$
\begin{equation}
\left\langle{v_\localz^2}\right\rangle^{1/2}_{2\,\af}=\left|\frac{dU_\poloidal}{dt}\right|\,\frac{\sqrt{12516735}\,\left|B_{\localz\mathrm{p}}\right|\,\rc^2}{2\,B_{\poloidal\mathrm{c}}^2\,\ac^2\,\left|2998\,B_{\localz\mathrm{p}}\rc^2-1155\,B_{\localz\mathrm{c}}\,\ac^2\right|}.\label{eqn:vtoroidal}
\end{equation}
\textbf{Case 3:} $v_\parallel=0$, $f=1/\sqrt{B_\poloidal^2+B_\localz^2}$\\
Solving for $\psi'$
\begin{mathletters}
\begin{eqnarray}
\psi'&=&\frac{B_\poloidal^2+B_\localz^2}{2}\,\left(\kappa_3+\lambda\,B_\poloidal\right),\\
\vm&=&\frac{\left(\kappa_3+\lambda\,B_\poloidal\right)}{2}\,\left(0,B_\localz,-B_\poloidal\right).
\end{eqnarray}
\end{mathletters}
Physical considerations require  $v_\poloidal\rightarrow0$ when
$r\rightarrow0$ which is equivalent to $B_\poloidal\rightarrow0$. This implies
$\kappa_3\equiv0$ and
\begin{mathletters}
\begin{eqnarray}
\psi'&=&\frac{\lambda\,B_\poloidal}{2}\,\left(B_\poloidal^2+B_\localz^2\right),\label{eqn:psi3}\\
\vm&=&\frac{\lambda\,B_\poloidal}{2}\,\left(0,B_\localz,-B_\poloidal\right).\label{eqn:vtotal}
\end{eqnarray}
\end{mathletters}
Substituting Equation~(\ref{eqn:psi3}) into Equation~(\ref{eqn:constraint})
determines the Lagrange multiplier
\begin{equation}
\lambda\equiv\frac{36960\,\constraint\,\rc^2}{9240\,\af^4\,B_{\poloidal\mathrm{p}}^4 - \af^2\,B_{\poloidal\mathrm{p}}^2\,\left(23984\,B_{\poloidal\mathrm{p}}^2 + 14421\,B_{\localz\mathrm{p}}^2\right)\,\rc^2},\label{eqn:lambda3}
\end{equation}
Substituting Equations~(\ref{eqn:lambda3}) and~(\ref{eqn:vtotal}) into
Equation~(\ref{eqn:I3}), integrating from $r=0\Longrightarrow\af$ and using
Equations~(\ref{eqn:krall1}) and (\ref{eqn:krall2}) the rms total plasma
velocity inside $r\le\af$ is
\begin{mathletters}
\begin{equation}
\left\langle{v^2}\right\rangle^{1/2}_{2\,\af}=\left|\frac{dU_\poloidal}{dt}\right|\,\frac{2\,B_{\localz\mathrm{p}}\,\rc^2\,\sqrt{2310}\,\sqrt{21674\,B_{\poloidal\mathrm{c}}^2 + 14421\,B_{\localz\mathrm{c}}\,B_{\localz\mathrm{p}}}}
 {\ac^2\,\left|B_{\poloidal\mathrm{c}}\,B_{\localz\mathrm{p}}\,(23984\,B_{\poloidal\mathrm{c}}^2 + 14421\,B_{\localz\mathrm{c}}\,B_{\localz\mathrm{p}})\,\rc^2-9240\,\ac^4\,B_{\poloidal\mathrm{c}}^3\,B_{\localz\mathrm{c}}\right|},\label{eqn:vall}
\end{equation}
the poloidal plasma velocity is
\begin{equation}
\left\langle{v_\poloidal^2}\right\rangle^{1/2}_{2\,\af}=\left|\frac{dU_\poloidal}{dt}\right|\,\frac{66\,\rc^2\,\sqrt{30590\,\left|B_{\localz\mathrm{c}}\,B_{\localz\mathrm{p}}^3\right|}}
 {\ac^2\,\left|B_{\poloidal\mathrm{c}}\,B_{\localz\mathrm{p}}\,(23984\,B_{\poloidal\mathrm{c}}^2 + 14421\,B_{\localz\mathrm{c}}\,B_{\localz\mathrm{p}})\,\rc^2-9240\,\ac^4\,B_{\poloidal\mathrm{c}}^3\,B_{\localz\mathrm{c}}\right|}
\end{equation}
and the vertical plasma velocity is
\begin{equation}
\left\langle{v_\localz^2}\right\rangle^{1/2}_{2\,\af}=\left|\frac{dU_\poloidal}{dt}\right|\,\frac{4\,B_{\localz\mathrm{p}}\,\rc^2\,\sqrt{12516735}}
 {\ac^2\,\left|B_{\localz\mathrm{p}}\,(23984\,B_{\poloidal\mathrm{c}}^2 + 14421\,B_{\localz\mathrm{c}}\,B_{\localz\mathrm{p}})\,\rc^2-9240\,\ac^4\,B_{\poloidal\mathrm{c}}^2\,B_{\localz\mathrm{c}}\right|}.
\end{equation}
\end{mathletters}
\section{THE MINIMUM PHOTOSPHERIC VELOCITIES CONSISTENT WITH THE HELICITY BUDGET\label{app:helicity}}
The vector potential in the upper half plane $\left(z\ge0\right)$ for an
azimuthally symmetric vertical magnetic field is
\begin{equation}
\Ap\left(r,z\right)=\frac{2\,\BesselJ_1\left(\gamma_n\,r\right)\,e^{-\gamma_n\,\localz}\,\int_0^a\,dr\,r\,\BesselJ_0\left(\gamma_n\,r\right)\,B_\localz\left(r\right)}{\zero_n\,a\,\BesselJ_1^2\left(\zero_n\right)}\,\widehat{\poloidal}
\end{equation}
where $\zero_{n=1,2,\ldots}=2.40483,5.52008,\ldots$ is the $n$th zero of the
zeroth\--order Bessel function $\BesselJ_0\left(\zero_n\right)=0$. This
potential field satisfies
\begin{mathletters}
\begin{eqnarray}
\grad\cdot\Ap\left(r,\localz\right)&=&0,\\
\nhat\cdot\Ap\left(r,0\right)&=&0,\\
\Bp\left(r,\localz\right)&=&\grad\cross\Ap\left(r,\localz\right)=\grad\Psi_\mathrm{R}\left(r,\localz\right),\\
\nhat\cdot\Bp\left(r,0\right)&=&B_\localz\left(r\right).
\end{eqnarray}
\end{mathletters}
\begin{equation}
\Ap\left(r,0\right)=\frac{a\,B_{\localz\mathrm{a}}}{2}\,\widehat{\poloidal}\,\left\lbrace\begin{array}{lr}
\displaystyle3\,\frac{r}{a}\,\left(1-\frac{r^2}{a^2}+\frac{r^4}{3\,a^4}\right)&r\le{a},\\
\noalign{\vskip0.1in}
\displaystyle\frac{a}{r}&r>a.
\end{array}\right.
\end{equation}
\begin{equation}
\CONSTRAINT\equiv\frac{d\Delta\Helicity}{dt}=-8\,\pi\,\int_0^{\rc}{dr}\,{r}\,A_{\poloidal\mathrm{R}}\left(r\right)\,\psi'\left(r\right).
\end{equation}
The minimum photospheric velocities consistent with the helicity
budget may be found by following the procedure outlined in
Appendix~\ref{app:power} using Equation~(\ref{eqn:functional}) with
$B_\poloidal\Longrightarrow{A}_{\poloidal\mathrm{R}}$.
\begin{equation}
\mathcal{H}\equiv\left[\psi'\left(r\right)\right]^2\,{f}^2\left(r\right)-\lambda\,A_{\poloidal\mathrm{R}}\left(r\right)\,\psi'\left(r\right).
\end{equation}
\textbf{Case 1:} $v_\parallel\neq0$, $v_\localz=0$,
and $f=B_\localz^{-1}$.\\
\begin{equation}
\lambda\equiv-\frac{560\,\CONSTRAINT}{437\,\pi\,\af^4\,B_{\localz\mathrm{p}}^4}
\end{equation}
\begin{equation}
\left\langle{v_\localz^2}\right\rangle^{1/2}_{2\,\af}=\left|\frac{d\Phi_\poloidal}{dt}\right|\,\frac{\sqrt{1155\,\left|B_{\localz\mathrm{c}}\right|}}{2\,B_{\poloidal\mathrm{c}}\,\sqrt{\left|2998\,B_{\localz\mathrm{p}}\rc^2-1155\,B_{\localz\mathrm{c}}\,\ac^2\right|}}.\label{eqn:helicity:vpoloidal}
\end{equation}
\textbf{Case 2:} $v_\parallel\neq0$, $v_\poloidal=0$, and $f=B_\poloidal^{-1}$ \\
\begin{equation}
\lambda\equiv\frac{2310\,\rc^2\,\CONSTRAINT}{B_{\poloidal\mathrm{p}}^2\,B_{\localz\mathrm{p}}^2\,\pi\,\af^4\,\left(2998\,\rc^2-1155\,\af^2\right)}
\end{equation}
\begin{equation}
\left\langle{v_\localz^2}\right\rangle^{1/2}_{2\,\af}=\left|\frac{d\Phi_\poloidal}{dt}\right|\,\frac{\sqrt{1155\,\left|B_{\localz\mathrm{c}}\right|}}{2\,B_{\poloidal\mathrm{c}}\,\sqrt{\left|2998\,B_{\localz\mathrm{p}}\rc^2-1155\,B_{\localz\mathrm{c}}\,\ac^2\right|}}.\label{eqn:helicity:vtoroidal}
\end{equation}
\textbf{Case 3:} $v_\parallel=0$ and $f=1/\sqrt{B_\poloidal^2+B_\localz^2}$ \\
\begin{equation}
\lambda\equiv-\frac{18480\,\CONSTRAINT\,\rc^2}
 {B_{\localz\mathrm{p}}^2\,\pi\,\af^4\,\left[(23984\,B_{\poloidal\mathrm{p}}^2 + 14421\,B_{\localz\mathrm{p}}^2)\,\rc^2-9240\,B_{\poloidal\mathrm{p}}^2\,\af^2\right]},
\end{equation}
\begin{equation}
\left\langle{v^2}\right\rangle^{1/2}_{2\,\af}=\left|\frac{d\Phi_\poloidal}{dt}\right|\,\frac{B_{\localz\mathrm{p}}\,\rc^2\,\sqrt{2310}\,\sqrt{21674\,B_{\poloidal\mathrm{c}}^2 + 14421\,B_{\localz\mathrm{c}}\,B_{\localz\mathrm{p}}}}
 {2\,\ac\,\left|B_{\localz\mathrm{p}}\,(23984\,B_{\poloidal\mathrm{c}}^2 + 14421\,B_{\localz\mathrm{c}}\,B_{\localz\mathrm{p}})\,\rc^2-9240\,\ac^4\,B_{\poloidal\mathrm{c}}^2\,B_{\localz\mathrm{c}}\right|},\label{eqn:vall:helicity}
\end{equation}
\section{ROTATIONAL TRANSFORM\label{app:Krall}}
\cite{Krall2000} assert that footpoint twisting modifies the \textit{net}
current and poloidal magnetic field at the edge of the current channel based
on a geometrical argument. Generally, the twist of a flux rope is estimated
from the field line equations
\begin{equation}
\frac{dr}{B_r}=\frac{r\,d\poloidal}{B_\poloidal\left(r\right)}=\frac{\mathcal{R}\,d\toroidal}{B_\toroidal\left(r\right)}.
\end{equation}
For a circular torus, the amount of rotation of the poloidal field about the
toroidal axis during the transit along the flux rope above the photosphere is
estimated from the rotational transform (effectively the reciprocal of the
safety factor calculation for Tokamaks)
\begin{mathletters}
\begin{eqnarray}
\Delta\poloidal\left(r\right)&\simeq&\int_0^{2\,\pi\,\arclength}{d\toroidal}\,\frac{\mathcal{R}\,B_\poloidal\left(r\right)}{r\,B_\toroidal\left(r\right)},\\
&\simeq&2\,\pi\,\arclength\,\frac{\mathcal{R}\,B_\poloidal\left(r\right)}{r\,B_\toroidal\left(r\right)}
\end{eqnarray}
\end{mathletters}
The amount of twist is a function of minor radius $r$. To add an amount of
twist $\delta\twist\left(r\right)$ to the flux rope, the equation becomes
\begin{equation}
\Delta\poloidal\left(r\right)+\delta\twist\left(r\right)=2\,\pi\,\arclength\,\frac{\mathcal{R}\,B_\poloidal\left(r\right)}{r\,B_\toroidal\left(r\right)}+\delta\twist\left(r\right)=2\,\pi\,\arclength\,\frac{\mathcal{R}\,\widetilde{B}_\poloidal\left(r\right)}{r\,B_\toroidal\left(r\right)}
\end{equation}
where ${B}_\poloidal$ is the initial poloidal field and
\begin{equation}
\widetilde{B}_\poloidal\left(r\right)=B_\poloidal\left(r\right)+\frac{\delta\twist\left(r\right)}{2\,\pi\,\arclength}\,\frac{r}{\mathcal{R}}\,B_\toroidal\left(r\right),\label{eqn:rotation}
\end{equation}
$\widetilde{B}_\poloidal$ is the final poloidal field. This equation has a
similar form  to Equation~(31) in \cite{Krall2000}
\begin{equation}
\widetilde{B}_{\poloidal\mathrm{c}}=B_{\poloidal\mathrm{c}}+\frac{\delta\twist_0}{2\,\pi\,\arclength}\,\frac{\ac}{\mathcal{R}}\,B_{\toroidal\mathrm{c}},\label{eqn:krall_wrong}
\end{equation}
where $\delta\twist_0=\pi$ is a uniform twist.  However, there are important
mathematical and conceptual differences between Equations~(\ref{eqn:rotation})
and~(\ref{eqn:krall_wrong}). The former Equation~(\ref{eqn:rotation}) implies
that the \textit{poloidal magnetic field profile} changes as a result of the
twist whereas the latter Equation~(\ref{eqn:krall_wrong}) relates the
coefficient of Equation~(\ref{eqn:chen:Bp}) before and after the
twist. \cite{Krall2000} substitute Equation~(\ref{eqn:chen:Bp}) into
effectively Equation~(\ref{eqn:tcurrent}) to produce a new current
\begin{equation}
\widetilde{I}_\toroidal=\frac{\ac\,c}{2}\,\widetilde{B}_{\poloidal\mathrm{c}}=\frac{\ac\,c}{2}\,\left(B_{\poloidal\mathrm{c}}+\frac{\delta\twist_0}{2\,\pi\,\arclength}\,\frac{\ac}{\mathcal{R}}\,B_{\toroidal\mathrm{c}}\right),
\end{equation}
where $\delta\vartheta_0=v_{\theta_0}\,t$.
However, the correct relationship Equation~(\ref{eqn:rotation}) implies that
$\widetilde{B}_\poloidal\left(a\right)=B_\poloidal\left(a\right)$ and
$\widetilde{I}_\toroidal=I_\toroidal=B_\poloidal\left(a\right)\,\ac\,c/2$
because $B_\toroidal\left(a\right)=0$, i.e., there is no change in net current
as a consequence of the twisting! The effect of twisting is to modify the
current distribution in the current channel and change the internal inductance
\begin{equation}
\widetilde{\xi}\equiv\frac{2\,\int_0^a{dr}\,{r}\widetilde{B}_\poloidal^2\left(r\right)}{a^2\,B_{\poloidal{a}}^2}.
\end{equation}
\bibliographystyle{apj}
\singlespace
\bibliography{bibliography}
\end{document}